\documentclass[12pt,preprint]{aastex}

\newcommand{\mbol} {M_{\rm bol}}
\newcommand{\msun} {$M_{\odot}$}

\newcommand{\lsun} {$L_{\odot}$}
\newcommand{\halpha} {H$\alpha$}

\newcommand{\Te} {T_{\rm eff}}
\newcommand{\logg} {\log g}
\newcommand{\nh} {\log{\rm H/He}}

\begin{document}


\title{Know Your Neighborhood: A Detailed Model Atmosphere Analysis of Nearby White Dwarfs}

\author{N. Giammichele\footnote{Visiting Astronomer, Kitt Peak National Observatory,
National Optical Astronomy Observatory, which is operated by the
Association of Universities for Research in Astronomy (AURA) under
cooperative agreement with the National Science Foundation.}, P.
Bergeron$^1$, \& P. Dufour} \affil{D\'epartement de Physique,
Universit\'e de Montr\'eal, C.P.~6128, Succ.~Centre-Ville,
Montr\'eal, Qu\'ebec H3C 3J7, Canada.}
\email{noemi.giammichele@astro.umontreal.ca, pierre.bergeron@astro.umontreal.ca,
patrick.dufour@astro.umontreal.ca}

\begin{abstract}

We present improved atmospheric parameters of nearby white dwarfs
lying within 20 pc of the Sun. The aim of the current study is to
obtain the best statistical model of the least-biased sample of the
white dwarf population. A homogeneous analysis of the local population
is performed combining detailed spectroscopic and photometric analyses
based on improved model atmosphere calculations for various spectral
types including DA, DB, DC, DQ, and DZ stars. The spectroscopic
technique is applied to all stars in our sample for which optical
spectra are available. Photometric energy distributions, when
available, are also combined to trigonometric parallax measurements to
derive effective temperatures, stellar radii, as well as atmospheric
compositions.  A revised catalog of white dwarfs in the solar
neighborhood is presented. We provide, for the first time, a
comprehensive analysis of the mass distribution and the chemical
distribution of white dwarf stars in a volume-limited sample.

\end{abstract}

\keywords{Solar neighborhood -- stars: luminosity function, mass function -- 
techniques: photometric -- techniques: spectroscopic -- white dwarfs}

\section{Introduction}

White dwarf stars represent a significant contribution to the global
stellar population and an important indicator of the evolutionary
history of the Galaxy. As such, it becomes crucial to characterize and
understand the white dwarf population as fully as possible. Mass
distribution, space density, and chemical composition are most
valuable pieces of information as we attempt to better constrain the
evolutionary history of these stars. Similarly, the white dwarf
luminosity function, defined as the number of stars as a function of
their intrinsic luminosity, becomes a valuable tool to narrow down the
age of the Galactic disk. But in order to take advantage of these
indicators, the white dwarf population sampled must be as close as
possible to being statistically complete.

The white dwarf population is composed mainly of low-luminosity stars
that are rather difficult to study as we move further away from the
Sun. Candidates are mainly discovered from either
proper-motion-limited or ultraviolet (UV) excess
surveys. Proper-motion surveys are characterized by the discovery of
high-proper motion stars through a comparison of identical fields
observed at two different epochs. By further combining these proper
motions with color indices, white dwarfs can be distinguished
successfully from other stellar populations. Despite its wide use in
the building of large surveys, important flaws remain.  Proper-motion
surveys naturally present a high kinematic bias. Moreover, high
density regions, such as the galactic plane, are usually avoided.

Additional contributions to our knowledge of the white dwarf
population were made through UV-excess surveys, a technique primarily
used among photometric surveys to identify hot white dwarf
candidates. Studies based on UV-excess, magnitude-limited surveys lead
to the building of the bright end of the luminosity function, like
those recently derived from the Palomar-Green (PG;
\citealt{liebert05,bergeron11}) and the Kiso surveys
(\citealt{lim10}). However, being restricted to the detection of only
bluer and thus hotter objects represents a significant bias, and does
not allow for a better understanding of the faint end of the
luminosity function. The goal of compiling a complete sample is thus
highly compromised.

Proper-motion and UV-excess surveys overlap very little and are both
incomplete in their own way. The one possibility for obtaining a less
biased sample is to use a complete volume-limited sample, centered
around the Sun. Such a local sample can provide an accurate
statistical model as long as the right balance between high
completeness and small number statistics is achieved. A precise
picture of this local sample can, if large enough, be extended to the
rest of the Galaxy to assess the importance of the white dwarf
population to the overall mass budget of the galactic disk.

Numerous studies were performed aiming to complete and to characterize
the sample of nearby white dwarfs. The first study dedicated to
building a complete census of the local sample of white dwarfs was
performed by \citet{holberg02}. The distance of 20 pc was chosen to
correspond to the volume of the NSTARS database, whose goal is to
compile information on all possible stellar sources near the Sun, in
order to achieve a better understanding of the local stellar
population. As \citet{holberg02} further discussed in their work, the
volume defined by this 20 pc limit was assumed to be reasonably
complete. The determination of the white dwarf candidates was entirely
based on photometric magnitudes collected from the Villanova White
Dwarf Catalog \citep[][hereafter WD Catalog]{mccook99}. Two main
selection criteria were used to determine white dwarfs within 20
pc. First, the WD Catalog was searched for objects with trigonometric
parallaxes $\pi \geq 0$\farcs$05$. For stars with no available
parallaxes, the cut was made with photometrically determined distances
based on $V-M_V \leq 1.505$.  Careful attention was paid to remove
manually obvious known anomalies.  As \citet{holberg02} reported, data
retrieved in this manner were far from uniform in quality and not
homogeneous in any way. A priority scheme was adopted to cope with the
different data sources, giving a higher priority to Johnson $B-V$,
Str\"{o}mgren $b-y$, and multichannel $g-r$ color indices, when
possible. The direct effect of this selection led to major
inhomogeneities in the calculations of absolute visual magnitudes and
resulting distances. The subsequent analysis, based on this local
sample, mainly focused on the calculation of the local space density
and the estimation of the completeness of the sample. Based on the
assumption that the 13 pc sample was entirely known, \citet{holberg02}
estimated the 20 pc sample to be only 65\% complete. Few details
showing the atmospheric properties of the white dwarfs in the solar
neighborhood were presented at that time.

The quest for completeness of the local white dwarf sample, based on
the 109 candidates determined by \citet{holberg02}, was pursued
through the contributions of \citet{ven03}, \citet{kaw04}, and
\citet{kaw06}, who surveyed the revised NLTT catalog of
\citet{salim03}. By using color-color and reduced proper motion
diagrams, as well as a spectroscopic follow-up of white dwarf
candidates, several stars were added to the original local sample,
while some others were removed. In particular, \citet{kaw06} extended
the search for possible candidates by spectroscopically identifying 8
new white dwarfs lying within 20 pc. Other contributions from
\citet{far05}, \citet{sub07}, and \citet{sub08} are also worth
mentioning in this effort.

\citet{holberg08} and \citet{sion09} reanalyzed the white dwarfs in
the solar neighborhood by updating the local sample of
\citet{holberg02} with the recent discoveries mentioned above, to form
a sample composed of 132 stars. \citet{holberg08} also compiled atmospheric
parameters of the local sample candidates, collected from numerous
sources, to calculate the mean mass of the sample, and used a
variety of spectroscopic, photometric, and trigonometric distances
to better estimate the local space density. \citet{sion09}
strictly focused on the kinematical properties and the
distribution of spectroscopic subtypes of the white dwarf
population within 20 pc.

The lack of consistency between the different model atmospheres and
analysis methods makes previous determinations of the ensemble
properties of the local white dwarf sample quite uncertain, and may
lead to erroneous estimates. Obviously, the quest for completeness of
the local sample is still a central preoccupation, but a proper
analysis of such a sample has been left aside, and to this day, there
has not been any systematic model atmosphere analysis of all available
data. It thus appears appropriate, at this time, to revisit the nearby
white dwarf population by performing a rigorous analysis, in an
homogeneous fashion, of every star in the sample.

In this paper, we present improved atmospheric parameters of all
possible nearby white dwarfs lying within 20 pc of the Sun. A
homogeneous and complete analysis of the local population is performed
combining detailed spectroscopic and photometric analyses based on
improved model atmosphere calculations for various spectral types
including DA, DC, DQ, and DZ stars. Our photometric and spectroscopic
observations are presented in Section \ref{obs}, while the theoretical
framework is discussed in Section \ref{theory}. The photometric and
spectroscopic data are then analyzed in Sections \ref{phot} and
\ref{spec}, respectively.  Selected astrophysical results from our
analysis, including the mass distribution and luminosity function, are
examined in Section \ref{resu}, and our conclusions follow in Section
\ref{conclusion}.

\section{Observational Data}\label{obs}

\subsection{Definition of the Local Sample}

The sample we analyze in this study is composed of spectroscopically
identified white dwarfs that lie in the solar neighborhood, within the
approximate limit defined at 20 pc. It is mostly drawn from the
complete list presented in \citet{sion09}, an updated version of the
local population defined by \citet{holberg02,holberg08}. As mentioned
earlier, significant additions to this initial sample have been made
by \citet{kaw04}, \citet{kaw06}, and \citet{sub07,sub08}, with some
contributions from other studies (see references in
\citealt{sion09}). We increased the sample size by taking into account
all possible white dwarfs that could lie within the uncertainties
inside the 20 pc region, which means including all objects from Table
4 of \cite{holberg08}. We also include the peculiar DQ star LHS 2229
(1008+290) since a new trigonometric parallax was made available to us
by H.~C. Harris. (2010, private communication) places the star inside
the 20 pc region, and the DC star LHS 1247 (0123$-$262) since its
distance, estimated from the photometric observations of
\citet[][hereafter BRL97]{BRL97}, places this candidate inside our
region of interest, within the uncertainties. 

We also include two DA stars from the spectroscopic analysis of
\citet{gianninas11}, L796-10 (0053$-$117) and GD 25 (0213+396), whose
distances are estimated to be inside the 20 pc region, within the
uncertainties; G138-31 (1625+093) also falls into this category
although the trigonometric parallax measurement available for this
star places it beyond 20 pc (note that we considered here only stars
that were hot enough for the spectroscopic method to be reliable,
$\Te\gtrsim6500$~K). There are also several DB stars in the spectroscopic
analysis of \citet{bergeron11} with distances within 20 pc, although 
all of these lie in a temperature regime where the
physics of line broadening becomes more questionable, van der Waals
broadening in particular. The closest DB star with reliable
atmospheric parameters lies at a distance of $\sim 30$ pc.

Finally, an initial sample of 169 white dwarf candidates has been
retained for this analysis. The complete list of objects is presented
in Table~\ref{tb:t1} where we give for each star the WD number from
the WD Catalog as well as an alternate name; whenever possible we used
the LHS or the Giclas names unless the object is better known under
another name in the literature. The additional entries for each object
are described in the next section.

\subsection{Spectroscopic Observations}

One of the original goals of this project was to characterize, as best
we could, the white dwarf population in the solar neighborhood. This
requires us to first provide a spectroscopic snapshot of this
population in the form of an atlas similar to that published by
\citet{wes93}. All in all, we managed to secure high signal-to-noise
ratio spectroscopic observations for 166 objects in our local sample
(the three objects missing are the Sirius-like systems 0208$-$510,
0415$-$594, and 1132$-$325). We describe these observations in turn.

Most of the blue spectra ($\lambda\sim3700-5200$ \AA) for the DA white
dwarfs were already available to us from of our numerous studies of
these stars \citep[see, e.g.,][]{BSL92,liebert05,gianninas11}, while
several spectra covering the region near H$\alpha$ --- required to
constrain the atmospheric composition of the coolest degenerates ---
were taken from the studies of BRL97 and \citet[][hereafter
BLR01]{BLR01}. Additional spectra were also acquired for the specific
purpose of this project.  For instance, blue spectra were secured
during several observing runs at the Steward Observatory 2.3 m
telescope equipped with the Boller $\&$ Chivens spectrograph and a
Loral CCD detector. The 4$\farcs$5 slit together with the 600 l
mm$^{-1}$ grating in first order provided a spectral coverage from
about 3200 to 5300 \AA\ at an intermediate resolution of $\sim 6$ \AA\
FWHM. Additional spectra covering the red portion of the spectrum were
acquired with the same setup but with the 400 l mm$^{-1}$ grating in
first order allowing a spectral coverage from about 3700 to 6900 \AA\
at a resolution of $\sim 9$ \AA\ FWHM. Similar spectra with a coverage
from 3800 to 6700 \AA\ were also secured at the Kitt Peak National
Observatory 2.1 m and 4 m telescopes equipped with the Goldcam and RC
spectrographs, respectively. Both used a 2$\farcs$0 slit with a
resolution of $\sim 6$ \AA\ FWHM, but different gratings of 316 l
mm$^{-1}$ and 500 l mm$^{-1}$, respectively. Details of our observing
and reduction procedures can be found in \citet{saf94}.

We also make use of several spectra already published in the
literature, and generously made available to us by the authors.  This
is the case for white dwarfs discovered in the ongoing survey by
J.~Subasavage: LHS 1243 (0121$-$429), LP 593-56 (0344+014), SCR
0821$-$6703 (0821$-$669), SCR 2012$-$5956 (2008$-$600), and L570-26
(2138$-$332) from \citet{sub07}; LP 50-73 (0011$-$721), L454-9
(0655$-$390), SCR 0708$-$6706 (0708$-$670), SCR 0753$-$2524
(0751$-$252), SCR 0818$-$3110 (0816$-$310), and SCR 1118$-$4721
(1116$-$470) from \citet{sub08}; LEHPM 2$-$220 (1009$-$184) from
\citet{sub09}; L40-116 (1315$-$781) from J.~P.~Subasavage (2010,
private communication). From the NLTT Survey: LHS 1421 (NLTT 8435,
0233$-$242), LP 522-46 (NLTT 56805, 2322+137), and G36-29 (NLTT 8581,
0236+259) from \citet{ven03}; LP 872-20 (NLTT 49985, 2048$-$250) from
\citet{kaw04}; LP 294-61 (NLTT 3915, 0108+277) from \citet{kaw06}. And
from other studies: LSR 1817+1328 (1814+134) from \citet{lepine03}; PM
J13420$-$3415 (1339$-$340) from \citet{lepine05}; SSPM J2231$-$7514
(2226$-$754) and SSPM J2231$-$7515 (2226$-$755) from \citet{scholz02};
LEHPM 1-4466 (2211$-$392) from \citet{oppenheimer01}; LHS 1008
(0000$-$345) from \citet{reimers96}.  Finally, optical spectra for LHS
2229 (1008+290) and GD 184 (1529+141) were taken from previous Data
Releases of the Sloan Digital Sky Survey.

The spectral type of each white dwarf in our nearby sample is reported
in Table~\ref{tb:t1}. In summary, this sample breaks down into the
following spectral types: 113 DA, 26 DC, 19 DQ, 10 DZ and 1 DBQA.
Strangely enough, there is not a single warm ($\Te>13,000$~K) DB star
in this nearby sample. As discussed above, according to the
spectroscopic analysis of relatively bright DB white dwarfs of
\citet{bergeron11}, the closest DB star lies at a distance of $\sim
30$ pc.

Figure \ref{fg:f1} presents the blue spectra of DA and DAZ stars
in our sample in order of decreasing effective temperature
(determined below in Section \ref{spec}). The spectrum of LHS 1660
(0419$-$487) is obviously contaminated by the presence of an M dwarf
companion.  GR 431 (0939+071; last object in Figure \ref{fg:f1}),
also known as PG 0939+072, is a problematic object. Classified DC7 in
the PG catalog, it was not included in the spectroscopic analysis of
the DA white dwarfs identified in the PG survey \citep{liebert05}, despite the
fact that it had been reclassified as DA2 in
\citet{holberg02}. However, it appeared again as DC7 in Table 4 of
\citet{holberg08}, a list of {\it possible} white dwarfs within 20
pc. Our spectrum shows that GR 431 is simply not a white dwarf star
(reclassified as a main sequence dF star by \citealt{gianninas11}),
and it is therefore excluded from our analysis. Our nearby sample thus
includes 168 genuine white dwarf stars, some of which are unresolved
double degenerate systems.

The blue spectra of DA and DAZ stars too cool to be analyzed using
line profile fitting techniques are displayed in Figure
\ref{fg:f2}. In some cases, these objects are completely
featureless in the spectral region shown here, and the presence of
hydrogen can only be inferred by the detection of H$\alpha$. DAZ stars
in Figures \ref{fg:f1} and \ref{fg:f2} can be easily
recognized by the presence of the Ca~\textsc{ii} H and K lines, the
most notable in this sample being GD 362 (1729+371;
Figure~\ref{fg:f1}) whose spectrum also shows spectral lines from
Ca~\textsc{i}, Mg~\textsc{i}, and Fe~\textsc{i} \citep{gianninas04};
high-resolution spectroscopy actually reveals the presence of 17
different elements, including helium \citep{zuc07}. Note that our blue
spectrum of LP 294-61 (0108+277; Figure~\ref{fg:f2}) does not
reveal any metallic feature and we have thus reclassified this star as
DA (see also Section 3.6 of \citealt{far09}).

The DA (and DZA) stars in our sample for which spectra at H$\alpha$
are also available are presented in Figure \ref{fg:f3} as
a function of decreasing equivalent widths. The presence of H$\alpha$
in the coolest white dwarfs is crucial to better constrain the
atmospheric parameters using the photometric method described in
Section \ref{phot}; the coolest DA stars in which H$\alpha$ can be
detected in our sample have photometric temperatures around $\Te\sim
5000$~K. The Zeeman triplet is clearly visible in five magnetic DA
stars displayed in the right panel of Figure
\ref{fg:f3}. \citet{putney97} classified G234-4 (0728+642)
as DAP based on polarization data at H$\alpha$, which indicated a 3
$\sigma$ detection with $B_e=+39.6\pm11.6$ kG, and a suggestion of
Zeeman features. Our spectrum of G234-4 shows no evidence of magnetic
splitting due to the lower spectral resolution of our data.

Our set of DC white dwarfs are displayed in Figure \ref{fg:f4} in
order of right ascension. These have featureless spectra, even at
H$\alpha$. The absence of any absorption feature implies that they are
either cool ($\Te\lesssim10,000$~K) helium-atmosphere white dwarfs, or
hydrogen-atmosphere white dwarfs too cool to show H$\alpha$
($\Te\lesssim5000$~K). Only a detailed photometric analysis of these
stars can resolve this ambiguity. LHS 1008 (0000$-$345), classified DC
by BRL97 but DAH by \citet{reimers96}, is not displayed here and is
further discussed below.

Spectroscopic data of normal DQ stars in our sample are displayed in
Figure \ref{fg:f5}. Because of the spectral classification scheme
devised by \citet{mccook99}, DQ white dwarfs may have carbon features
detectable only in the ultraviolet. The spectrum of L97-3
(0806$-$661), for instance, appears featureless in the optical.
G47-18 (0856+331) is a unique DQ star (excluding those discovered in the
SDSS) that shows both C$_2$ Swan bands and C~\textsc{i} atomic
lines. Note also the presence of the CH G-band ($\sim4300$ \AA) in the spectrum of G99-37
(0548$-$001) and BPM 27606 (2154$-$512), the only two such stars
known; both stars are also magnetic and show circular polarization in
the CH band \citep[][and references therein]{vornanen10}.  BPM 27606
also turns out to be one of the normal DQ stars with the strongest
C$_2$ Swan bands known.  In addition to a very weak C$_2$ absorption
feature, the spectrum of GD 184 (1529+141) also shows narrow hydrogen
lines (see also Figure~\ref{fg:f3}), up to H$\gamma$, which would make
this star a unique DAQ white dwarf. Our detailed analysis of this
object, described below, reveals instead that GD 184 is in fact a DA +
DQ double degenerate system.

Peculiar DQ stars with shifted C$_2$ Swan bands are displayed in
Figure \ref{fg:f6}; a normal DQ star is also reproduced at the
top of the figure for comparison. Note how the molecular bands in the
other objects appear shifted and more symmetrical with respect to this
normal DQ star, with the extreme case of LHS 2229 (1008+290) displayed
at the bottom. This phenomenon has recently been explained by
\citet{kowalski10} as a result of pressure shifts of the carbon bands
that occurs in cooler, helium-dominated atmospheres. These stars are
now being classified as DQpec (rather than C$_2$H stars). The presence
of a very strong magnetic field has also been reported in the bottom
two objects (see, e.g., \citealt{schmidt99}).

Our DZ spectra, showing the presence of metal lines, mainly the
Ca~\textsc{ii} H \& K doublet, are displayed in Figure
\ref{fg:f7}. L745-46A (0738$-$172), GD 95 (0843+358), and Ross 640 (1626+368)
are actually DZA stars with very shallow H$\alpha$ absorption lines
(shown in Figure~\ref{fg:f3}), resulting from the presence
of a trace of hydrogen in a helium-dominated atmosphere.  LP 701-29
(2251$-$070) is a heavily blanketed DZ star, the only known case where
Ca~\textsc{i} $\lambda$4226 appears stronger than the Ca II doublet.

Finally, additional spectra of miscellaneous white dwarfs are
displayed in Figure \ref{fg:f8}. These include LDS 678A
(1917$-$077), the only cool and relatively bright DBQA star known
(carbon is observed only in the UV), GW+70 8247 (1900+705), a heavily
magnetic DA white dwarf, and G240-72 (1748+708) whose spectrum is
characterized by a deep yellow sag of unknown origin in the $4400-6300$
\AA\ region (see also \citealt{wes93}).  Our spectrum of LDS 678A
shows a significant absorption feature at H$\alpha$, although
another spectrum taken several years earlier showed only a hint of
H$\alpha$, which might indicate some spectroscopic variability; the
only other features present in the spectrum are weak neutral helium
lines (the feature observed at $\sim 6300$ \AA\ is a night sky feature). Also
shown in Figure \ref{fg:f8} are two spectra of LHS 1008
(0000$-$345), classified DC by BRL97 based on the top spectrum
displayed here. However, spectroscopic observations analyzed by
\citet{reimers96}, and reproduced at the bottom
of Figure \ref{fg:f8}, clearly indicate the presence of magnetic
features, suggesting that LHS 1008 might be variable. Reimers et al.~(see their Figure 6) actually managed to reproduce
the absorption feature near 4600 \AA\ with a magnetic offset dipole, hydrogen-atmosphere model, and a polar field
strength of 86 MG. Finally, G227-35 (1829+547) is another magnetic white dwarf
with strong magnetic polarization
\citep{angel75}. \citet{cohen93} also reported a prominent feature
near 7450 \AA\ attributed to the stationary point in a transition of
H$\alpha$. In the wavelength range displayed in Figure
\ref{fg:f8}, however, the spectrum of G227-35 appears
featureless, with the exception perhaps of a weak absorption feature
near 6000 \AA, which is actually predicted in the models for this star
shown in Figure 7 of \citet{putney95}.

A final quick glance at the spectra shown in the figures above reveals
that the local population of white dwarf stars includes some of the
strangest and more unique objects we know. We definitely live in a strange
neighborhood!

\subsection{Photometric Observations}

Optical $BVRI$ photometry was retrieved from the studies of BRL97 and
BLR01 for 82 cool white dwarfs in our sample. Additional
$VRI$ photometry for 21 objects was taken from the various studies of
nearby white dwarfs in the southern hemisphere by J.~Subasavage. For
other white dwarfs in our sample with no available $(B)VRI$ photometry,
we relied instead on Str\"{o}mgren $ubvy$ photometry (3 objects), SDSS
$ugriz$ photometry (7 objects), multichannel data (1 object), or additional
sources given in Table 1.  In order to characterize the complete
energy distribution of cool white dwarfs, the optical photometry was
combined with infrared $JHK$ photometry from BRL97 and BLR01 (75
objects), $JHK_S$ photometry extracted from the online version of the
Two Micron All Sky Survey (2MASS) survey (39 objects), or $JHK$
photometry on the MKO system taken from
\citet[][2 objects]{kilic06}. For LHS 2229 (1008+290), we used the $BVIJHK$
photometry from \citet{schmidt99}.

Our adopted optical and infrared photometric data are reported in
Table \ref{tb:t1} together with references given in the last column; a
special column also indicates which optical photometric set is
used. For stars analyzed only spectroscopically, we also provide $V$
magnitudes taken mostly from the online version of the WD Catalog, which will
serve to estimate spectroscopic distances. Photometric uncertainties
are taken from the appropriate references, with the exception of the
$V$ magnitudes retrieved from the WD Catalog, for which we
assume 5\% for simplicity.

The ($V-I$, $V-K$) two-color diagram is displayed in Figure
\ref{fg:f9} for 99 white dwarfs in our sample with these color indices
available. DA and non-DA
stars are represented by filled and open circles, respectively.  Also
shown are the predictions from pure hydrogen and pure helium cooling
sequences at $\logg = 8.0$ (described in Section \ref{theory}). DA and non-DA
stars form two, nearly distinct sequences in this diagram, which
follow closely the behavior of the model sequences at higher
temperatures, although both sequences are shown to cross at lower
temperatures. Hence, the atmospheric composition of the coolest non-DA
stars in this diagram cannot be interpreted easily. For instance, ER 8
(1310$-$472; labeled in the figure) is the coolest and oldest white
dwarf identified in the analysis of BRL97, and it has a pure hydrogen
atmospheric composition despite its DC nature. Worth mentioning here
is the strong deficiency of non-DA stars in a particular range of
$V-K$ colors between $\sim$1.2 and 1.7, which corresponds to the
so-called non-DA gap first discussed by BRL97. The only non-DA star
lying in this interval is L40-116 (1315$-$781; labeled in the figure);
this object will be further discussed in Section \ref{phothhe}.
There are also several outliers identified in this
diagram, all of the non-DA type: LHS 2229 (1008+290) and LP 790-29
(1036$-$204) are DQpec white dwarfs with strong shifted C$_2$ Swan
bands (see Figure \ref{fg:f6}) that affect the colors; LHS 1126 (0038$-$226) and SCR
2012$-$5956 (2008$-$600) both show a strong infrared flux deficiency
that has been interpreted in terms of the H$_2$-He collision-induced
absorptions in a mixed H/He atmosphere \citep[see,
e.g.,][]{bergeron94}; G195-19 (0912$+$536) is a 
magnetic white dwarf with a $\sim 100$ MG field that probably affects the $I$ passband
(see our photometric fit below).

\subsection{Trigonometric Parallax Measurements}

BRL97 found that even though the model energy distributions are
somewhat sensitive to surface gravity, it is practically impossible to
determine $\logg$ from the observed photometry alone. Only for stars
with available trigonometric parallax measurements is it possible to
determine the stellar radius, and thus the mass through the
mass-radius relation. In our sample, 125 stars have trigonometric
parallax measurements taken mostly from the Yale Parallax Catalog
\citep[][hereafter YPC]{ypc} and the Hipparcos Catalog
(\citealt{perry97}). The parallax values and corresponding
uncertainties are reported in Table \ref{tb:t1} together with the
appropriate reference.

The $M_{V}$ versus $(V-I)$ color-magnitude diagram obtained using
these trigonometric parallaxes is displayed in Figure
\ref{fg:f10} for 85 stars in our sample with available $V$ and
$I$ colors. Again, white dwarfs are distinguished in terms of their DA
or non-DA spectral types, and the predictions from pure hydrogen and
pure helium cooling sequences at $\logg = 8.0$ are superimposed on the
observed data; note that the data already suggest a slightly higher mean $\logg$ value
for this sample. As mentioned by BLR01, DA and non-DA stars form
well-defined narrow sequences in this diagram, although not as narrow
as those observed in the previous figure, most likely because the
trigonometric parallax measurements come from inhomogeneous parallax
samples. Also, all overluminous white dwarfs hot enough for H$\alpha$
to be detected are of the DA spectral type. As discussed in BRL97,
most, if not all, of these objects are unresolved binaries --- e.g.,
L870-2 (0135$-$052) --- and their luminosity reflects the contribution of
two white dwarfs with probably average masses.

\section{Theoretical Framework}\label{theory}

Our model atmospheres and synthetic spectra are derived from the LTE
model atmosphere code originally described in \citet{bergeron95a} and
references therein, with recent improvements discussed in
\citet{TB09}. In particular, we now rely on their improved
calculations for the Stark broadening of hydrogen lines with the
inclusion of nonideal perturbations from protons and electrons ---
described within the occupation probability formalism of \citet{hum88}
--- directly inside the line profile calculations.  Convective energy
transport is taken into account following the revised ML2/$\alpha =
0.8$ prescription of the mixing-length theory (see
\citealt{bergeron95b} and \citealt{TB09} for details).  Non-LTE
effects are also included at higher effective temperatures but these
are totally irrelevant for the purpose of this work. More details
regarding our helium-atmosphere models are provided in
\citet{bergeron95a,bergeron11}.

Our model grid covers a range of effective temperature between $\Te=
1500$ K and 45,000 K in steps of 500 K for $\Te< 15,000$ K, 1000 K up
to $\Te= 18,000$ K, 2000 K up to $\Te=30,000$ K, and by steps of 5000
K above. The $\logg$ ranges from 6.5 to 9.5 by steps of 0.5 dex, with
additional models at $\logg=7.75$ and 8.25.  Additional models, in
particular for cool stars, have been calculated with mixed hydrogen
and helium compositions of $\nh = -1.0$ to 3.0 (by steps of 0.5).

Synthetic colors are then obtained using the procedure outlined in
\citet{holberg06} based on the Vega fluxes taken from
\citet{bohlin04}. Since the photometric technique described below
relies heavily on the flux at the $B$ bandpass, we now include in our
models the opacity from the red wing of Ly$\alpha$ calculated by
\citet{kowalski06} and kindly provided to us by P.~Kowalski, which is
known to affect the flux in the ultraviolet region of the energy
distribution. These improved synthetic colors were used in Figures
\ref{fg:f9} and \ref{fg:f10}, and are available from our
Website\footnote{See http://www.astro.umontreal.ca/\~{
  }bergeron/CoolingModels.}.

The photometric analysis of DQ and DZ white dwarfs requires improved
models since strong carbon or other metallic features may not be the
only elements that affect the flux in some photometric bands.  Indeed,
the presence of heavier elements in helium-rich atmospheres may also
provide enough free electrons to affect the atmospheric structure
significantly, and thus the predicted energy distributions. To
circumvent these problems, we rely on the LTE model atmosphere
calculations developed by \citet{duf05,duf07} for the study of DQ and
DZ stars, respectively, based on a modified version of the code
described in \citet{bergeron95a}. The main addition to the models is
the inclusion of metals and molecules in the equation of state and
opacity calculations.

\section{Photometric Analysis}\label{phot}

\subsection{General Procedure}

Atmospheric parameters, $\Te$ and $\logg$, and chemical compositions
of cool white dwarfs can be measured accurately using the photometric
technique developed by BRL97.  We first convert optical and infrared
photometric measurements into observed fluxes and compare the
resulting energy distributions with those predicted from our model
atmosphere calculations. To accomplish this task, we first transform
every magnitude $m$ into an average flux $f^{m}_{\lambda}$ using the
equation

\begin{equation}
m=-2.5\log f^{m}_{\lambda} + c_m\ , \label{eq:1}
\end{equation}

\noindent where

\begin{equation}
f^{m}_{\lambda}=\frac{\int_{0}^{\infty}f_{\lambda}S_m(\lambda)\lambda\,d\lambda
}{\int_{0}^{\infty}S_m(\lambda)\lambda\,d\lambda}\label{eq:2}
\end{equation}

\noindent and where $S_m(\lambda)$ is the transmission function of
the corresponding bandpass, $f_{\lambda}$ is the monochromatic flux
from the star received at Earth, and $c_m$ is a constant to be
determined. The transmission functions in the optical for the $BVRI$,
$ugriz$, and Str\"{o}mgren photometric systems are described in
\citet{holberg06} and references therein, while the multichannel
photometric system is discussed in \citet{green76}. Note that the
$ugriz$ and multichannel photometry are defined on the AB magnitude
system, which requires a slightly different definition of the above
equations (see equation 3 of \citealt{holberg06} for instance). The
transmission functions for the $JHK$ or $JHK_S$ filters on the
Johnson-Glass, 2MASS, and Mauna Kea Observatories (MKO) photometric
systems are taken respectively from \citet{bes88}, \citet{cohen03},
and \citet{MKO}.  Since the $JHK$ magnitudes in Table 1 from BRL97 and
BLR01 are on the CIT system, they first need to be transformed on the
Johnson-Glass system using the equations given by
\citet{leg92}. 

The constants $c_m$ in equation (1) for each passband are determined using the
improved calibration fluxes from \citet{holberg06}, defined with the
{\it Hubble Space Telescope} absolute flux scale of Vega
\citep{bohlin04}, and appropriate magnitudes on a given system. Note that the Str\"{o}mgren $u$ magnitude
in Table 1 of \citet{holberg06} is erroneous and should read $1.437$
instead of $1.357$ \citep{hauck98}, which yields a corrected zero
point of $c_u=-19.80809$ (instead of $-19.8882$).

For each star in Table \ref{tb:t1}, a minimum set of four average
fluxes $f^{m}_\lambda$ is obtained, which can be compared with model
fluxes. Since the observed fluxes correspond to averages over given
bandpasses, the monochromatic fluxes from the model atmospheres need
to be converted into \textit{average fluxes} as well,
$H^{m}_{\lambda}$, by substituting $f_{\lambda}$ in equation
\ref{eq:2} for the monochromatic Eddington flux $H_{\lambda}$. We
can then relate the average observed fluxes $f^{m}_{\lambda}$ and
the average model fluxes $H^{m}_{\lambda}$ --- which depend on
$\Te$, $\logg$, and chemical composition --- by the equation

\begin{equation}
f^{m}_{\lambda} = 4\pi(R/D)^2H^{m}_{\lambda} \label{eq:3}
\end{equation}

\noindent where $R/D$ defines the ratio of the radius of the star
to its distance from Earth. We then minimize the $\chi^2$ value
defined in terms of the difference between observed and model
fluxes over all bandpasses, properly weighted by the photometric
uncertainties.  Our minimization procedure relies on the nonlinear
least-squares method of Levenberg-Marquardt \citep{press86}, which
is based on a steepest decent method. Only $\Te$ and the solid
angle $\pi (R/D)^2$ are considered free parameters, while the
uncertainties of both parameters are obtained directly from the
covariance matrix of the fit. 

For stars with known trigonometric parallax measurements, we first
assume a value of $\log g=8$ and determine the effective temperature
and the solid angle, which combined with the distance $D$ obtained
from the trigonometric parallax measurement, yields directly the radius
of the star $R$.  The radius is then converted into mass using
evolutionary models similar to those described in \citet{fon01} but
with C/O cores, $q({\rm He})\equiv \log M_{\rm He}/M_{\star}=10^{-2}$
and $q({\rm H})=10^{-4}$, which are representative of
hydrogen-atmosphere white dwarfs, and $q({\rm He})=10^{-2}$ and
$q({\rm H})=10^{-10}$, which are representative of helium-atmosphere
white dwarfs.  In general, the log $g$ value obtained from the
inferred mass and radius ($g=GM/R^2$) will be different from our
initial guess of $\log g=8$, and the fitting procedure is thus
repeated until an internal consistency in log $g$ is reached.  For
white dwarfs with no parallax measurement, we simply assume a value of
$\logg=8.0$.

\subsection{Analysis with Hydrogen- and Helium-Atmosphere Models}\label{phothhe}

We first perform a detailed analysis of all the objects in our sample
with available photometry using hydrogen- and helium-atmosphere
models; a more detailed analysis of the DQ and DZ stars in this sample
will be presented in the following section.  We exclude LHS 1660
(0419$-$487) from this analysis because of a bright, nearby star
that contaminates the photometric observations, and we simply
rely on the spectroscopic solution for this object.

Our results for the analysis of the optical and infrared photometry
are presented in Figure \ref{fg:f11}.  Average observed fluxes
in the left panels are represented by error bars, while the
corresponding model fluxes are shown as open or filled circles,
depending on the atmospheric composition. The photometric bandpasses
used in the fitting procedure and the atmospheric parameters of each
solution are indicated in each panel. On the right panels are shown
the spectroscopic observations near H$\alpha$ compared to the model
predictions {\it assuming the pure hydrogen solution} (the spectral
type is also given); these only serve as an internal check of our
photometric solutions and are not used in the fitting procedure. For
instance, cases where an H$\alpha$ absorption feature is predicted but
is {\it not} observed clearly suggest that the pure helium solution is
more appropriate. In cases where the star is too cool to show
H$\alpha$ ($\Te\lesssim5000$~K), however, one has to rely on the
predicted energy distributions to decide which atmospheric composition
best fit the photometric data. Based on our inspection of these fits,
we adopt the solutions shown in red in the left panels. In general,
the fits to the energy distributions, and the internal consistency
with the presence or absence of H$\alpha$, are excellent. Note that
the absorption feature seen in G47-18 (0856+331) is a neutral carbon
line and not H$\alpha$. We discuss some of these photometric fits in turn.

LHS 1008 (0000$-$345) and G83-10 (0423+120) belong to this strange
class of objects, first identified by BRL97, whose energy
distributions are better fit with pure hydrogen models (or something
intermediate between hydrogen and helium), while their spectra are
featureless near the H$\alpha$ region. Note how the photometric
fits for these two objects are qualitatively identical. BRL97 discuss some exotic scenarios to
account for this behavior, although none have been successful so far
\citep[see, e.g.,][]{malo99}. The analysis of \citet{reimers96}
suggests, however, that the feature observed in their spectrum of LHS
1008 (reproduced at the bottom of our Figure \ref{fg:f8}) is
due to hydrogen in the presence of a high magnetic field ($\sim 86$
MG). The presence of a strong magnetic field could thus provide a
natural explanation for the discrepancy observed here, and magnetic
models are probably required to reproduce the photometry more
adequately. If our interpretation is correct, G83-10 could perhaps
harbor a strong magnetic field, although 
the spectropolarimetric survey of DC stars by \citet{putney97}
reports a null result for this star.

There are five weakly magnetic white dwarfs displayed in Figure
\ref{fg:f11} that exhibit the Zeeman triplet: LHS 1044 (0011$-$134),
LHS 1243 (0121$-$429), LHS 1734 (0503$-$174), G99-47 (0553+053), and
G256$-7$ (1309+853). The predicted H$\alpha$ profiles shown here do
not include the magnetic field, however; synthetic spectra with offset
dipole models are illustrated in \citet{brl92} for instance. In all
these cases, the presence of a weak magnetic field does not appear to
affect significantly the energy distributions, which are well
reproduced by our pure hydrogen models. The same conclusion applies to
LHS 1038 (0009+051), one of the first white dwarfs discovered with a
magnetic field below 100 kG \citep{schmidt94}, which manifests itself
here as a small discrepancy in the core of H$\alpha$. Other magnetic
white dwarfs in our sample include G240-72 (1748+708) and G227-35
(1829+547), both classified as DXP stars, with broad, unidentified
absorption features.  Since we do not have appropriate models for
these stars, we simply assume a pure helium composition and we also
neglect the $V$ bandpass for G240-72, which is affected by the deep
yellow sag observed in Figure \ref{fg:f8}.

LHS 1126 (0038$-$226) and SCR 2012$-$5956 (2008$-$600) are the only
two objects in our sample that show an infrared flux deficiency (see
also Figure \ref{fg:f9}). As explained above, this flux deficiency is
the result of the collision-induced absorption by molecular hydrogen
due to collisions with {\it neutral helium}, and mixed compositions
of $\log {\rm H}/{\rm He}\sim-1.3$ and $-3$, respectively, are
required to reproduce their energy distributions. Interestingly
enough, both stars appear to have the same $\log g$ values, and thus stellar masses
($\sim 0.44$ \msun).

Our pure helium models obviously fail to reproduce the flux in the $B$
bandpass of the strongest DZ stars in our sample, due to the strong
Ca~\textsc{ii} H \& K absorption lines, or other absorption
features in the blue. This is the case for vMa 2 (0046+051),
G139-13 (1705+030), and LP 701-29 (2251$-$070). For these stars in
particular, we have simply neglected the $B$ magnitude in our fitting
procedure. Note the higher $\log g$ values (up to $\sim 8.4$) measured for
these stars based on our pure helium models. A similar conclusion
applies to the normal DQ stars --- for instance LHS 1227 (0115+159),
Wolf 219 (0341+182), L879-14 (0435$-$088), although in most cases, the
C$_2$ Swan bands are not strong enough to affect significantly the
flux in any particular bandpass. The presence of additional free
electrons is affecting the atmospheric structure, however. Note again
the high $\log g$ values measured for the DQ stars based on our pure
helium models.  DZ and DQ white dwarfs will be analyzed in greater
detail in the next section using more appropriate models.

Four objects in our sample have only photographic magnitudes in the
optical --- LP 294-61 (0108+277), G36-29 (0236+259), LP 872-20
(2048$-$250), and LP 872-20 (2215+368) --- and the fits to their
energy distributions are not particularly good in this spectral
region, although the match at H$\alpha$ is excellent, with the glaring
exception of LP 294-61, which shows a much weaker H$\alpha$ absorption
feature than predicted. \citet[][see their Section 3.6]{far09}
describes this object as an optical pair of stars composed of a white
dwarf and a background red dwarf. The two components are still well
separated, however, and it is therefore unlikely that the optical and/or
infrared magnitudes of the white dwarf are contaminated by the red
dwarf. Actually, experiments with only optical or infrared magnitude
sets yield the same value of $\Te$. Most likely, LP 294-61 is an
unresolved double degenerate DA + DC system, but given the problems
discussed above, the nature of this object remains uncertain.

About 14 white dwarfs in Figure \ref{fg:f11} have 
$\logg$ values and corresponding masses too low to have evolved as
single stars within the lifetime of our Galaxy ($M\lesssim0.47$
\msun). Since with the photometric technique, it is really the radius
of the star that is being measured (see equation 3), the most obvious
explanation is that these objects are in fact unresolved double
degenerates, with two stars contributing to the total
luminosity. Indeed, some are well confirmed double degenerate systems:
G1-45 (0101+048), L870-2 (0135$-$052), and L587-77A (0326$-$273).  For
this last object, the binary nature can also be inferred from the
discrepant fit at H$\alpha$. Such discrepancies are also observed for
L532-81 (0839$-$327) and G187-8 (2048+263), although the binary nature
of these stars has not yet been confirmed. Further insights into the
nature of these low mass white dwarfs will be gained by combining the
photometric analysis with the results obtained from spectroscopy
(see Section \ref{mdistr}).

For L745-46A (0738$-$172), GD 95 (0843+358), and
Ross 640 (1626+368), we simply assumed for the sake of simplicity a
pure helium composition, although these stars are DZA white dwarfs
with a small trace of hydrogen on the order of ${\rm
H/He}\sim10^{-4}-10^{-3}$. H$\alpha$ in these cases is heavily
broadened through van der Waals interactions in a helium dominated
environment (see Figure 12 of BLR01). These two objects will be
analyzed in greater detail in the next section.

\citet{kowalski06} suggested that most, if not all, cool DC stars
probably have hydrogen-rich atmospheres, based on the analysis of the
BRL97 and BLR01 photometry with their improved model atmospheres that
include the previously missing red wing opacity from Ly$\alpha$. This
does not necessarily imply that all cool white dwarfs have hydrogen
atmospheres, however, since cool DZ stars, such as LP 701-29
(2251$-$070), obviously have helium-dominated atmospheres. We further
examine the suggestion of Kowalski \& Saumon below. We have 20 non-DA
stars in our sample, most of which appear in Figure \ref{fg:f9}; the
only two stars missing are LP 207-50 (0749+426; $ugriz$ data) and LP
287-39 (2215+368; $V$ magnitude only).  Four of these are strong DZ,
DQpec, or even DXP stars.  Of the remaining 16 objects, 9 are indeed
better fitted with pure hydrogen models. Worth mentioning in this
category is L40-116 (1315$-$781), the only non-DA stars in the
so-called ``non-DA gap'' observed in Figure \ref{fg:f9} and discussed
above. The predicted H$\alpha$ profile would suggest that this star
has a helium-atmosphere, but the spectrum is rather noisy, and we can
even catch a glimpse of Zeeman splitting at H$\alpha$.  It is
therefore quite possible that L40-116 is a magnetic DA white dwarf;
improved spectroscopy at higher signal-to-noise ratio would be
required for this star.

The remaining 7 objects were {\it originally} classified in our
analysis as helium-atmosphere white dwarfs. However, a new spectrum of
LHS 1421 (0233$-$242) obtained by Kawka, Vennes, Arazimova, \& Nemeth
(2011, in preparation; private communication by S. Vennes) reveals
that this star is actually a magnetic DA star with Zeeman splitting
observed at H$\alpha$; we thus decided to adopt instead the hydrogen
solution for this object. Also, SCR 0708$-$6706 (0708$-$670) has a bad
photometric fit, and improved photometry could perhaps change our
conclusion about the atmospheric composition of this star.  This is
also the case for LP 593-56 (0344+014), for which the optical
photometry is better fitted with the hydrogen solution while the
infrared data favor the helium solution. There is even a hint of
H$\alpha$ in the spectrum. We have thus decided to switch to the
hydrogen solution for this object as well. Similarly for LP 287-39
(2215+368), for which we have only a $V$ magnitude in the optical. The
pure hydrogen solution could be equally as good, and again, there is
even a hint of Zeeman splitting at H$\alpha$; we thus switched to the
hydrogen solution for this star. vB 3 (0743$-$336) and LHS 378
(1444$-$174) had ambiguous fits (not shown here) when using the $JHK$
photometry taken from BRL97 and BLR01, but the 2MASS $JHK_S$
photometry available for these stars clearly improved our fits using
pure hydrogen models.  We thus adopted the 2MASS photometry and the
pure hydrogen solutions for these 2 objects. Finally, the weak DZ star
LP 658-2 (0552$-$041) is better fitted with a helium-atmosphere model,
although the fit is admittedly not perfect; the use of 2MASS
photometry did not improve our fit either. However, the sharp calcium
lines observed in Figure \ref{fg:f7} clearly indicate that this star
has a hydrogen atmosphere, otherwise our calculations (not shown here)
predict that the calcium line profiles are much shallower and broader
than those observed here (contrast LP 658-2 with the other DZ stars
displayed in Figure \ref{fg:f7}). Again, we switched to the hydrogen
solution for this object. Based on the previous discussion, we are
thus inclined to agree with the conclusions of \citet{kowalski06} that
most, if not all, cool DC stars probably have hydrogen-rich
atmospheres.

\subsection{Photometric Analysis of DQ and DZ Stars}

Even though the photometric fits to the DQ and DZ white dwarfs
discussed in the previous section appear reasonable, they do require
improved model atmospheres when analyzed with the photometric
technique. As discussed above, the presence of heavier elements in
helium-rich models provides enough free electrons to affect the
atmospheric structure significantly, and thus the predicted energy
distributions.  

\subsubsection{DQ Stars}

The photometric technique used to fit the energy distributions of DQ
stars is described in \citet{duf05}, and is similar to that
described above, with the exception that a third fitting parameter,
the carbon abundance, is also taken into account. Spectroscopic
observations in the optical (see Figure
\ref{fg:f5}) are used to determine the carbon abundance by
fitting the C$_2$ Swan bands at the values of $\Te$ and $\logg$
obtained from a first fit to the energy distribution with an arbitrary
carbon abundance. This improved carbon abundance is then used to
obtain new estimates of the atmospheric parameters from the energy
distribution, and so forth. This iterative procedure is repeated until
$\Te$, $\logg$, and the carbon abundance converge to a consistent
photometric and spectroscopic solution. Two objects in our sample have
carbon features detectable only in the ultraviolet, L97-3 (0806$-$661)
and LDS 678A (1917$-$077).  For these two stars, we replace the
optical spectra in our fitting procedure with low-dispersion {\it IUE}
({\it International Ultraviolet Explorer}) spectra taken from
\citet{holberg03}.

Since our models do not yet include the opacity from the
pressure-shifted C$_2$ Swan bands required to properly analyze the
peculiar DQ stars (DQpec; Figure \ref{fg:f6}), we adopt the results
based on our previous analysis with pure helium models, with the
exception of LHS~290 (1043$-$188) and G225-68 (1633+572), for which we
simply fix the carbon abundance at $\log {\rm C}/{\rm He}=-7.0$ and
$-7.5$, respectively.  The corresponding model spectra reproduce the
depth of the C$_2$ molecular bands fairly well, and the fact that the
bands do not appear at the proper wavelengths is less crucial than
obtaining the correct amount of free electrons in our models. We were
unable to achieve similar qualitative fits for the other peculiar DQ
white dwarfs in our sample, and this is why we relied on our previous
analysis with pure helium models for these stars. Note that for LHS
2229 (1008+290), we neglected the $B$ and $V$ photometric
measurements, which are affected by the strong carbon features present
in this star. A similar problem arises for LP 790-29 (1036$-$204), for
which the strong molecular bands make it extremely difficult to model,
even approximately.

The results for the DQ stars in our sample are presented in Figure
\ref{fg:f12}. As before, observed fluxes are represented by error
bars, while model fluxes are shown as filled circles. The
spectroscopic observations used to constrain the carbon abundance in
the fitting procedure are displayed in the right panels. Overall, the
photometric energy distributions and the optical or UV spectra are
well reproduced by our models for the normal DQ stars, with the
exception of BPM 27606 (2154$-$512), for which the predicted depth of
the C$_2$ Swan bands is a bit too shallow, because we give a higher
weight to the photometry in our fitting procedure.  For the DBQA star
LDS 678A (1917$-$077), we measure a hydrogen abundance of $\log {\rm
  H/He}\sim-5$ based on our fit (not shown here) to the weak H$\alpha$
absorption feature (see Figures \ref{fg:f8} and \ref{fg:f11}).  Since
the photometric fits discussed here assume ${\rm H/He}=0$, we used a
smaller model grid with $\log {\rm H/He}=-5$ and found atmospheric
parameters that are almost identical to those obtained here without
hydrogen.

As discussed by \citet{duf05}, even though the quality of the
photometric fits displayed in Figure \ref{fg:f12} have not
improved with respect to the results obtained with pure helium models
(i.e., Figure \ref{fg:f11}), the values of $\Te$ and $\logg$
have been significantly reduced ($\sim 0.3$ dex in $\logg$) when
carbon is properly included in the model atmospheres.

GD 184 (1529+141; NLTT 40489), an extremely rare DAQ white dwarf,
was discovered by \cite{kaw06} who
assigned a temperature of $\Te=5250$~K based on a spectroscopic fit to
the weak hydrogen lines (no carbon feature had been reported,
however). Our photometric fit based on pure hydrogen models suggests,
on the other hand, a much higher temperature around $\Te\sim8900$~K,
although in this case, the predicted H$\alpha$ profile is much too
strong, as if the absorption line was diluted by a featureless DC
companion. Given the simultaneous presence of weak C$_2$ absorption
features in the optical spectrum, the simplest explanation is that GD
184 is an unresolved double degenerate system composed of a cool DA
star and a much hotter DQ star. This interpretation can be tested
further by attempting to fit simultaneously the photometric and
spectroscopic measurements with composite models.  Given the
photometric and spectroscopic constraints as well as the large number
of fitting parameters, we simply assume $\logg=8$ for both components
of the system, and only allow the temperature and the carbon abundance
of the DQ white dwarf to vary. The temperature of the DA star is then
varied until a proper match to H$\alpha$ is reached. Our best
photometric and spectroscopic fits are displayed in Figure
\ref{fg:f13}; the contribution of each component of the system is also
shown in the top panel. Even though this solution represents an
excellent match to the observations (the 2MASS photometry has
admittedly large uncertainties), it is obviously not a unique
solution, however. A trigonometric parallax measurement could help to
further constrain the parameters of both stars. This is the second
such DA+DQ system ever identified, the first being NLTT 16249
discovered by \citet{ven12}, who also reported the presence of
photospehric traces of nitrogen in the spectrum of the DQ component.

\subsubsection{DZ Stars}

The fitting procedure for DZ stars is similar to that described above
for DQ stars with the exception that spectroscopic observations of the
Ca~\textsc{ii} H \& K doublet are used to determine the metal abundance
\citep[see also][]{duf07}. For the abundance of other heavier elements, 
not visible spectroscopically, we assume solar ratios for relative
abundances with respect to calcium.  Also, since invisible traces of
hydrogen may affect the predicted metallic absorption features (see
\citealt{duf07} for details), we study the influence of this
additional parameter by using model grids calculated with hydrogen
abundances of $\nh= -3$, $-4$, and $-5$. For the three DZA stars in
our sample --- L745-46A (0738$-$172), GD 95 (0843+358), and Ross 640
(1626+368) --- the hydrogen abundance is measured directly by fitting
H$\alpha$. For vMa 2 (0046+051), LEHPM 2-220 (1009$-$184), and G139-13
(1705+030), our fit to the calcium lines could be greatly improved by
including hydrogen in our models, and we thus determined {\it
indirectly} the hydrogen abundance in these stars by fitting the
calcium lines with H/He as a free parameter; the hydrogen abundance in
LEHPM 2-220 was actually forced to $\nh=-3$ to avoid extrapolation
outside of our model grid. Finally, we assumed a value of $\nh= -5$
for the remaining DZ stars in our sample, with the exception of LP 701-29
(2251$-$070) for which we used hydrogen-free models to avoid
collision-induced absorption by molecular hydrogen in the infrared, 
which is not observed.

The results for the DZ stars in our sample are presented in Figure
\ref{fg:f14}. The spectroscopic observations used to determine the
metal and hydrogen abundances in the fitting procedure are shown in
the right panels. The predicted energy distributions for all DZ stars
in our sample agree extremely well with the observed photometry.  Note
that our temperature estimate for LP 701-29, $\Te=4000$~K, is at the
limit of our grid, and the observed energy distribution suggests a
somewhat lower temperature. We can also notice that the Ca~\textsc{ii}
H \& K doublet is fairly well reproduced in most DZ stars, but small
discrepancies still occur. As before, we assign a higher weight to
properly fitting the energy distributions rather than the
spectroscopic data.

One problematic DZ star in our sample is LP 726-1
(0840$-$136). Clearly, our best solution totally fails to reproduce
the calcium lines. Both the Ca~\textsc{ii} H \& K doublet and the
Ca~\textsc{i} $\lambda$4226 line are predicted too shallow. The
calcium doublet can actually be made deeper by increasing the hydrogen
abundance in our models. However, since LP 726-1 is extremely cool,
even small amounts of hydrogen creates a significant infrared flux
deficiency through collision-induced absorptions, which is not
observed. An alternative would be to lower the $\logg$ value.  This
solution (not shown here) does indeed improve our fit to the
Ca~\textsc{ii} H \& K doublet, but it still fails to match the
Ca~\textsc{i} $\lambda$4226 line. As discussed in \citet{sub07}, it is
possible that our failure to simultaneously reproduce the photometric
and spectroscopic data is related to additional pressure effects
neglected in our equation of state \citep[see also][]{duf07}.

As was the case for DQ stars, the effective temperature and surface
gravity are significantly reduced when metals are included in the
model atmosphere calculations ($\sim 1000$~K in $\Te$ and $\sim 0.3$
dex in $\logg$). For objects with trigonometric parallax measurements,
the mean surface gravity is now much closer to the canonical $\logg=8$
value.

\section{Spectroscopic Analysis}\label{spec}

The atmospheric parameters of DA stars with well-defined Balmer lines
($\Te\gtrsim 6500$~K) can be determined precisely from the optical
spectra using the so-called spectroscopic technique developed by
\citet{BSL92}. A similar approach can of course be used for (hot) DB
stars, although none have been identified in our nearby sample. The
technique relies on detailed fits to the observed normalized Balmer
line profiles with model spectra, convolved with the appropriate
Gaussian instrumental profile. We use the same Levenberg-Marquardt
nonlinear least-squares fitting method described above.  In this case
the $\chi^2$ minimization procedure uses all Balmer lines
simultaneously to determine the atmospheric parameters $\Te$ and
$\logg$.  In the case of contamination by an unresolved main-sequence
companion, usually an M dwarf, we simply exclude from the fit the
absorption lines that are contaminated (usually H$\beta$). For two DAZ
stars in our sample, G74-7 (0208+396) and G180-63 (1633+433), we rely
on the fitting technique and model atmospheres described in
\citet{gianninas11}, which include the opacity from Ca~\textsc{ii} H \& K 
known to contaminate the H$\epsilon$ line at the spectral resolution
used here. For GD 362 (1729+371), we adopt directly the solution of
\citet{tre10} based on mixed hydrogen and helium model atmospheres.

The results of or our spectroscopic fits are displayed in Figure
\ref{fg:f15} in order of right ascension. In all cases,
the model spectra match perfectly the observations, including the two
DAZ stars discussed above. There is an obvious contamination at
H$\beta$ in LHS 1660 (0419$-$487) from the M dwarf companion, and this
line has been omitted from our fit. We also note a small
discrepancy in the line cores of LP 907-37 (1350$-$090) due to the
presence of a relatively weak ($\sim 100$ kG) magnetic field
\citep{schmidt94}.

Even though the spectroscopic technique is arguably the most accurate
method for measuring the atmospheric parameters of DA stars, it has an
important drawback at low effective temperatures ($\Te\lesssim
13,000$~K) where spectroscopic values of $\logg$ are significantly
larger than those of hotter DA stars. This so-called high-$\logg$
problem has been discussed at length in \citet{tre10} and references
therein. The most promising solution for this problem was a mild and
systematic contamination of the atmospheric regions by helium brought
to the surface by convective mixing, a process that would mimic the
high $\logg$ values inferred from the spectroscopic technique based on
pure hydrogen models \citep{bergeron90}. However, this suggestion has
been refuted by \citet{tre10}, who convincingly showed, using Keck
high-resolution spectra of cool DA stars, that helium is simply not present
in these stars. More recently, \citet{tremblay11b} showed that this
high-$\logg$ problem is instead related to the limitations of the
mixing-length theory used to describe the convective energy transport in DA
stars, and that more realistic, 3D hydrodynamical model atmospheres
are required in order to obtain a surface gravity distribution that
resembles that of hotter radiative-atmosphere DA stars. 

The high-$\logg$ problem is particularly problematic for the study of
the white dwarfs in the nearby sample. Indeed, out of the 165 white
dwarfs analyzed in this paper, 51 are DA stars in the $6000-13,000$~K
temperature range with optical spectra available. Since photometry is
available for only 22 of these objects, we must therefore rely heavily
on spectroscopic $\logg$ values to estimate distances.  Such {\it
spectroscopic distances} are obtained by combining $V$ magnitudes with
absolute visual magnitudes calculated from model atmospheres at the
spectroscopic values of $\Te$ and $\logg$, and more reliable estimates
of surface gravities thus become essential. Unfortunately, full 3D model
atmosphere grids are not available yet, and we must therefore rely on a
more approximate procedure.

In what follows, we attempt to derive an empirical correction based on a
statistically large and representative sample of DA white dwarfs.
The best characterization of the high-$\logg$ problem can be found in
the recent study of \citet{tremblay11a} who analyzed the DA stars identified in
the Data Release 4 of the Sloan Digital Sky Survey.  In particular,
the $\logg$ distribution of DA stars as a function of effective temperature, shown in
their Figure 18 and reproduced here in the top panel of Figure
\ref{fg:f16}, shows a significant increase in the $\logg$ values
at low temperatures, with a distinctive triangular shape (see Section
4.2 of \citealt{tremblay11a} for a more elaborate discussion). We next fit
a third order polynomial through the SDSS
data points with $\Te<14,000$~K in this $\Te$-$\logg$ diagram, using average bins of 500~K in
temperature. The result of this polynomial fit is displayed in Figure
\ref{fg:f16}, and is given by the expression
$\logg=1.292+1.862\times10^{-3}\,\Te-1.598\times10^{-7}\,\Te^2+4.368\times10^{-12}\,\Te^3$,
valid between $\Te=7000$~K and 14,000~K.  A low order polynomial was
preferred in order to ensure a certain smoothness in our correction
procedure, and to get rid of any possible large variations due to the
inhomogeneous distribution of stars between consecutive bins. Careful
attention was also given to remove all stars outside one standard
deviation from the mean $\logg$ value in each bin. By doing so, we
want to make sure that we eliminate any possible bias that could
result from any excess of high- or low-mass stars in a given bin. Also
shown in Figure \ref{fg:f16} is a 0.594 \msun\ evolutionary
track, which corresponds to the median mass of DA stars above 13,000~K
determined by \citet[][see their Table 4]{tremblay11a}. Finally, the
empirical correction we apply to our spectroscopic $\logg$ values is
simply given by the difference between the black and red curves in
Figure \ref{fg:f16}, at any given temperature.

The spectroscopic $\logg$ values for the DA stars in the SDSS
corrected in this fashion are displayed as a function of effective temperature
in the bottom panel of Figure \ref{fg:f16}. The continuity of the
$\logg$ distribution observed here through the entire temperature
range suggests that our correction procedure is reasonably
sound. We point out that our $\logg$ correction procedure directly
results in larger spectroscopic distances, and as such, the number of
stars in our local sample, defined within a given volume of space,
might eventually be reduced. From this point on, all spectroscopic
$\logg$ values given below take into account the correction
discussed above. 

\section{Selected Results}\label{resu}

\subsection{Comparison of Fitting Techniques}

We compare in Figure \ref{fg:f17} the effective
temperature and absolute magnitude for the 23 white dwarfs in our
sample, all of the DA type, that have both spectroscopic and
photometric measurements. In general, the $\Te$ and $M_V$ values agree
well with both techniques, with a few exceptions labeled in the figure
and discussed below. We also notice, for the
hotter objects, a small but systematic offset of
$\sim300$~K between spectroscopic and photometric temperatures --- with $\Te({\rm spec})>\Te({\rm phot})$ --- which might
suggest that the high-$\logg$ problem probably affects the
spectroscopic temperatures as well, and that the expected $\logg$ and $\Te$
corrections are somehow correlated.

Two objects, G1-45 (0101+048) and L587-77A (0326$-$273), stand out in
both panels of Figure \ref{fg:f17}. As discussed above, these are well
known unresolved double degenerate systems, and the discrepancy
between spectroscopic and photometric temperatures suggests that both
components of the system probably have different effective
temperatures or even spectral types --- a DA+DC system for
instance. This would provide a natural explanation for the discrepant
fit observed at H$\alpha$ for L587-77A (see Figure \ref{fg:f11}).
L870-2 (0135$-$052), labeled in the bottom panel, is also a well known
unresolved double degenerate binary. The fact that it does not stand
out in the upper panel is due to the fact that both components of the
system have comparable effective temperatures \citep{bergeron89}. Two
other objects in the bottom panel, LHS 3163 (1609+135) and L24-52
(2105$-$820), have higher than average $\logg$ values based on the
photometric analysis, and there is no indication that these stars are
overluminous and thus binaries. Interestingly enough, if we adopt
instead the {\it uncorrected} spectroscopic $\logg$ values for these
two white dwarfs, the spectroscopic and photometric surface gravities
agree almost perfectly, suggesting that perhaps the high-$\logg$
problem does not affect every cool DA star in the same fashion.

The last object labeled in the bottom panel of Figure \ref{fg:f17} is
GD~1212 (2236$-$079), a relatively cool ($\Te\sim11,000$~K) ZZ Ceti
white dwarf with a dominant period around 1200~s \citep{gianninas06},
which shows a difference of 0.32 dex in $\logg$ between both fitting
techniques. We have an excellent trigonometric parallax measurement
from \citet{sub09} for this object, so there is no reason to expect
our photometric $\logg$ determination to be too far off. The
photometric and spectroscopic fits displayed in Figures \ref{fg:f11}
and \ref{fg:f15}, respectively, also appear perfectly normal. The only
peculiarity is our fit at H$\alpha$ where the observed line profile is
significantly shallower than predicted by our models. Such a
discrepancy is usually the signature of the dilution of a DA star by
an unresolved DC companion (see the extreme case of 1529+141 in Figure
\ref{fg:f11}). If this interpretation is correct, the contribution of
this featureless companion to the total luminosity of the system could
perhaps account for the small $\sim0.5$\% amplitude variations
observed in GD 1212 \citep{gianninas06}, which are at odds with the
general tendency to observe large amplitudes in cooler, long-period ZZ
Ceti stars.

\subsection{Adopted Atmospheric Parameters}

Data sets appropriate for a photometric or spectroscopic analysis
could not be found for 3 stars in our sample --- GJ 86 B (0208$-$510),
HD 27442 B (0415$-$594), and vB 4 (1132$-$325) --- all three of which
are Sirius-like systems. Of the 165 remaining white dwarfs analyzed in
this paper, 68 have spectroscopic $\Te$ and $\logg$ determinations
while 118 have photometric determinations, 29 of which have no
trigonometric parallax measurements and a value of $\logg=8$ had to be
assumed.

The final parameters for all white dwarfs in our sample are selected
using the following criteria. For stars with $\Te>12,000$~K, we adopt
the spectroscopic solution since these are the only ones available
anyway. To circumvent the high-$\logg$ problem below $\Te=13,000$~K,
we favor the photometric solution over the spectroscopic solution,
when both are available, except for stars with no trigonometric
parallax measurement, in which case we adopt the spectroscopic
solution, if available. As already mentioned above, for GD 362
(1729+371), we substitute the solution of \citet{tre10} based on mixed
hydrogen and helium model atmospheres. 

Our final results are presented in Table \ref{tb:t2} where we give for
each object the effective temperature ($\Te$), surface gravity
($\logg$), stellar mass ($M/$\msun), atmospheric composition (H- or
He-dominated, or mixed H/He), absolute visual magnitude ($M_V$), luminosity
($L/$\lsun), distance ($D$), white dwarf cooling time, and the method
employed to obtain the atmospheric parameters. For the spectroscopic
solutions, we provide here the corrected $\logg$ values, and those differ
from the uncorrected values given in Figure \ref{fg:f15}.
Whenever necessary to derive the above quantities, we rely on the same evolutionary models as before.
For the photometric solutions, the distance given in Table \ref{tb:t2}
is obtained directly from the trigonometric parallax measurement, if
available, or from the fitted solid angle $\pi(R/D)^2$ assuming a
radius corresponding to $\logg=8$ (the latter are noted with $\sigma_{\log g}=0$ 
in Table \ref{tb:t2}).  For the spectroscopic solutions,
the distance is again obtained directly from the trigonometric
parallax measurement, if available, otherwise it is derived from the
distance modulus, $V-M_V=5\log D-5$, with the $V$ magnitudes taken
from Table~\ref{tb:t1} and the $M_V$ values calculated from model
atmospheres at the spectroscopic $\Te$ and $\logg$ values.

We finally note that the atmospheric parameters for the DA+DQ system
GD 184 (1529+141) given in Table \ref{tb:t2}, and in particular the distance of $D=60.7$ pc, are
obtained under the assumption of a single star. The deconvolved
atmospheric parameters are provided in Figure \ref{fg:f13};
assuming that both components of the system have $\logg=8$ yields a
revised distance of $D\sim73$ pc, way beyond our 20 pc limit.

\subsection{Distances}

Based on the adopted distances presented in Table \ref{tb:t2}, we
obtain a list of 131 white dwarfs that are within 20 pc, taking into
account the uncertainties. We compare in Figure \ref{fg:f18} our
distance estimates with those from Table 1 of \citet{sion09}. The
largest differences occur for the objects that were not previously
included in Table 2 of \citet{holberg08}, and for which the distances
in Sion et al.~represent only crude estimates. We also note that 7
white dwarfs lie between 20 and 21 pc in Table \ref{tb:t2}, some of
which rely on the assumption of $\logg=8$ for the photometric
method. To be more conservative, we decided to include these objects in
our sample as well, for a grand total of 138 nearby white dwarfs.  

A more detailed comparison with
Sion et al.~reveals that 9 white dwarfs have been removed 
from their sample --- 
GD 5 (0008+424), 
LP 294-61 (0108+277), 
G84-26 (0457$-$004), 
LP 207-50 (0749+426), 
G49-33 (0955+247), 
SDSS 1124+595 (1124+595), 
LP 276-033 (1653+385), 
LHS 3254 (1655+215), 
and LP 522-46 (2322+137)
--- while 20 represent new additions --- 
L796-10 (0053$-$117), 
G1-45 (0101+048), 
LHS 1247 (0123$-$262),
GD 25 (0213+396),
LHS 1442 (0243$-$026), 
LHS 1660 (0419$-$487), 
LHS 1734 (0503$-$174),
G111-64 (0810+489), 
SCR 0818-3110 (0816$-$310),
LHS 2022 (0827+328),
G47-18 (0856+331), 
LHS 2229 (1008+290), 
LHS 2522 (1208+576), 
PM J13420$-$3415 (1339$-$340), 
LHS 56 (1756+143), 
LTT 8189 (2039$-$202), 
LTT 8190 (2039$-$682), 
G261-43 (2126+734),
G128-7 (2248+293), 
and LHS 4019 (2347+292). 
Most of these additions were already included in the list
of possible white dwarfs within 20 pc from \citet[][see their Table
4\footnote{The entry for
0827+387 and 1339+259 in Table 4 of \citet{holberg08} are typographical errors
and actually correspond to 0827+328 and 1339$-$340, respectively.}]{holberg08}. Our improved distance estimates confirm their place
in the local sample.

We finally mention that even though the 3 white dwarfs without analyzable
data (0208$-$510, 0415$-$594, and 1132$-$325) are not included in our
analysis below, they are still part of the local
sample. From this point on, when we refer to the local sample, we
restrict ourselves to the list of 138 objects in Table
\ref{tb:t2} with distances inside the $\sim 20$ pc region, as define above.
All objects excluded from this local sample are marked with
an asterisk in Table \ref{tb:t2}.

\subsection{Mass Distribution}\label{mdistr}

Since photometric analyses of stars with no trigonometric parallax
measurements assume a value of $\logg=8.0$, these are not taken into
account in the analysis of the mass distribution discussed here, but
will be included in the calculation of the luminosity function
presented in the next section. Also, for the moment, all suspected or
confirmed double degenerate systems are considered as single objects
for reasons discussed below.

The mass distribution as a function of effective temperature for each
star in our sample is displayed in Figure \ref{fg:f19}.  Atmospheric
compositions and spectral types are indicated with different
symbols. In particular, filled and open symbols represent hydrogen-
and helium-rich compositions, respectively.  Also superposed in this
figure are the theoretical isochrones for our C/O core evolutionary
models with thick hydrogen layers, as well as the corresponding
isochrones with the main sequence lifetime added to the white dwarf
cooling age (for $\tau\geq2$ Gyr isochrones only); here we simply
assume \citep{lrb98} $t_{\rm MS}=10(M_{\rm MS}/M_\odot)^{-2.5}$ Gyr
and $M_{\rm MS}/M_\odot=8\ln[(M_{\rm WD}/M_\odot)/0.4]$. As can be
seen from these results, white dwarfs with $M\lesssim0.48$
\msun\ cannot have C/O cores, and yet have been formed from single
star evolution within the lifetime of the Galaxy. Some of these
low-mass objects must either be unresolved double degenerates, or
single white dwarfs with helium cores. In the former case, the stellar
masses inferred from these figures are underestimated --- especially
if the unresolved components have comparable luminosities, and the
corresponding cooling ages derived here become meaningless.  The
second possibility corresponds to single (or binary) helium-core
degenerates whose core mass was truncated by Case B mass transfer
before helium ignition was reached \citep{iben87}.

The objects displayed in red in Figure \ref{fg:f19} are double
degenerate binaries that have been confirmed through radial velocity
measurements: G1-45 (0101+048; \citealt{maxted00}), L870-2
(0135$-$052; \citealt{saffer88}), and L587-77A (0326$-$073;
\citealt{zuc03}). The low masses inferred for these systems
($M\sim0.25-0.45$ \msun) are a direct consequence of the fact that we
assumed that these are single objects, and thus overestimated their
stellar radius by a factor of $2^{1/2}$ (see equation \ref{eq:3}), if
both components are identical. Had we assumed two stars instead of
one, a simple calculation yields photometric masses in the range
$M\sim 0.48-0.76$ \msun, right in the bulk of average mass white
dwarfs. A good example of this calculation is for L870-2 with an
extremely low photometric mass of 0.24
\msun\ ($\Te\sim7100$~K). Assuming two identical DA stars yields
instead 0.46 \msun\ for both components of the system, in excellent
agreement with our spectroscopic mass of 0.47 \msun\ (note that the
spectroscopic mass is not affected by the presence of two DA stars if
they have comparable atmospheric parameters; see Figure~1 of
\citealt{liebert91}). It turns out, indeed, that the two DA components
in the L870-2 system are virtually identical
\citep{bergeron89}. Hence, most, if not all low-mass stars observed in
Figure \ref{fg:f19} are probably unresolved double degenerates.  These
include LHS 1243 (0121$-$429), LHS 1734 (0503$-$174), L532-81
(0839$-$327), G187-8 (2048+263), G128-7 (2248+293), and probably vB 11
(2054$-$050), although the latter is one of the coolest white dwarfs
in our sample, and the atmospheric parameters might be uncertain.  A
simple mass redetermination for these objects, as prescribed above, is
not so simple, however. For instance, L532-81, with a photometric mass
of 0.44 \msun\ ($\Te\sim9100$~K), is already in perfect agreement with
its spectroscopic mass of 0.42 \msun. In this particular case,
assuming two DA stars would be the wrong thing to do, as this object
appears to be a single DA star. We note, however, that the photometric
fit for this object, shown in Figure \ref{fg:f11}, reveals a small but
significant discrepancy at H$\alpha$, which can easily be explained if
the companion star is a DC white dwarf. It is thus not a simple task
to deconvolve the individual components of these unresolved binary
systems, and their mass determinations should therefore be taken with
caution.

Going back to Figure \ref{fg:f19}, we can see that, not unexpectedly,
the local sample is predominantly composed of cool white dwarfs. The
first striking feature we observe is the complete absence of white
dwarfs with helium-rich atmospheres above $\Te\sim 12,000$~K. To
illustrate this more quantitatively, we show in the left panel of
Figure \ref{fg:f20} the total number of stars as a function of
effective temperature per bin size of 2000~K, as well as the
individual contribution of the hydrogen-atmosphere white dwarfs. Also,
in the right panel, we show the ratio of helium-atmosphere white
dwarfs to the total number of stars. The number of helium-rich white
dwarfs increases dramatically below 13,000~K, and keeps increasing
to a ratio value above 40\% in the range $\Te\sim 7000-9000$~K.  Clearly, these
results indicate that some physical mechanism is transforming the
surface composition of white dwarf stars as they cool off. The most
obvious mechanism in this temperature range is convective mixing,
where the thin convective hydrogen atmosphere is mixed with the deeper
and more massive helium convection zone.  Since the depth of the
hydrogen convection zone increases at lower temperatures, the
temperature at which mixing occurs becomes a function of the mass of
the hydrogen envelope --- the thicker the hydrogen layer, the cooler the
mixing temperature. If this interpretation is correct, our results
suggest that convective mixing occurs for a significant fraction
($\sim 40$\%) of DA stars below $\Te\sim10,000$~K or so. This confirms
the results obtained by \citet{tre07} who performed a model atmosphere
analysis of cool white dwarfs with 2MASS infrared photometry
available, to show that the ratio of helium- to hydrogen-atmosphere
white dwarfs increases gradually from a constant value of $\sim0.25$
between $\Te=15,000$~K and 10,000~K to a value twice as large in the
range $10,000 > \Te > 8000$~K.  Our results and those of Tremblay et
al.~imply that a significant fraction of DA stars, as much as 40\%
according to our analysis, may have hydrogen mass layers in the range
$M_{\rm H}/M_{\rm tot}=10^{-10}$ to $10^{-8}$ (see \citealt{tre07} for
details).

The gradual drop of helium- to hydrogen-atmosphere white dwarfs below
7000~K in Figure \ref{fg:f20} is more difficult to explain. This
actually corresponds to the location of the non-DA gap, or deficiency,
between roughly 5000~K and 6000~K where virtually all white dwarfs are
DA stars (see Figure~\ref{fg:f19}).  As already discussed above,
several exotic mechanisms have been proposed by BRL97 to explain this
deficiency, although none appears completely satisfactory, and we
simply refer the reader to their discussion (see Figure~35 and Section
6.3.2 of BRL97). Perhaps convective mixing cannot be sustained
indefinitely, and the hydrogen and helium layers eventually become
well separated again at lower effective temperature.

We also notice in Figure \ref{fg:f19} that there appears to be a
deficiency of massive stars ($M\gtrsim1$ \msun) below $\Te\sim7000$~K with respect to
what is observed at higher temperatures. Since these objects have
not yet reached the stage of crystallization (which we see here as
the rapid reduction of the cooling timescales at high mass), we
should expect such cool and massive white dwarfs to be present
in the local sample. Another way of saying this, the
oldest objects in Figure \ref{fg:f19} have a total age of roughly
9.5 Gyr or so, while there are no massive white dwarfs older than
4 Gyr. One could argue that since massive white dwarfs have
smaller radii and are thus less luminous, perhaps they have simply
gone undetected in proper motion surveys. However, the coolest
massive white dwarf in Figure \ref{fg:f19} is G108-26 (0644+025),
and its luminosity ($\log L/L_{\odot}=-3.79$) is not particularly
low compared to other objects in Table \ref{tb:t2}. A more
plausible explanation is that all the massive white dwarfs
observed here are the results of mergers. In this case their
cooling ages cannot be interpreted directly from the isochrones
shown here since these are based on single star evolution; the cooling
ages inferred here would represent lower estimates at best. If this
interpretation is correct, our results suggest that such mergers
in the solar neighborhood simply did not have enough time to cool off to
low temperatures ($\Te\lesssim 7000$~K) within the age of the
local galactic disk.

A similar situation can be observed at the bottom of the mass
distribution in Figure \ref{fg:f19}, where all low-mass white
dwarfs are found only at low effective temperatures, and they all have 
hydrogen-rich atmospheres. There are no low-mass white dwarfs found at high
temperatures, despite the fact these are abundant in spectroscopic
analyses of DA stars discovered in UV-excess surveys, such as the
PG survey (see, e.g., Figure~12 of \citealt{liebert05}). First, even
though the low-mass white dwarfs in Figure \ref{fg:f19}
are relatively cool, their cooling ages are
less than 2 Gyr. As discussed above, most of these objects
are probably unresolved double degenerates, 
and they are likely to be the result of common envelope evolution.
As such, it is difficult to interpret their location in this
diagram. Also, as discussed in BLR01, we are forced to conclude that
this particular evolutionary channel does not produce helium-atmosphere 
white dwarfs, presumably because the objects that go
through this close-binary phase end up with hydrogen layers too
massive to allow the DA to DB conversion near $\Te\sim30,000$~K,
or below.

The mass distribution of all white dwarfs in our sample,
regardless of their effective temperature, is displayed in Figure
\ref{fg:f21}. We also show the separate contributions of
hydrogen- and helium-rich stars.  The mean mass of the local sample is
$\langle M\rangle=0.650$ \msun\ with a standard deviation of
$\sigma_M=0.161$ \msun; \citet{holberg08} obtained $\langle M\rangle=0.665$ \msun\ 
for the local white dwarfs in their sample, a value only slightly
larger than that reported here. The corresponding mean masses for the
hydrogen-atmosphere white dwarfs are $\langle M\rangle=0.647$ \msun\
and $\sigma_M=0.171$ \msun, and for the helium-atmosphere white
dwarfs, $\langle M\rangle=0.660$ \msun\ and $\sigma_M=0.132$
\msun. The larger dispersion of the hydrogen stars is a direct result
of the presence of both low- and high-mass tails, which are clearly
less pronounced in the helium-rich sample.  BLR01 found for their
hydrogen-rich subsample $\langle M \rangle = 0.61$ \msun, a value
noticeably lower than that obtained here.  However, a comparison of
the mass distributions (see Figure~22 of BLR01) reveals that the
discrepancy comes mainly from the low-mass tail, which is far less
predominant in our local sample. 

The average and median masses for both hydrogen- and helium-atmosphere white
dwarfs in Figure \ref{fg:f21} are remarkably similar. This result is in
sharp contrast with the results obtained by \citet{kep07} in their
spectroscopic analyses of DA and DB stars identified in the Data
Release 4 of the SDSS. Their results, restricted to effective
temperatures above 16,000~K, yield $\langle M\rangle_{\rm DA}\simeq
0.593$ \msun\ and $\langle M\rangle_{\rm DB}\simeq 0.683$ \msun, a
significantly larger difference than that obtained here. Since Kepler
et al.~rely solely on the spectroscopic approach, there could be an
indication that this method still needs further investigation. Alternatively,
the problem may lie with the analysis of Kepler et al.~itself, or with the SDSS data, or
both. For instance, the mass distribution for brighter DB stars
recently obtained by \citet{bergeron11}, and based also on the spectroscopic technique, 
yields $\langle M\rangle_{\rm DB}\simeq 0.671$ \msun, a somewhat
lower value than that obtained from the SDSS data, and only $\sim 0.01$
\msun\ larger than that that quoted above for the local subsample.
Similarly, \citet{gianninas11} report a mean mass of $\langle M
\rangle = 0.638$ \msun\ for bright DA stars drawn from the WD Catalog,
in good agreement with the value obtained in our analysis.  However,
the mean mass quoted for the PG magnitude-limited survey, $\langle M
\rangle = 0.629$ \msun, is somewhat lower. But this result is actually
expected from such magnitude-limited surveys since massive white
dwarfs are intrinsically less luminous than their normal mass
counterparts, and therefore their number will be significantly
underestimated in a survey limited by the magnitude of the star. On
the other hand, less massive white dwarfs, with their larger radii and
higher luminosities, will be sampled at much larger distances in a
magnitude-limited survey, and will thus be overrepresented. Finally,
\citet{falcon10} measured a mean mass of 0.647 \msun\ for DA stars based
on a gravitational redshift determination of their mean mass, in
excellent agreement with our value obtained above for the
hydrogen-atmosphere white dwarfs.

Focusing again on the mass distribution of hydrogen-rich white dwarfs
displayed in Figure \ref{fg:f21}, we can see a distinctive high-mass
excess near 1 \msun, which appears as a very sharp distribution on its
own. Clearly, these cannot represent the descendants of single massive
progenitors on the main sequence.  Indeed, because of the
initial--final mass function, the number of massive white dwarfs
expected in such a small volume of space is probably close to zero.
Hence the massive white dwarfs in our local sample must be again
interpreted as mergers in this context. And since we have a
volume-limited sample, we are forced to conclude that the fraction of
mergers in the Galaxy must be enormous. According to our results,
$\sim6$\% of the hydrogen-rich white dwarfs in our sample are
mergers. The fact that the mass distribution of these mergers is so
narrow also suggests a common evolutionary scenario for these
systems. In simple terms, if we interpret these $\sim 1$ \msun\ white
dwarfs as the merger of two $\sim 0.5$ \msun\ components, we could
even explain the apparent deficiency of hydrogen-rich white dwarfs in
the $0.50-0.55$ \msun\ mass range observed in the red histogram shown
in Figure \ref{fg:f21}, which is also apparent from an examination of Figure
\ref{fg:f19}.  After all, the local disk is old enough to have formed
a sufficiently large number of white dwarfs in this particular mass
range, and yet, very few are present in our local sample.

\subsection{Luminosity Function}

The white dwarf luminosity function (WDLF) is, by definition, a
measure of the space density of white dwarfs as a function of
luminosity, expressed here as the number of stars per pc$^3$ per unit
of bolometric magnitude. In order to get an accurate picture of this
function, the exercise has to be rigorously performed with a
well-defined sample, and thus the most important requirement for a
proper determination rests on the completeness of the sample.  One of
the most recent determination of the WDLF, covering the entire range
of bolometric magnitudes ($7\lesssim \mbol\lesssim16$) was
presented by \citet{har06} using the magnitude-limited Sloan Digital
Sky Survey. Several corrections for completeness and contamination had
to be made, however, to counterbalance important selection effects
present in the SDSS sample. A similar situation arises in the case of
all magnitude-limited surveys such as the PG survey \citep{liebert05}
or the KUV survey \citep{lim10}. Using a volume-limited survey avoids
most of the problems inherent to these previous determinations, and
provides a non-biased sample where completeness issues are better
controlled. The most obvious drawback, unfortunately, is the small
size of the sample, which may lead to significant statistical
uncertainties.

The WDLF calculated here is based on the derived bolometric
magnitude of each white dwarf in our sample, obtained via the
spectroscopic or photometric results presented in Table
\ref{tb:t2}, with $\mbol=-2.5\log L/L_{\odot}+M_{\rm
bol}^{\odot}$, where $\mbol^{\odot}=4.75$ is the bolometric magnitude
of the Sun. Since a proper account of the number of individual stars
in each magnitude bin represents a crucial step in the determination
of the WDLF, we consider in our calculation all confirmed and
suspected double degenerate binaries, mentioned in Section
\ref{mdistr}, as two individual white dwarfs.  We end up this way with
a total of 147 individual nearby ($D\lesssim20$ pc) white dwarfs for
the WDLF calculation. Similarly, we deconvolve the individual masses
of each system by using the procedure described in Section \ref{mdistr},
even though we realize this approach is only approximate.

Our results are presented in Figure \ref{fg:f22}, where the total
number of white dwarfs in each bin of bolometric magnitude, divided by
the enclosed volume defined by the 20 pc limit, is given as a function
of $\mbol$. Also shown at the top is the corresponding effective
temperature scale. Our WDLF is also compared with that obtained by
\citet{har06} using the SDSS sample, and with that derived by
\citet{bergeron11} using the DA and DB stars identified in the PG
survey (see their Figure 24). The latter is different from
the WDLF published by \citet{liebert05} since all DA stars in the PG
survey have been refitted with the spectroscopic method using models
that include the new Stark broadening profiles discussed in Section
\ref{theory} \citep{gianninas11}. Also, both determinations assume
a Galactic disk scale height of 250 pc.

As discussed by \citet{har06}, the space density obtained from the
SDSS and the PG surveys agree very well for $7<\mbol<10$ --- even
better than what is shown in Figure 4 of Harris et al. --- but
significant deviations occur for fainter magnitude bins where the PG
survey is believed to be incomplete
\citep{liebert05}. In contrast, the WDLF obtained from the
local sample yields space densities in the same magnitude interval
that are larger by a factor of $\sim 2$.  This overdensity of white
dwarfs above $\Te=12,000$~K is readily apparent upon a closer examination of
the mass distribution shown in Figure \ref{fg:f19}. We emphasize
that all the stars in this temperature range, except one, have
distance estimates based on trigonometric parallax measurements, so
there is no doubt that these objects indeed belong to the local
sample. The only explanation we can come up with to account for this
discrepancy is related to the small number of white dwarfs in the
brightest magnitude bins. Hopefully, the small number statistics will
eventually be improved when the volume of the local sample is
increased to larger distances, such as the effort undertaken by
\citet{lim10_2} to define a complete sample of white dwarf stars
within a distance of 40 pc from the Sun.

Our results at the faint end of the luminosity function, where the
number of stars in each magnitude bin is more statistically
significant, are in remarkably good agreement with the SDSS, although
the detailed shape is somewhat different. Our result is likely to be more
realistic, however, since the fainter magnitude bins in the SDSS are
not as well sampled as the brighter bins. For instance, there are only
31 objects in the magnitude bin at the peak of the SDSS distribution, while
there are over 500 objects in the magnitude bins near $\mbol=11$.

We cannot help noticing a distinct feature present in all three
WDLF calculations displayed in Figure \ref{fg:f22}, near $\mbol=11$,
where the space density decreases slightly before rising again
towards lower luminosities. This slight drop in the luminosity function
occurs below $\Te\sim14,000$~K, which corresponds to the temperature
range where the atmospheres of DA stars become convective. Perhaps
improved treatment of convective energy transport along the lines
of the 3D hydrodynamical model atmospheres discussed in \citet{tremblay11b}
might shift the stars around from one magnitude bin to the other,
and make this peculiar feature disappear. This localized feature clearly
deserves further investigation.

Finally, by integrating the luminosity function over all magnitude
bins, we obtain a measure of the total space density of white dwarfs
of $4.39\times 10^{-3}\ {\rm pc}^{-3}$, and a corresponding mass
density of $2.9\times 10^{-3}\ M_{\odot}\ {\rm pc}^{-3}$.  If we
compare these results with the values obtained by \citet{holberg08} {\it
from the complete portion of the sample within 13 pc} --- i.e., a space
density of $4.8\pm0.5\times 10^{-3}\ {\rm pc}^{-3}$ and a mass density
of $3.2\pm0.3\times 10^{-3}\ M_{\odot}\ {\rm pc}^{-3}$ --- we can
assess the completeness of our sample to be slightly higher than 90\%.
Note that our estimate is not affected by the uncertainties in the
brighter magnitude bins since the contribution of these stars to the
space density is negligible (see Figure~\ref{fg:f22}).

\section{Conclusion}\label{conclusion}

A detailed photometric and spectroscopic analysis of 166 nearby white
dwarf candidates was presented. Homogeneous determinations
of the atmospheric parameters of the local population were
performed based on state-of-the-art model atmospheres and improved
photometric calibrations. We developed a method for correcting the
spectroscopic $\logg$ values at low effective temperatures ($\Te<13,000$~K)
where the so-called high-$\logg$ problem occurs, and which would
have prevented us from obtaining reliable mass and distance
estimates in this range of temperature.

We determined a mean mass of $\langle M\rangle=0.650$ \msun\ for the
complete sample, with corresponding values of 0.647 \msun\ and 0.660
\msun\ for the hydrogen- and helium-atmosphere white dwarfs,
respectively. There is thus no indication for differences in the mean
mass values between these two populations, although the mass
distribution of hydrogen white dwarfs contains a significantly larger
number of low- and high-mass stars. The mean mass for
hydrogen-atmosphere white dwarfs is entirely consistent with that obtained
for over 1300 bright DA stars in the WD Catalog, $\langle
M\rangle=0.638$ \msun\ \citep{gianninas11}. The large fraction of
massive stars observed in the local sample has been interpreted as the
result of mergers. The main argument for the merger hypothesis is that
these massive white dwarfs cannot have evolved from massive main
sequence progenitors since the Sun is not located in a region of
active star formation where the required massive progenitors could be found.

The local volume-limited sample represents a snapshot of what a
representative sample of white dwarf stars might look like. The small
spectral atlas displayed here reveals that peculiar white dwarfs are
not rare objects, on the contrary. It is thus not too surprising that
large surveys such as the SDSS revealed even stranger objects, such as
the carbon atmosphere ``Hot DQ'' stars \citep{dufour08} or white
dwarfs with oxygen dominated atmospheres \citep{gansicke10}.
Interestingly enough, most of what we know about the mean
properties of white dwarf stars comes from the spectroscopic
analysis of hot ($\Te\gtrsim15,000$~K) DA stars with their
comfortable radiative, pure hydrogen atmospheres. Our local sample
contains only $\sim 10$ of these DA stars. A more typical white
dwarf has instead a convective atmosphere, with quite often a
helium-dominated atmosphere, with more than occasional traces of
heavier elements. These objects represent a bigger challenge in terms of
the modeling of their energy distribution.

The white dwarf luminosity function we derived here for the local
sample follows the exact same trend as those previously obtained in
many studies, including the SDSS, except at higher luminosities
where the local sample contains only a few objects, from
a statistical point of view.  This portion of
the luminosity function will eventually be improved by surveys
aimed at pushing the local sample to a distance of 40 pc
\citep{lim10_2}, an increase by a factor of 8 in terms of volume.
However, the total space density we derived here, $4.39\times 10^{-3}$
white dwarfs per cubic parsec, is not affected by this deficiency
of hot stars in the local sample. Note that our space density is
comparable to the value obtained by \citet{har06} based of the
SDSS, $4.6 \times 10^{-3}\ {\rm pc}^{-3}$, but significantly
larger than the value derived by \citet{lrb98}, $3.39 \times
10^{-3}\ {\rm pc}^{-3}$, based on a model atmosphere analysis of
43 white dwarfs identified in the proper motion sample of \citet{ldm88}.

We wish to deeply thank S.~Vennes, A.~Kawka, J.~P.~Subasavage,
R.-D.~Scholz, S.~Salim, B.~R.~Oppenheimer, S. L{\'e}pine, and
S.~Jordan for sharing their spectroscopic data with us,
P.-E. Tremblay, A. Gianninas, and M.-M. Limoges for precious advice
and collaboration with some of the data acquisition and reduction,
P.~Kowalski for sharing with us his calculations of the Ly$\alpha$
opacity, and A.~Gianninas for a careful reading of our manuscript. We
also wish to thank the director and staff of Steward Observatory and
NOAO for the use of their facilities. PD is a CRAQ postdoctoral
fellow. This work is supported in part by the NSERC Canada and by the
Fund FQRNT (Qu\'ebec).

\clearpage
\clearpage
\begin{deluxetable}{llccccccccccccccc}
\tabletypesize{\scriptsize}
\tablecolumns{12}
\tablewidth{0pt}
\rotate
\tablecaption{OBSERVATIONAL DATA}
\tablehead{
\colhead{WD} &
\colhead{Name} &
\colhead{Spectral Type} &
\colhead{$\pi$ (mas)} &
\colhead{$\sigma_{\pi}$ (mas)} &
\colhead{$1$} &
\colhead{$2$} &
\colhead{$3$} &
\colhead{$4$} &
\colhead{Opt. Phot.} &
\colhead{$J$} &
\colhead{$H$} &
\colhead{$K$} &
\colhead{Notes} }
\startdata
0000$-$345    & LHS 1008                  &      DAH        & 75.7   & 9.0    & 15.46      & 15.02      & 14.72      & 14.45      & $BVRI$     & 14.17      & 14.02      & 13.87      & 1, 2, 3        \\
0008+424      & GD 5                      &      DA         & ...    & ...    & ...        & 15.23      & ...        & ...        & $V$        & ...        & ...        & ...        &                \\
0009+501      & LHS 1038                  &      DAP        & 90.6   & 3.7    & 14.81      & 14.36      & 14.08      & 13.77      & $BVRI$     & 13.41      & 13.26      & 13.21      & 1, 2, 3        \\
0011$-$134    & LHS 1044                  &      DAH        & 51.3   & 3.8    & 16.52      & 15.89      & 15.56      & 15.22      & $BVRI$     & 14.85      & 14.62      & 14.52      & 1, 2, 3        \\
0011$-$721    & LP 50-73                  &      DA         & ...    & ...    & ...        & 15.17      & 14.87      & 14.55      & $VRI$      & 14.21      & 13.97      & 13.92      & 4, 5, 6        \\
\\
0038$-$226    & LHS 1126                  &      DQpec      & 110.54 & 1.18   & 15.20      & 14.50      & 14.08      & 13.71      & $BVRI$     & 13.32      & 13.47      & 13.71      & 1, 2, 7        \\
0046+051      & vMa 2                     &      DZ         & 226.95 & 5.35   & 12.91      & 12.39      & 12.13      & 11.90      & $BVRI$     & 11.69      & 11.61      & 11.52      & 8, 9, 10       \\
\enddata
\tablecomments{
(1) $BVRI$ from \citet{BRL97}; 
(2) $JHK$ from \citet{BRL97}; 
(3) $\pi$ from YPC; 
(4) $VRI$ from \citet{sub08}; 
(5) $JHK_S$ from 2MASS; 
(6) Not in Villanova White Dwarf Catalog; 
(7) $\pi$ from \citet{sub09}; 
(8) $BVRI$ from \citet{BLR01}; 
(9) $JHK$ from \citet{BLR01}; 
(10) $\pi$ from \citet{perry97}; 
(11) $V$ from \citet{salim03}; 
(12) $BR$ from \citet{USNO-A1/A2}; 
(13) $VRI$ from \citet{sub07}; 
(14) Sirius-like star; 
(15) $\pi$ from \citet{smart09}; 
(16) $uvby$ from \citet{bes78}; 
(17) Unpublished $BVRI$ from \citet{BRL97}; 
(18) $BGVRI$ from \citet{green84}; 
(19) $BVRI$ from \citet{Prov97}; 
(20) $ugriz$ from SDSS; 
(21) $JHK$ from \citet{kil09}; 
(22) $VRI$ from \citet{sub09}; 
(23) Classified DC7 in the PG catalog; 
(24) $BVIJHK$ from \citet{schmidt99}; 
(25) $\pi$ from H.~C. Harris. (2010, private communication); 
(26) $BV$ from \citet{hin79}; 
(27) $\pi$ from \citet{gatewood09}; 
(28) $\pi$ from \citet{kilic08}; 
(29) $\pi$ from \citet{lepine09}; 
(30) $uvby$ from \citet{weg83}; 
(31) $BVRI$ from \citet{bergeron05}; 
(32) $\pi$ from \citet{duc07}. }
\label{tb:t1}
\end{deluxetable}

\clearpage
\clearpage
\begin{deluxetable}{lrcccccrcc}
\tabletypesize{\scriptsize}
\tablecolumns{12}
\tablewidth{0pt}
\tablecaption{ADOPTED ATMOSPHERIC PARAMETERS OF NEARBY CANDIDATES}
\tablehead{
\colhead{WD} &
\colhead{$\Te$ (K)} &
\colhead{log $g$} &
\colhead{$M/M_{\odot}$} &
\colhead{Comp.} &
\colhead{$M_{V}$} &
\colhead{log $L/L_{\odot}$} &
\colhead{$D ({\rm pc})$} &
\colhead{Age (Gyr)} &
\colhead{Notes}}
\startdata
  0000$-$345   &  6643  (149)  &8.45 (0.16)    &0.88 (0.10)    &              H&14.41          &$- 3.82$       &13.2 (1.6)     &4.07           &    2\\
  0008+424     &  7202  (107)  &8.07 (0.07)    &0.64 (0.04)    &              H&13.58          &$- 3.45$       &21.4 (1.1)     &1.56           &    1\\
  0009+501     &  6502  (143)  &8.23 (0.05)    &0.73 (0.04)    &              H&14.15          &$- 3.72$       &11.0 (0.5)     &3.02           &    2\\
  0011$-$134   &  5992  (114)  &8.21 (0.11)    &0.72 (0.07)    &              H&14.44          &$- 3.85$       &19.5 (1.5)     &3.73           &    2\\
  0011$-$721   &  6325  (203)  &8.00 (0.00)    &0.59 (0.00)    &              H&13.94          &$- 3.63$       &17.6 (0.7)     &1.96           &    2\\
\\
  0038$-$226   &  5229  ( 76)  &7.93 (0.02)    &0.53 (0.01)    &           H+He&14.72          &$- 3.94$       &9.05 (0.10)    &4.44           &    2\\
  0046+051     &  6215  (183)  &8.16 (0.03)    &0.68 (0.02)    &             He&14.17          &$- 3.77$       &4.41 (0.10)    &3.30           &    2\\
\enddata
\tablecomments{
(1) Spectroscopic technique; (2) photometric technique.
}
\label{tb:t2}
\end{deluxetable}

\clearpage

\figcaption[f1.ps] {Blue spectroscopic observations of DA
and DAZ stars, shown in order of decreasing effective temperature,
from top left to bottom right. All spectra are normalized at 4500 \AA\ and
offset from each other by a factor of 0.5. The last object,
0939+071, is not a white dwarf.\label{fg:f1}}

\figcaption[f2.ps] {Blue spectra of DA and DAZ stars
too cool to be analyzed using line profile fitting techniques,
shown in order of right ascension. All spectra are normalized to a
continuum set to unity and offset from each other by a factor of
0.5.\label{fg:f2}}

\figcaption[f3.ps] {White dwarfs in our sample whose
spectra show \halpha, shown in order of decreasing equivalent widths,
from top left to bottom right. All spectra are normalized to a
continuum set to unity, and offset vertically from each other by a
factor of 0.4. The five objects in the right panel are magnetic and
exhibit the characteristic Zeeman splitting.\label{fg:f3}}

\figcaption[f4.ps] {Spectroscopic observations of
featureless DC stars, shown in order of right ascension. All spectra
are normalized to a continuum set to unity and offset from each other
by a factor of 0.7.\label{fg:f4}}

\figcaption[f5.ps] {Spectroscopic observations of normal DQ
stars shown in order of approximate molecular band/line strength. All spectra are
normalized to a continuum set to unity and offset from each other by a
factor of 0.5. Some of these DQ stars show carbon features only in the
ultraviolet.\label{fg:f5}}

\figcaption[f6.ps] {Spectroscopic observations of peculiar DQ
stars shown in order of approximate line strength; a normal DQ star,
reproduced from Figure \ref{fg:f5}, is also shown at the top for
comparison. All spectra are normalized at 6600 \AA\ and offset from
each other by a factor of 0.3.\label{fg:f6}}

\figcaption[f7.ps] {Blue spectroscopic observations of DZ
stars, shown in order of approximate slope. The two objects at the top
are actually DZA stars. All spectra are normalized at 4600 \AA\ (4900
\AA\ for 2251$-$070) and offset from each other by a factor of
0.7.\label{fg:f7}}

\figcaption[f8.ps] {Spectroscopic observations of miscellaneous
white stars in our sample, discussed in the text. All spectra are
normalized at 6400 \AA\ and offset from each other by a factor of
0.2.\label{fg:f8}}

\figcaption[f9.ps] {($V-I$, $V-K$) two-color diagram for
the data set from Table \ref{tb:t1}. DA and non-DA stars are
represented by filled and open circles, respectively, and the
cross indicates the size of the average error bar. The pure
hydrogen (red line) and pure helium (blue line) model sequences at
$\logg = 8.0$ are superimposed on the observed data. Temperatures
are indicated by small filled circles every $10^{3}$ K on the
cooling sequence, starting at 12,000 K at the bottom left.\label{fg:f9}}

\figcaption[f10.ps] {$M_{V}$ vs. ($V-I$) color-magnitude
diagram for the data set from Table \ref{tb:t1}. Objects are split
into DA stars (filled circles) and non-DA stars (open circles), based
on the presence or absence of \halpha. The pure hydrogen (red line)
and pure helium (blue line) model sequences at $\logg = 8.0$ are
superimposed on the observed data. Temperatures are indicated by small
filled circles every $10^{3}$ K on the cooling sequence, starting at
12,000 K at the upper left of the diagram.  The object at the top
right of the diagram is LHS 1660 (0419$-$487), whose colors are
contaminated by the presence of an M dwarf companion.\label{fg:f10}}

\figcaption[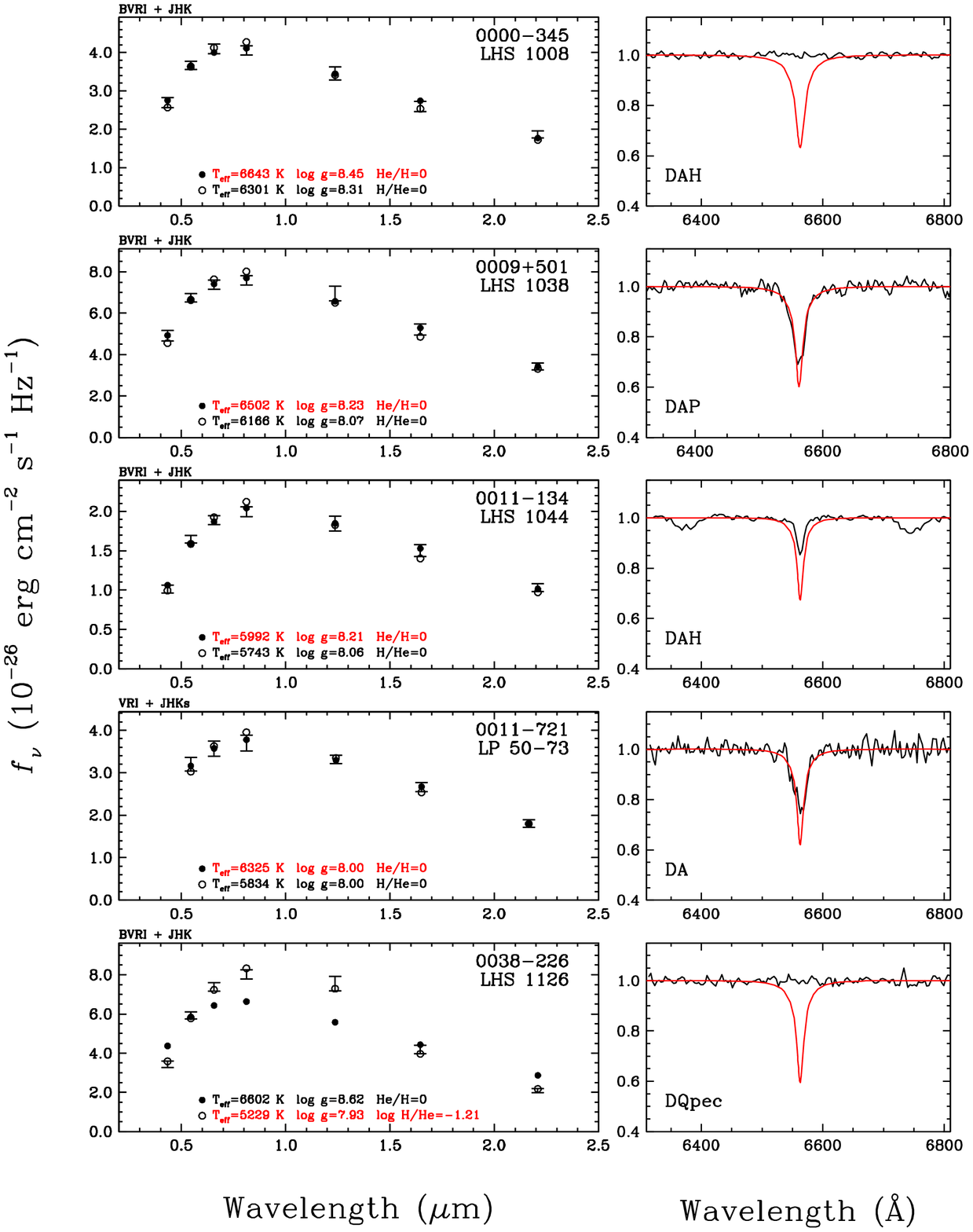] {Fits to the observed energy distributions
with pure hydrogen models (filled circles) and with pure helium models
or mixed hydrogen/helium (open circles), with abundances indicated in
each panel. Adopted atmospheric parameters and abundances are
emphasized in red. Here and in the following figures, the photometric
observations are represented by error bars. In the right panels are
shown the observed normalized spectra together with the synthetic line
profiles calculated with the atmospheric parameters corresponding to
the pure hydrogen solutions.\label{fg:f11}}

\figcaption[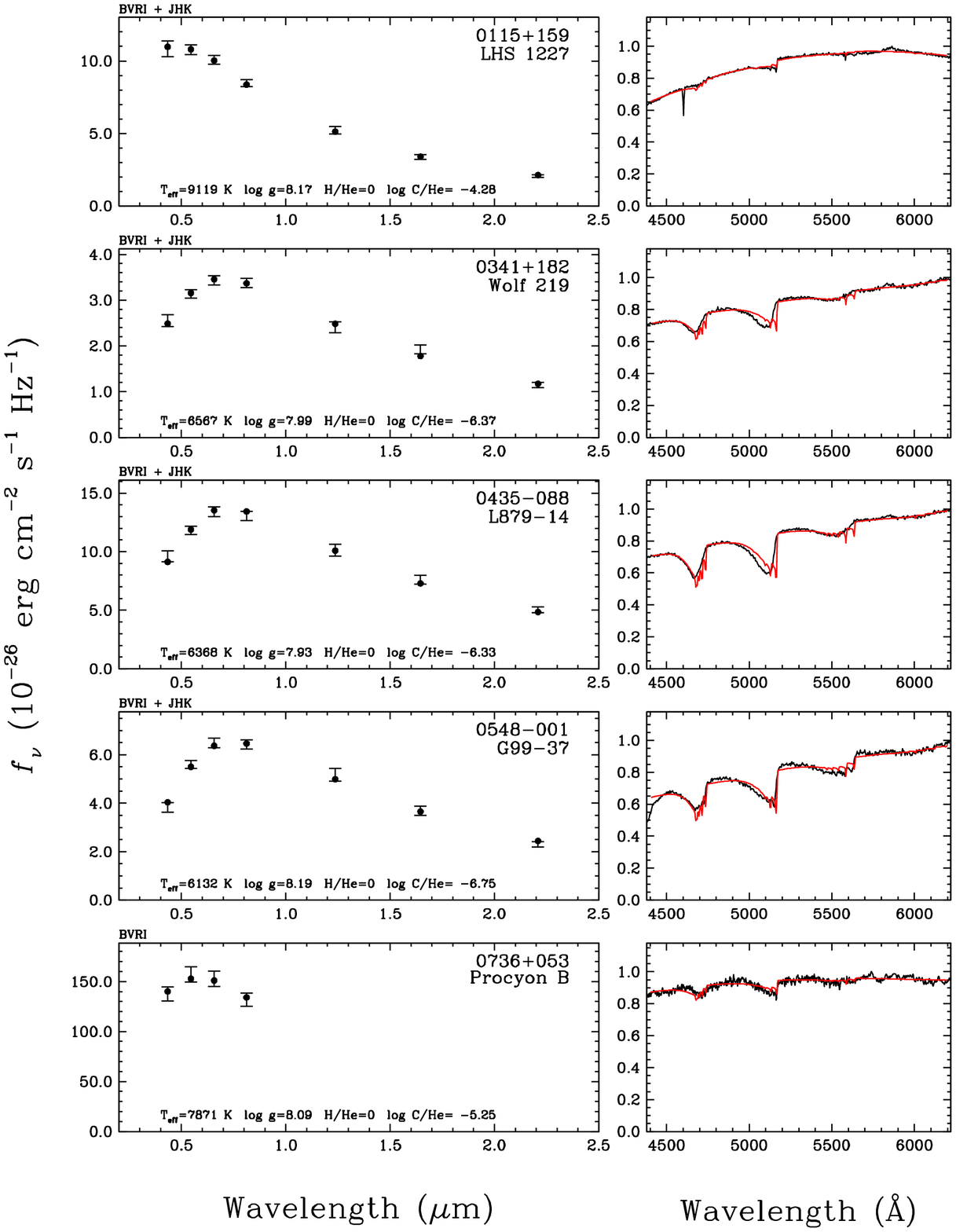] {Fits to
  the observed energy distributions of DQ stars.  The filled circles correspond to
  our best fit with the atmospheric parameters and carbon abundances
  given in each panel. In the right panels are shown the observed
  spectra together with the predicted model fit (in
  red).\label{fg:f12}}

\figcaption[f13.ps] {Best fit to the energy distribution (top) 
and optical SDSS spectrum (bottom) of GD 184 (1529+141) assuming an
unresolved double degenerate system composed of a DA and a DQ white
dwarf at $\logg=8$ for both components. The atmospheric parameters of
each component are given in the lower panel; we assume for simplicity
that both stars have $\logg=8$. In the top panel, the solid line
represents the combined monochromatic model flux, while the
contribution of each component is shown as dotted lines (DQ top, DA
bottom).\label{fg:f13}}

\figcaption[f14a.ps] {Fits to
  the energy distributions of DZ stars. The filled circles correspond to
  our best fit with the atmospheric parameters as well as hydrogen and calcium abundances
  given in each panel. In the right panels are shown the observed
  spectra together with the predicted model fit (in red); the insert
shows our fit to H$\alpha$, when detected.\label{fg:f14}}

\figcaption[f15.ps] {Fits to the optical spectra of
the DA stars in our sample. The lines range from H$\beta$
(bottom) to H8 (top), each offset vertically by a factor of 0.2.
Theoretical line profiles shown in green are not used in the
fitting procedure.\label{fg:f15}}

\figcaption[f16.ps] {Top panel: atmospheric parameters for the DA
stars in the SDSS, taken from \citet{tremblay11a}. The red curve
corresponds to a third order polynomial fit through the cool ($\Te<
14,000$~K) objects using temperature bins of 500 K (see text for
details). The dashed line shows a 0.594 \msun\ evolutionary track,
which corresponds to the median mass of the hotter DA stars
in this sample. Bottom panel: same results but with the $\logg$ correction applied
to the coolest DA stars.\label{fg:f16}}

\figcaption[f17.ps] {Comparison of effective
temperatures and absolute magnitudes for the 22 white dwarfs in our
sample (all of the DA type) that have both spectroscopic and
photometric determinations. The objects labeled in the figure are
discussed in the text.\label{fg:f17}}

\figcaption[f18.ps] {Comparison of our
distance estimates with those from \citet{sion09}. The dotted line
indicates the 1:1 correspondence, while the dashed line represents the
20 pc limit of the nearby sample defined in this study. Error bars are
shown only for stars beyond this distance.\label{fg:f18}}

\figcaption[f19.ps] {Masses of all stars in the local
sample ($D\lesssim20$ pc) as a function of effective temperature. 
The various symbols are defined in the legend; known
unresolved double degenerate binaries (DD) are shown in
red. Also shown are theoretical isochrones labeled in Gyr;
solid lines correspond to white dwarf cooling ages only, while the
dotted lines also include the main sequence lifetime. \label{fg:f19}}

\figcaption[f20.ps] {Left panel: total
number of white dwarfs (solid-line histogram) and hydrogen-atmosphere white
dwarfs (hatched histogram) as a function of effective temperature. Right
panel: ratio of helium-atmosphere white dwarfs to the total number
of stars as a function of effective temperature.\label{fg:f20}}

\figcaption[f21.ps] {Mass distribution for
the white dwarf stars in the local sample. The individual
contributions of the hydrogen-atmosphere (red) and helium-atmosphere (blue)
white dwarfs are shown in the left and right panels, respectively.
Mean values and standard deviations of the three distributions are
indicated in the figure.\label{fg:f21}}

\figcaption[f22.ps] {Luminosity function for our
sample of nearby white dwarfs as a function of $M_{\rm
bol}$ (solid line), compared to the luminosity function obtained
by \cite{har06} for white dwarfs in the SDSS (dashed line). For
the local sample, the number of stars in each magnitude bin is
also given. The temperature scale 
assuming $M=0.6$ \msun\ is also shown at the top
of the figure.\label{fg:f22}}

\clearpage
\begin{figure}[p]
\plotone{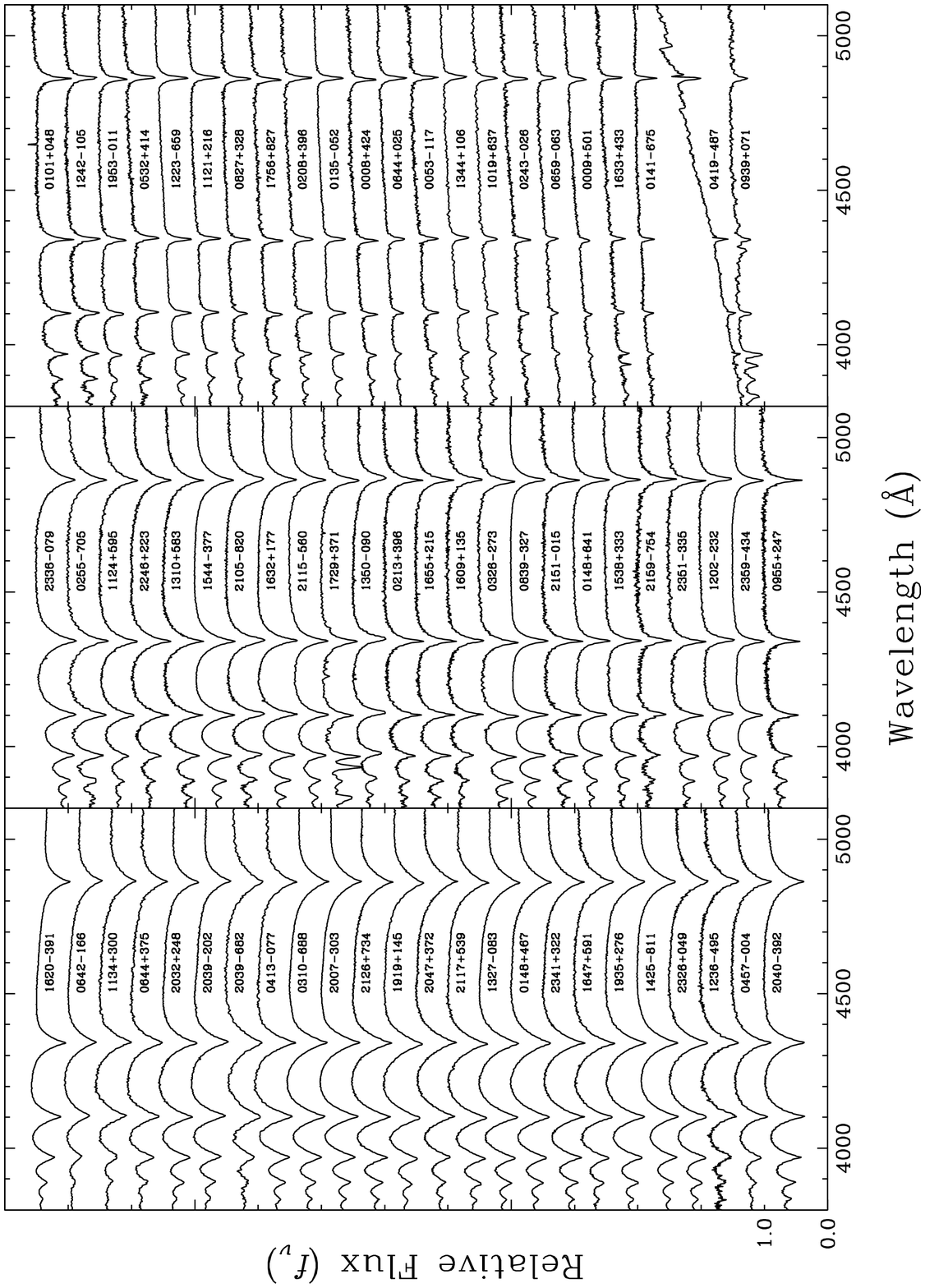}
\begin{flushright}
Figure \ref{fg:f1}
\end{flushright}
\end{figure}

\clearpage
\begin{figure}[p]
\plotone{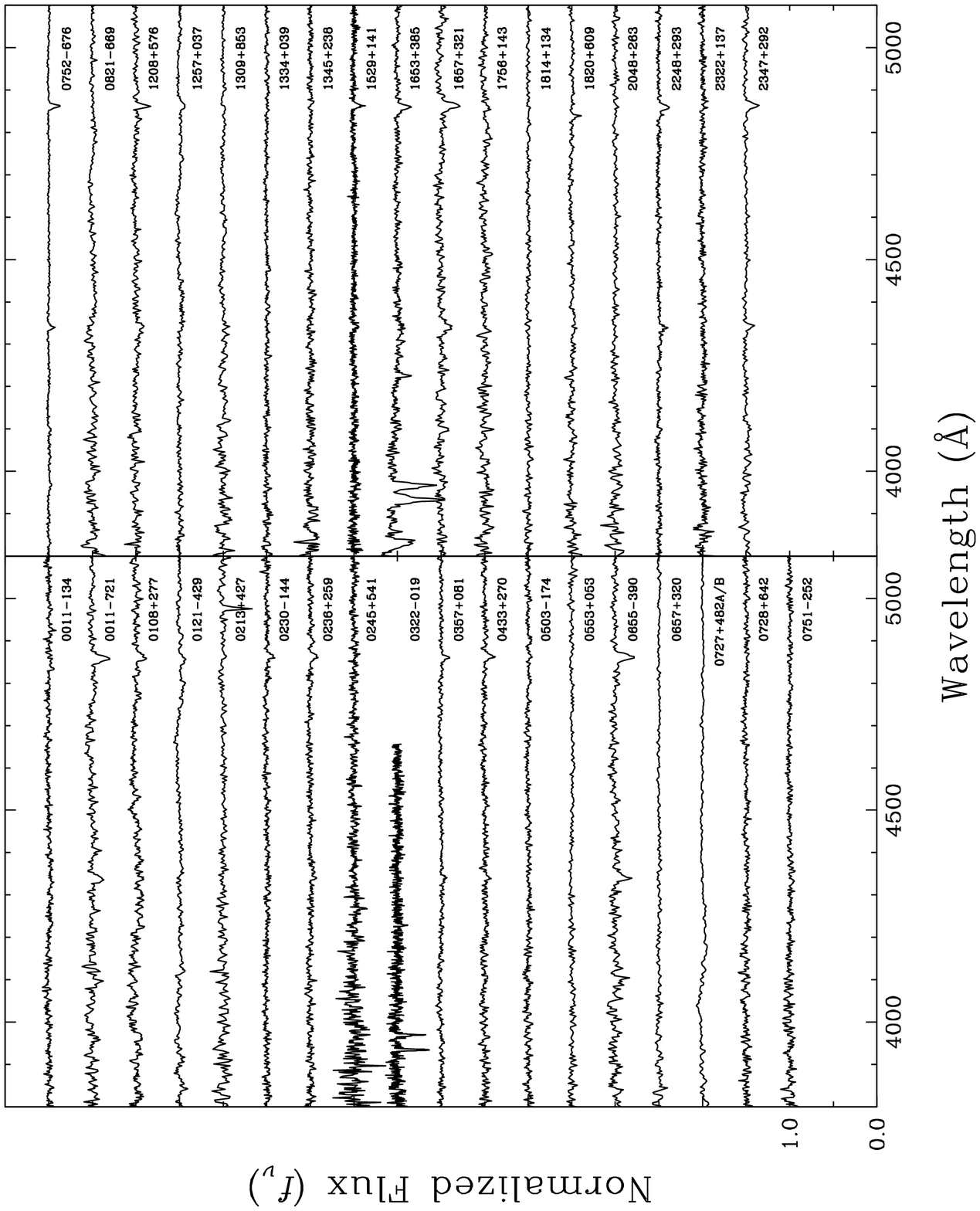}
\begin{flushright}
Figure \ref{fg:f2}
\end{flushright}
\end{figure}

\clearpage
\begin{figure}[p]
\plotone{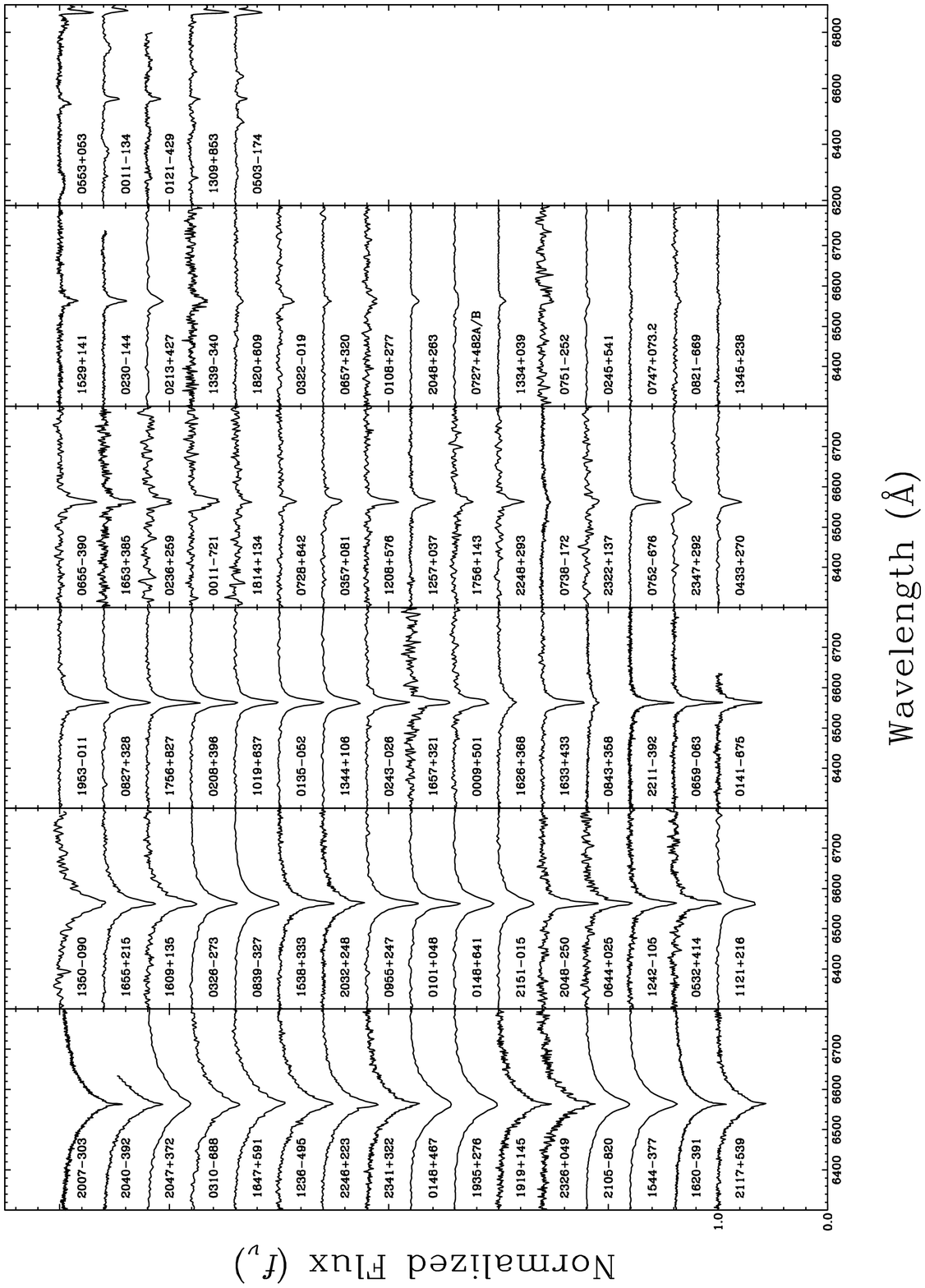}
\begin{flushright}
Figure \ref{fg:f3}
\end{flushright}
\end{figure}

\clearpage
\begin{figure}[p]
\plotone{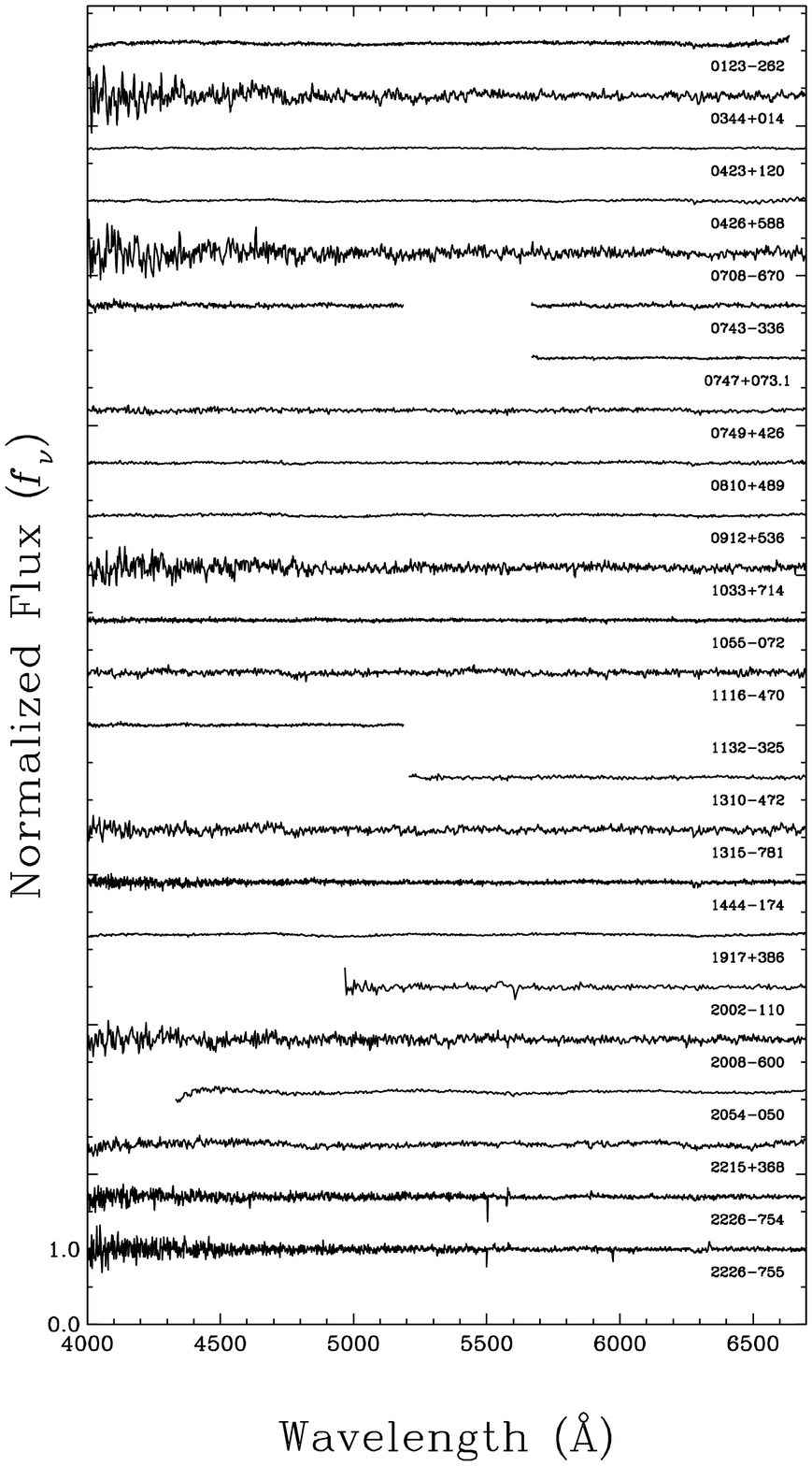}
\begin{flushright}
Figure \ref{fg:f4}
\end{flushright}
\end{figure}

\clearpage
\begin{figure}[p]
\plotone{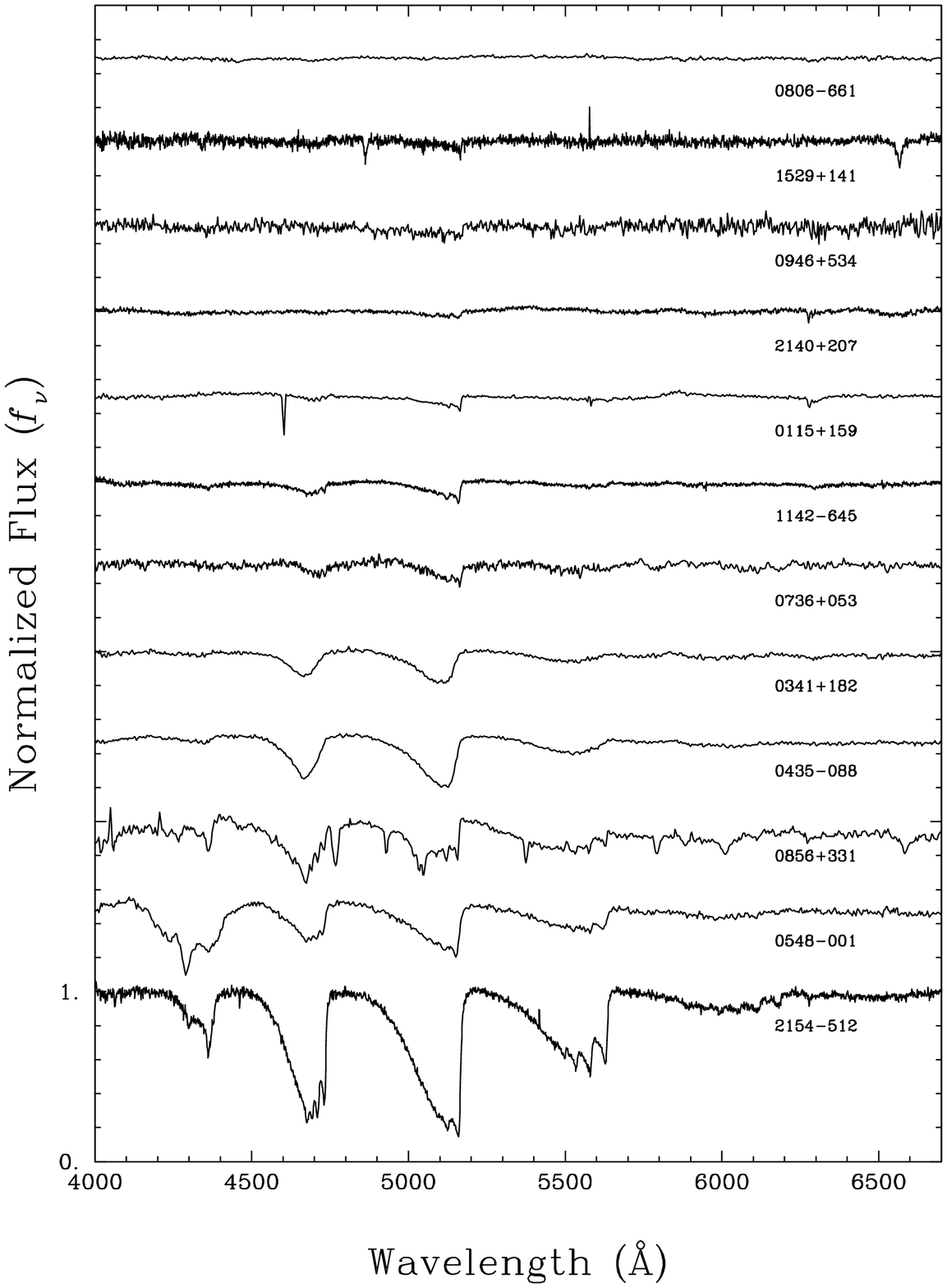}
\begin{flushright}
Figure \ref{fg:f5}
\end{flushright}
\end{figure}

\clearpage
\begin{figure}[p]
\plotone{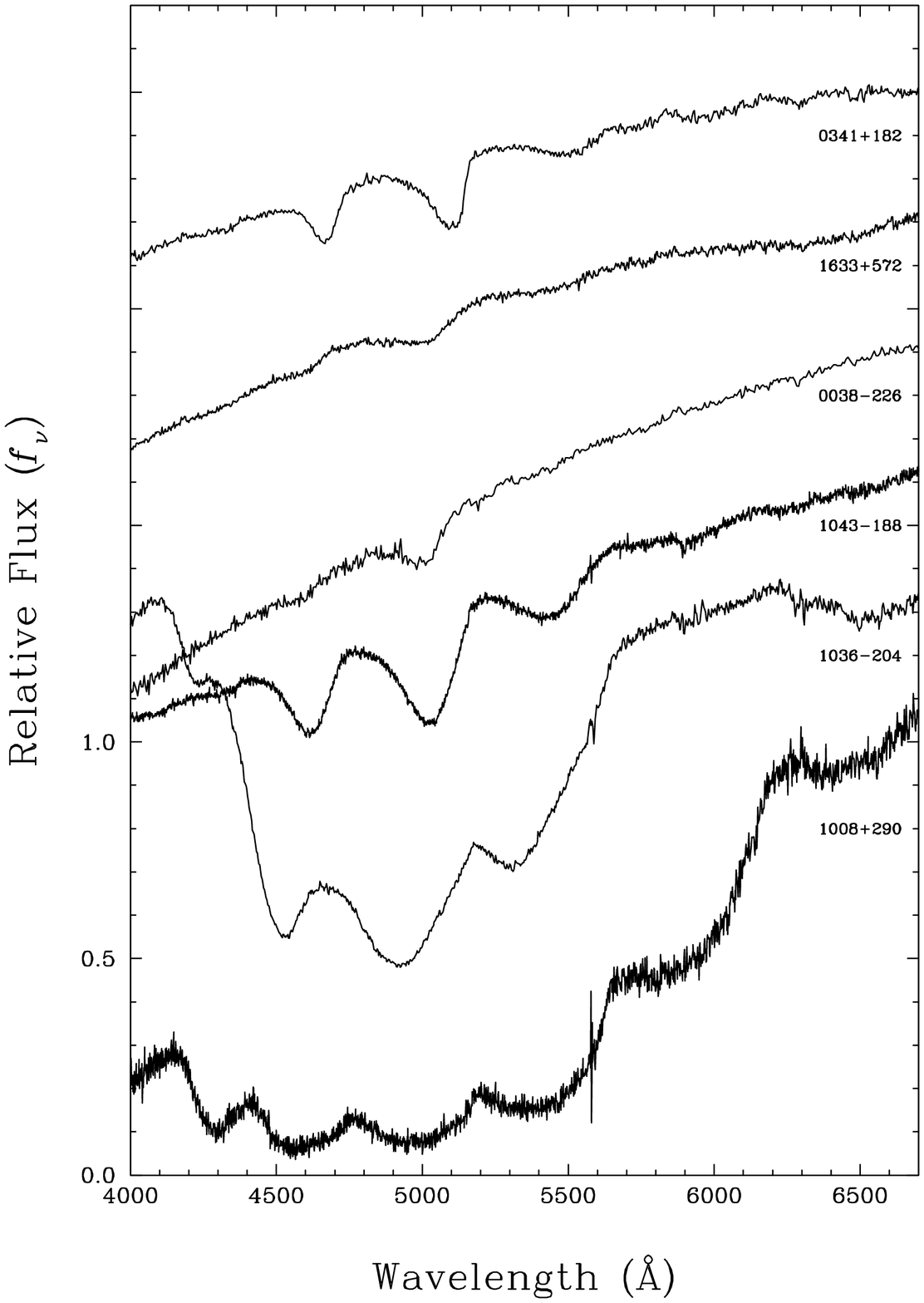}
\begin{flushright}
Figure \ref{fg:f6}
\end{flushright}
\end{figure}

\clearpage
\begin{figure}[p]
\plotone{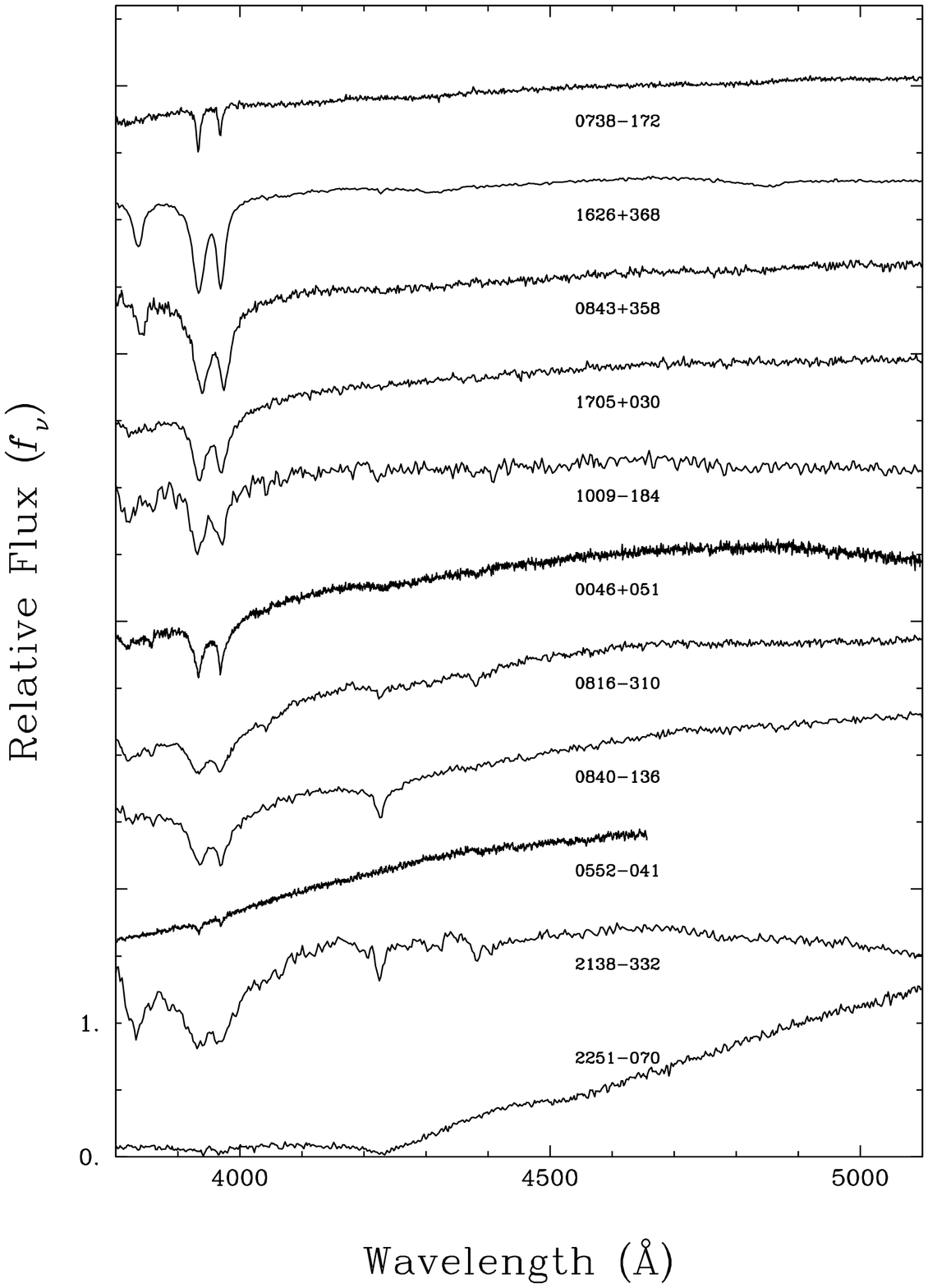}
\begin{flushright}
Figure \ref{fg:f7}
\end{flushright}
\end{figure}

\clearpage
\begin{figure}[p]
\plotone{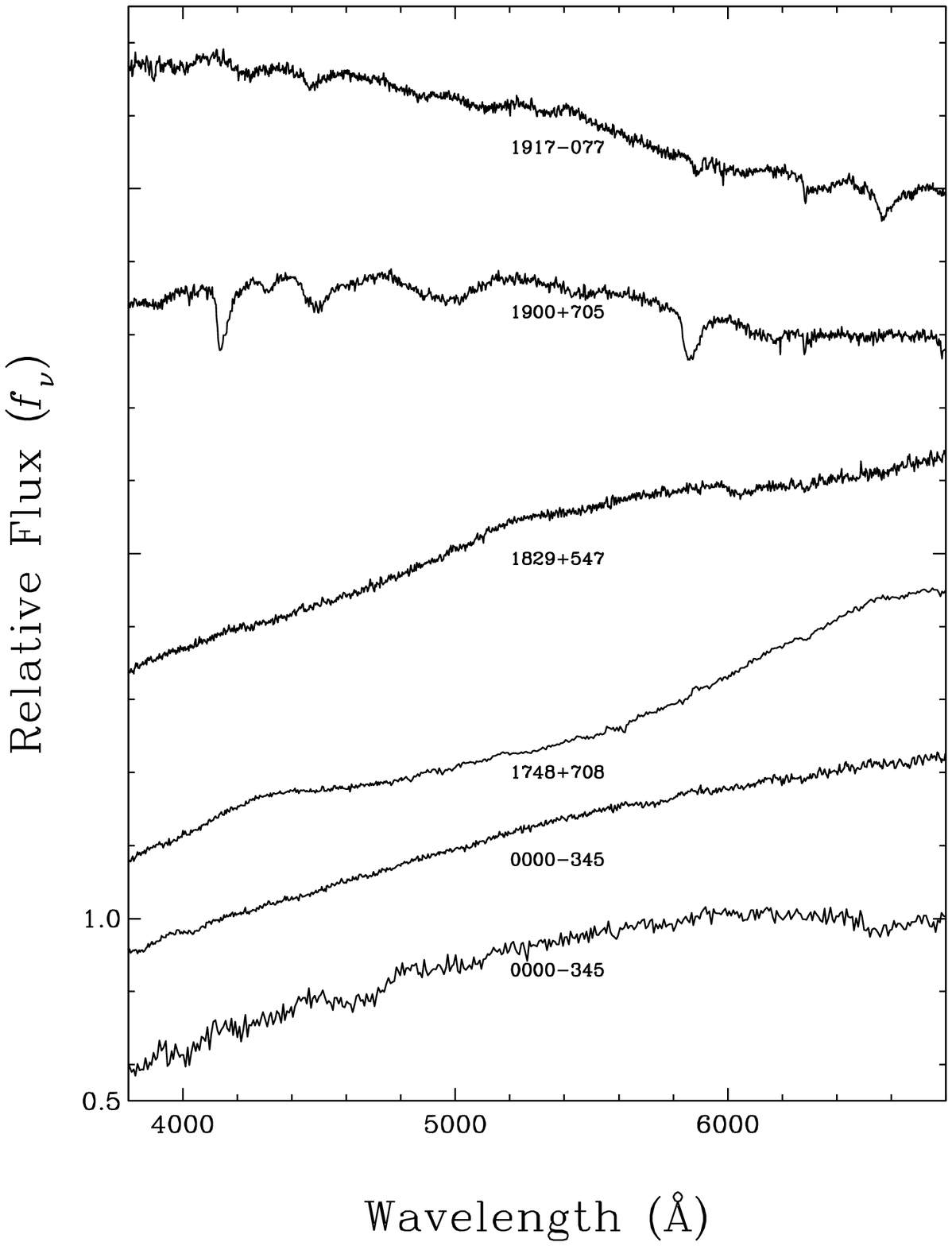}
\begin{flushright}
Figure \ref{fg:f8}
\end{flushright}
\end{figure}

\clearpage
\begin{figure}[p]
\plotone{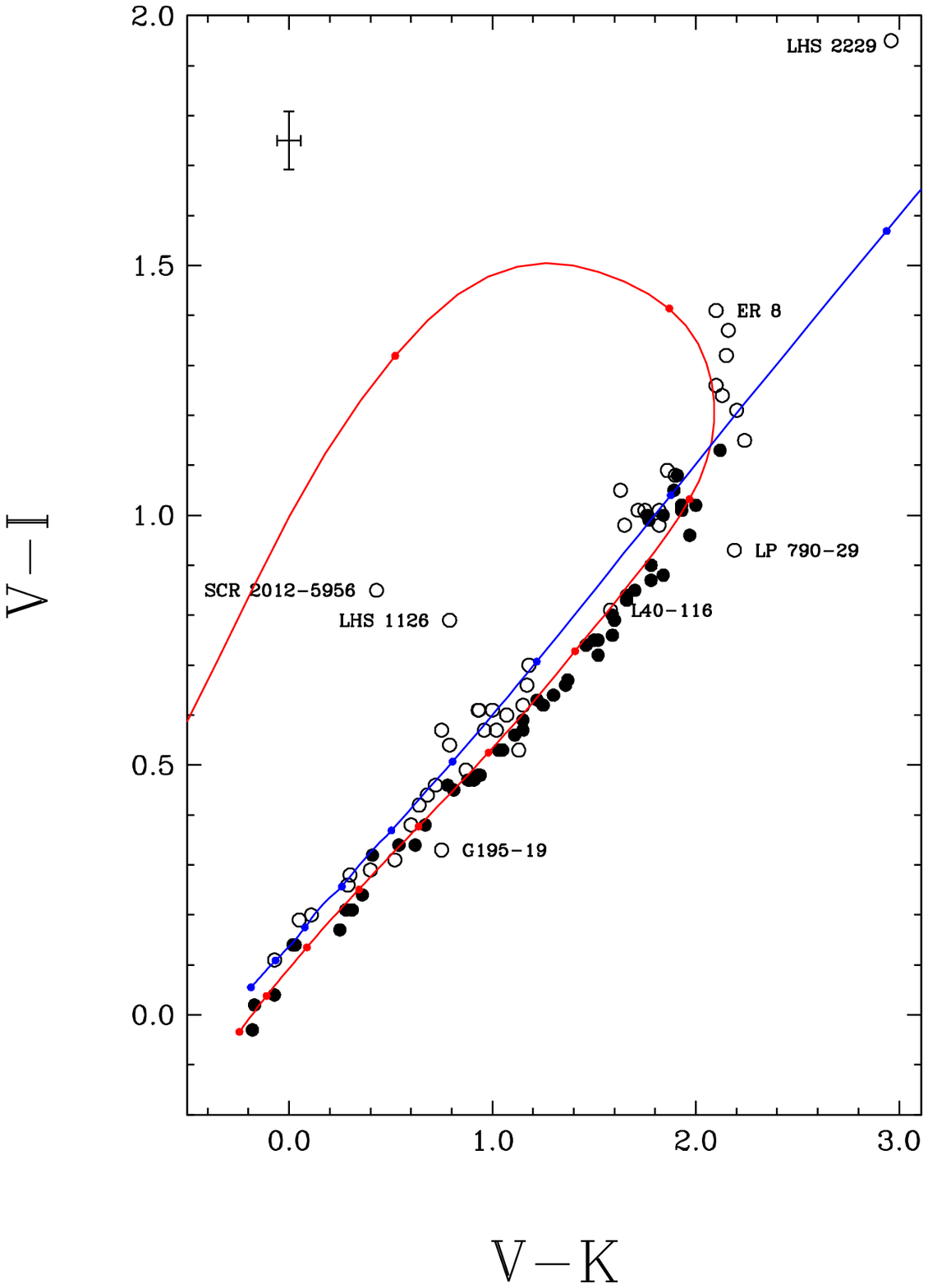}
\begin{flushright}
Figure \ref{fg:f9}
\end{flushright}
\end{figure}

\clearpage
\begin{figure}[p]
\plotone{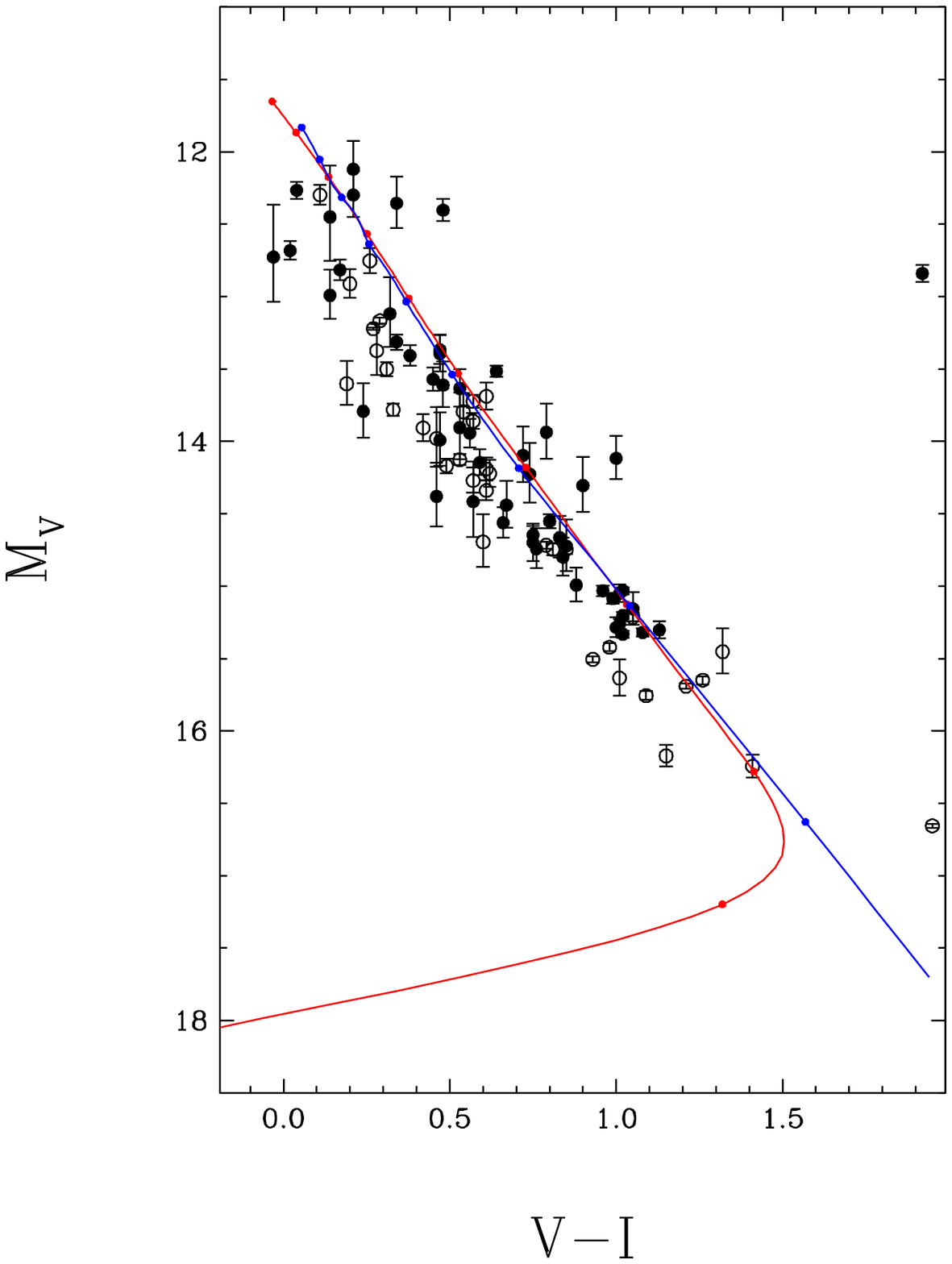}
\begin{flushright}
Figure \ref{fg:f10}
\end{flushright}
\end{figure}

\clearpage
\begin{figure}[p]
\plotone{f11a}
\begin{flushright}
Figure \ref{fg:f11}a
\end{flushright}
\end{figure}

\clearpage
\begin{figure}[p]
\plotone{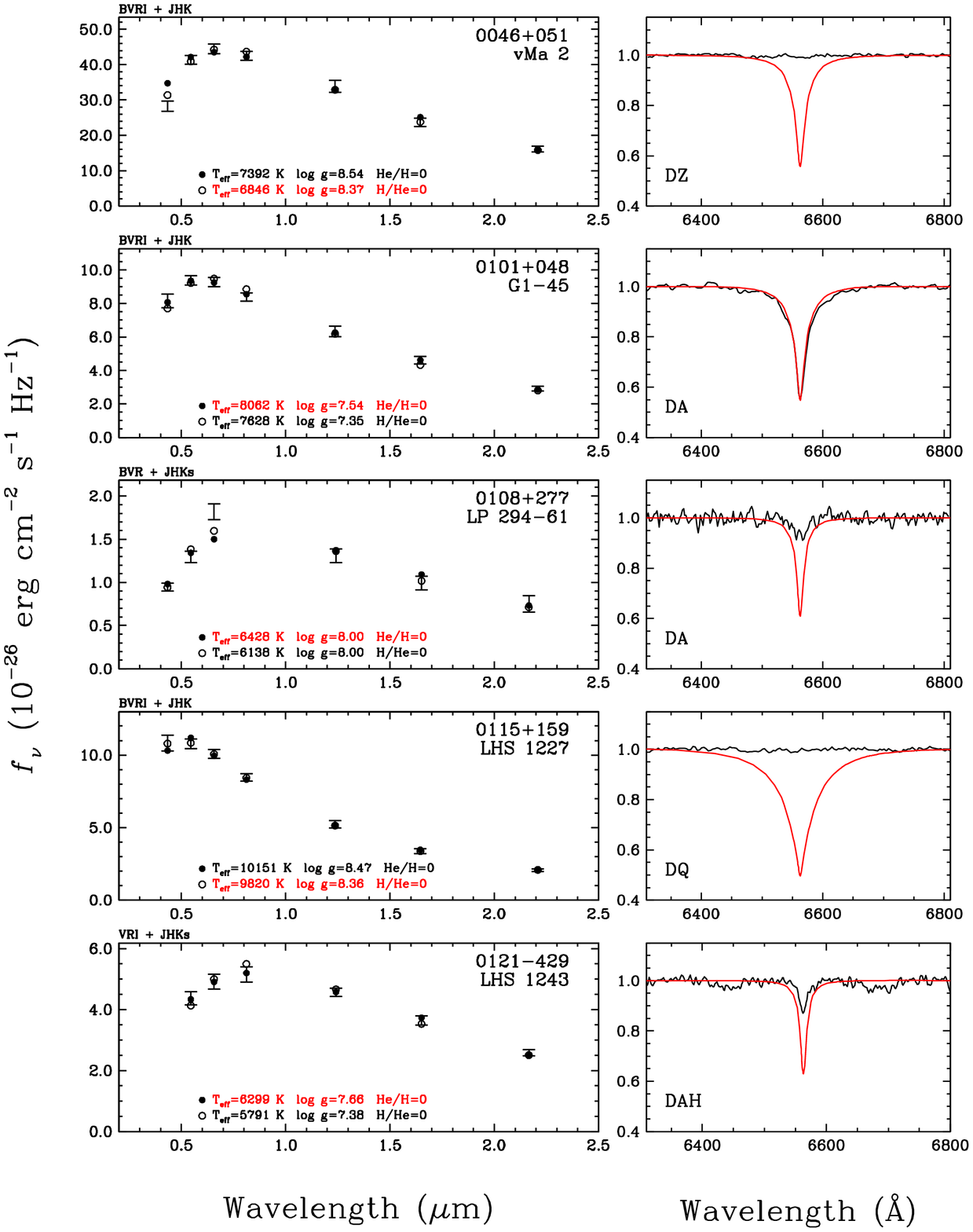}
\begin{flushright}
Figure \ref{fg:f11}b
\end{flushright}
\end{figure}

\clearpage
\begin{figure}[p]
\plotone{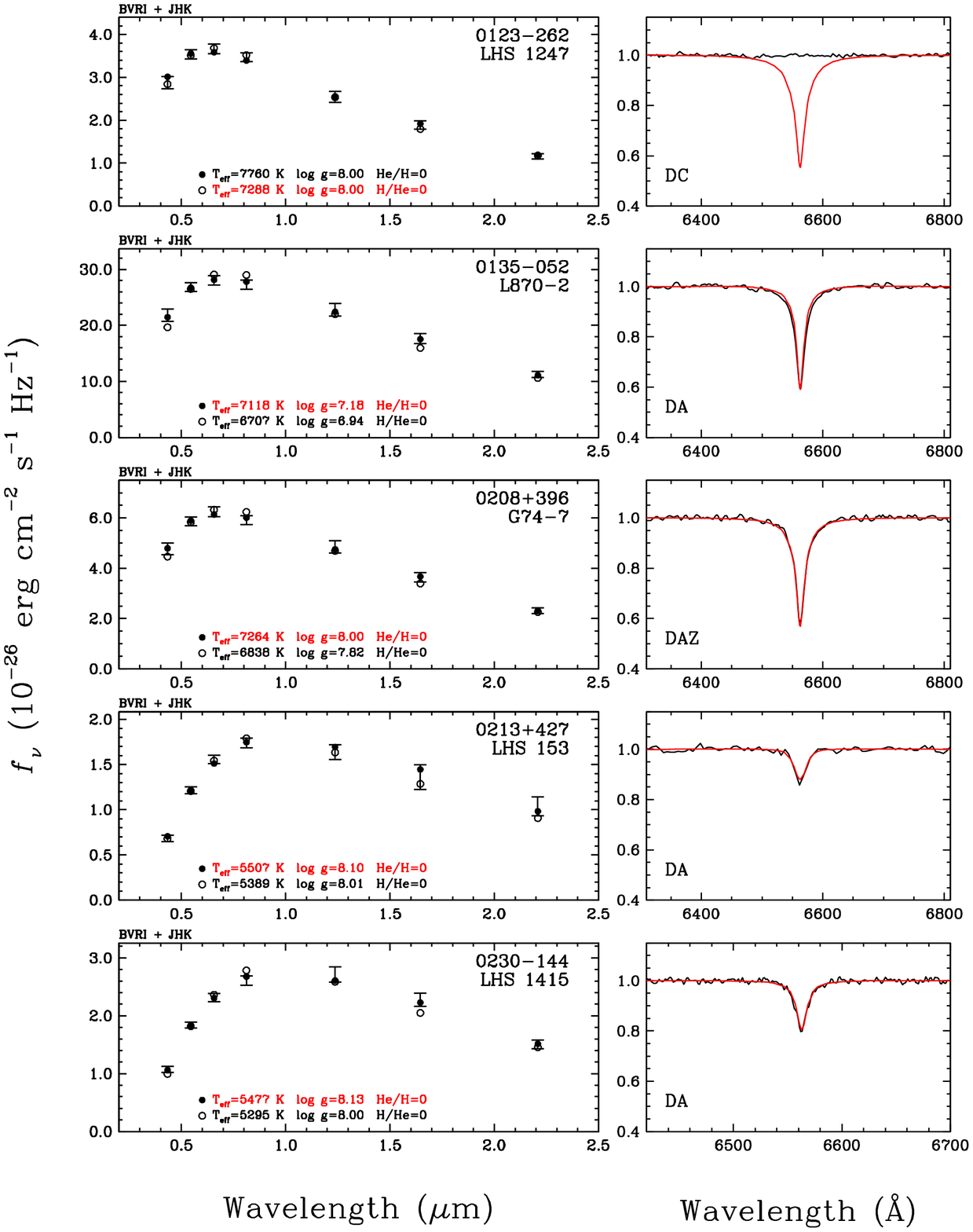}
\begin{flushright}
Figure \ref{fg:f11}c
\end{flushright}
\end{figure}

\clearpage
\begin{figure}[p]
\plotone{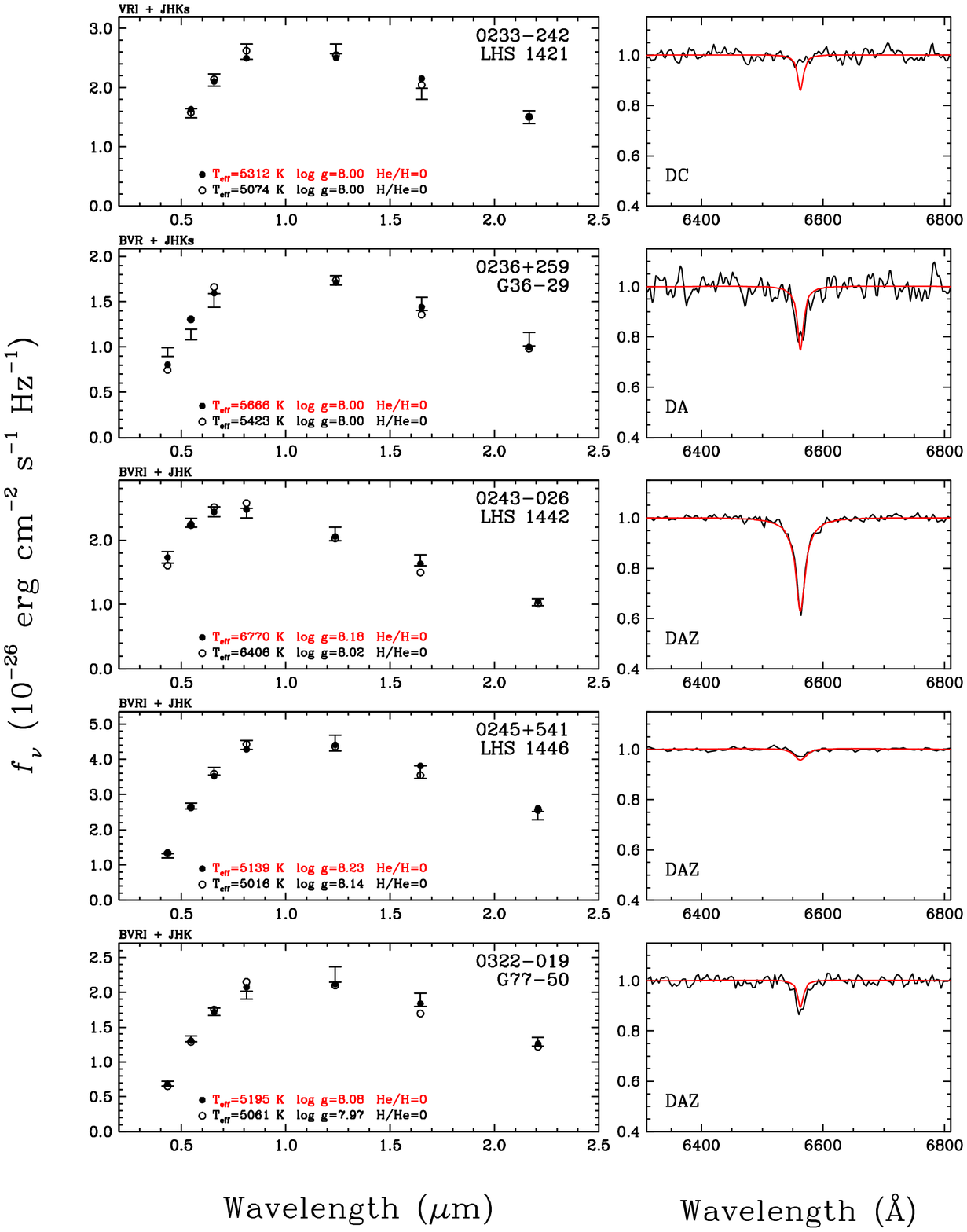}
\begin{flushright}
Figure \ref{fg:f11}d
\end{flushright}
\end{figure}

\clearpage
\begin{figure}[p]
\plotone{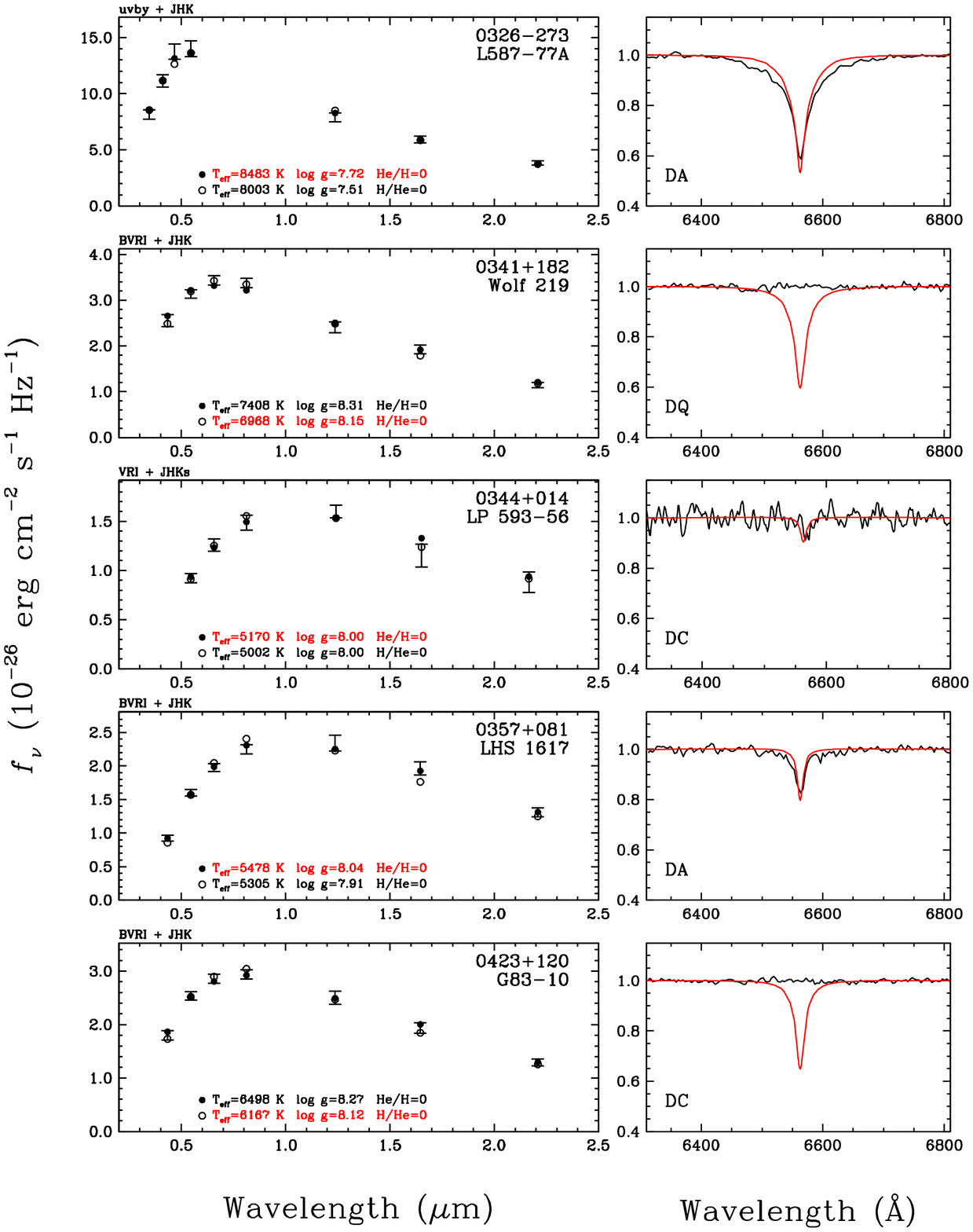}
\begin{flushright}
Figure \ref{fg:f11}e
\end{flushright}
\end{figure}

\clearpage
\begin{figure}[p]
\plotone{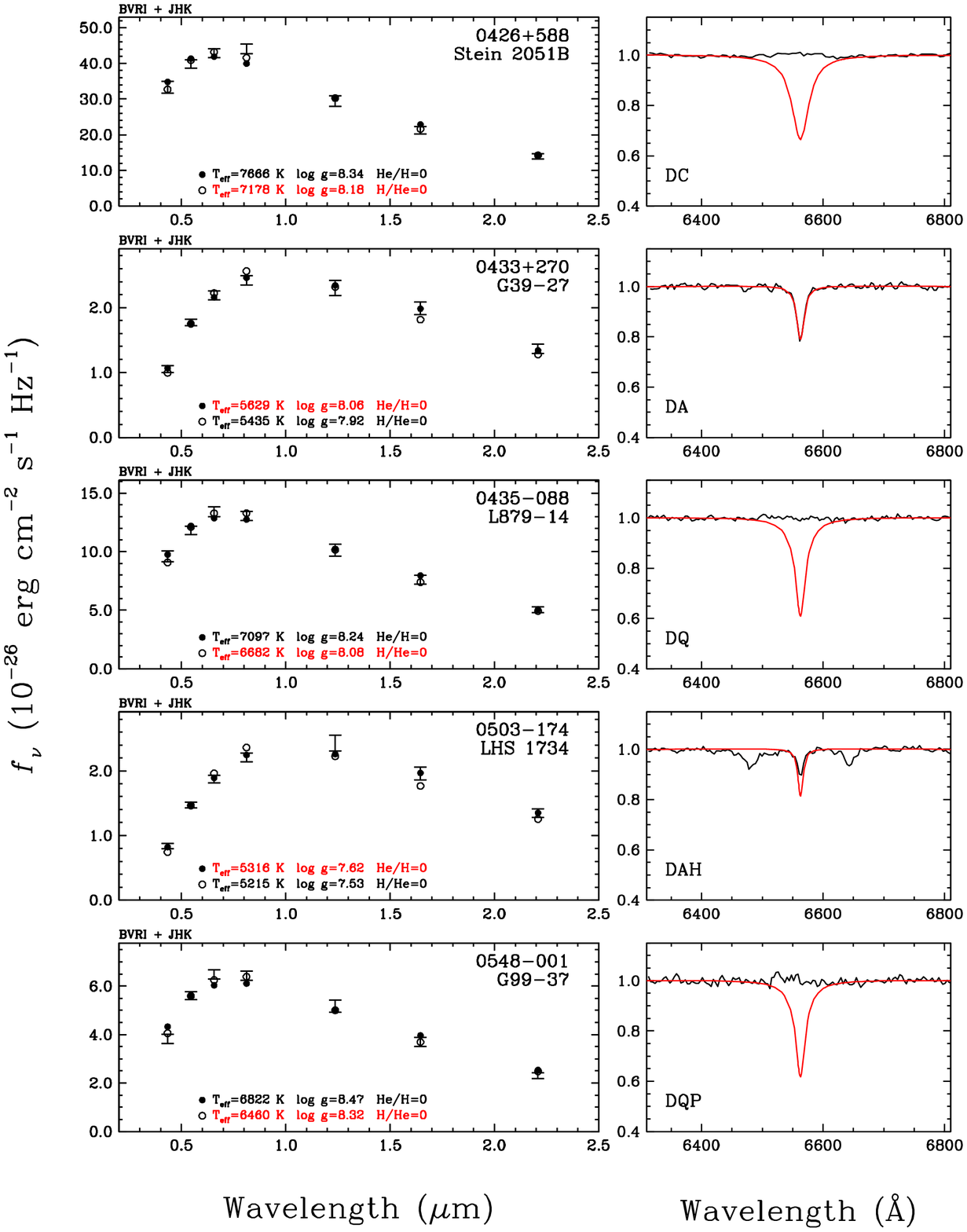}
\begin{flushright}
Figure \ref{fg:f11}f
\end{flushright}
\end{figure}

\clearpage
\begin{figure}[p]
\plotone{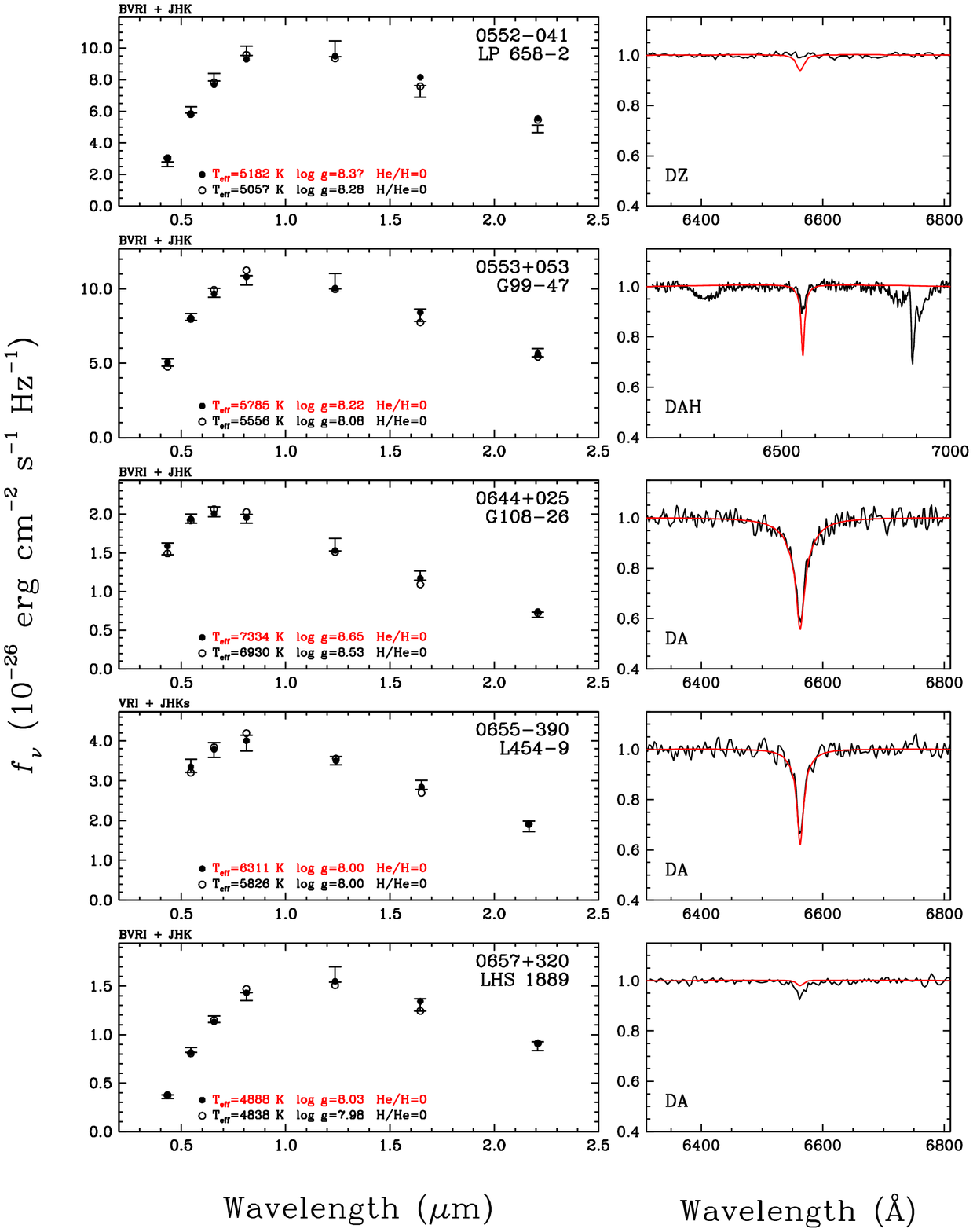}
\begin{flushright}
Figure \ref{fg:f11}g
\end{flushright}
\end{figure}

\clearpage
\begin{figure}[p]
\plotone{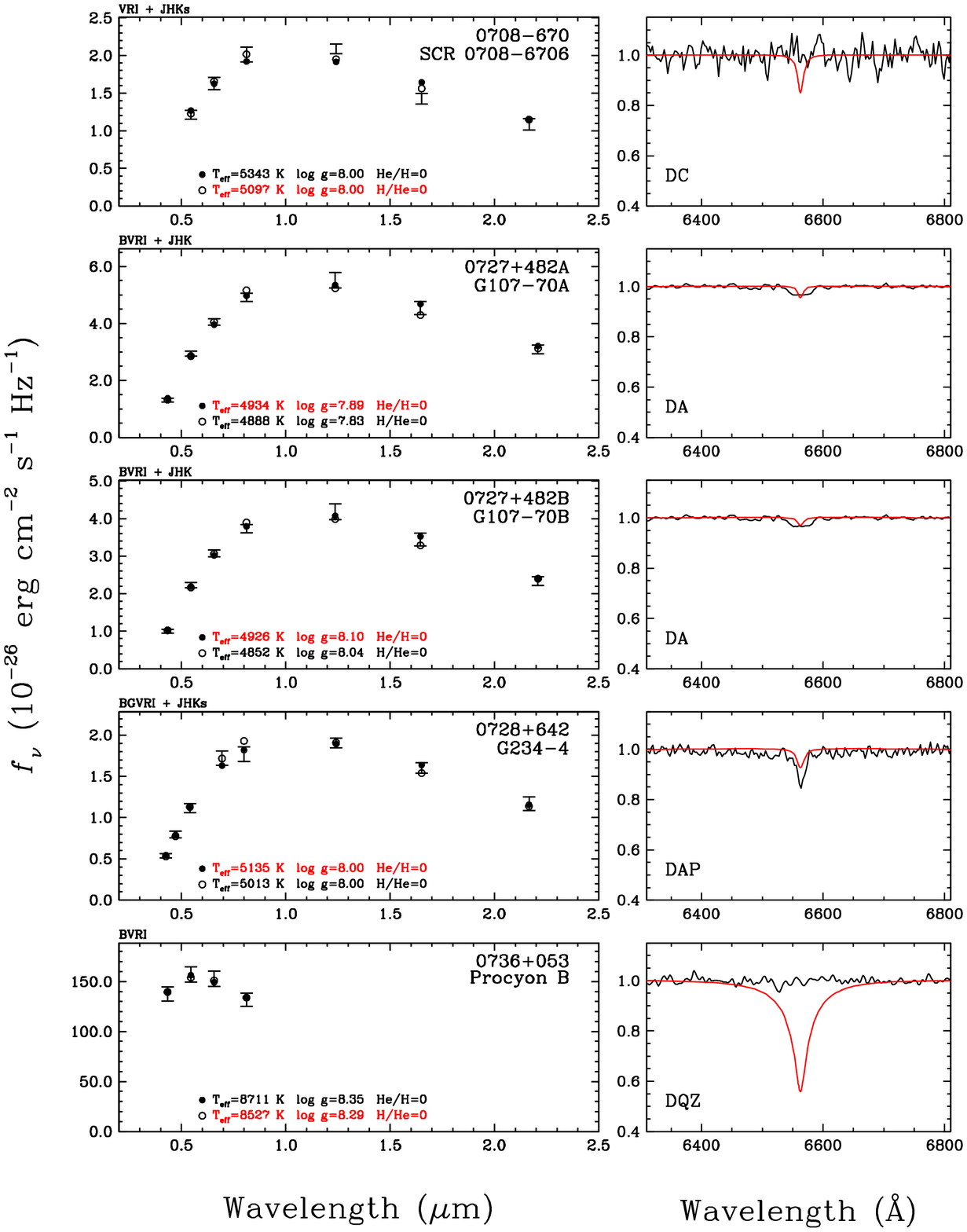}
\begin{flushright}
Figure \ref{fg:f11}h
\end{flushright}
\end{figure}

\clearpage
\begin{figure}[p]
\plotone{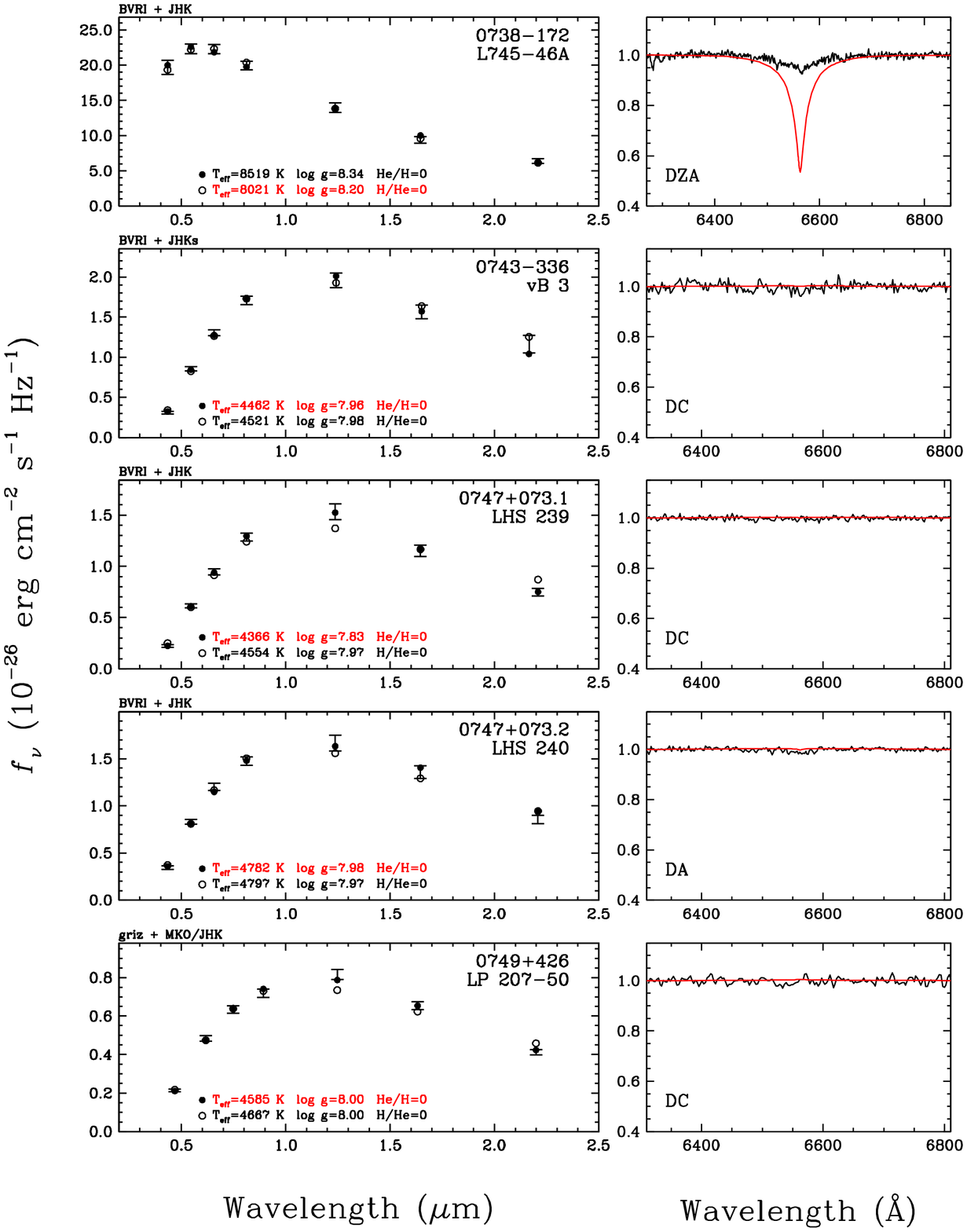}
\begin{flushright}
Figure \ref{fg:f11}i
\end{flushright}
\end{figure}

\clearpage
\begin{figure}[p]
\plotone{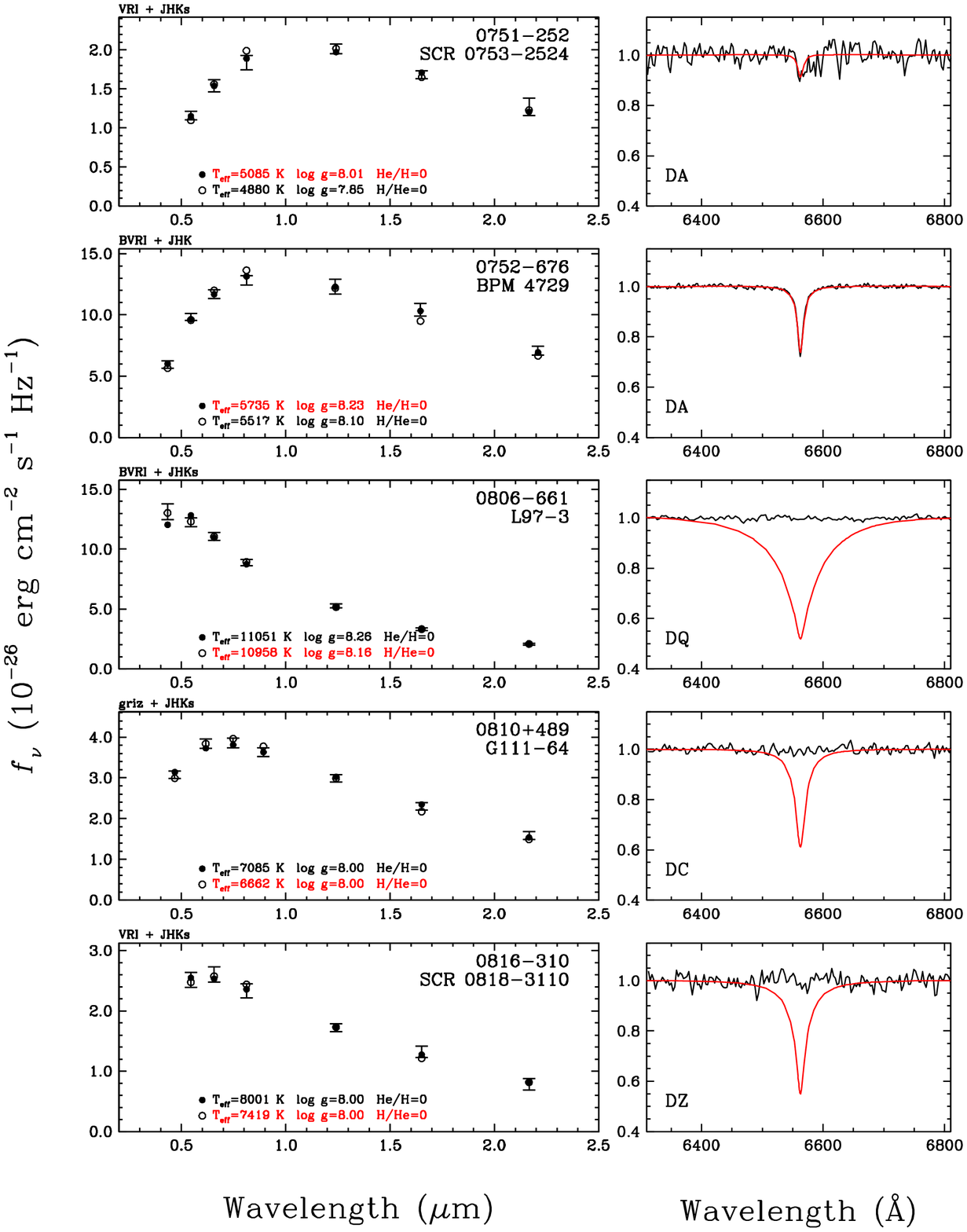}
\begin{flushright}
Figure \ref{fg:f11}j
\end{flushright}
\end{figure}

\clearpage
\begin{figure}[p]
\plotone{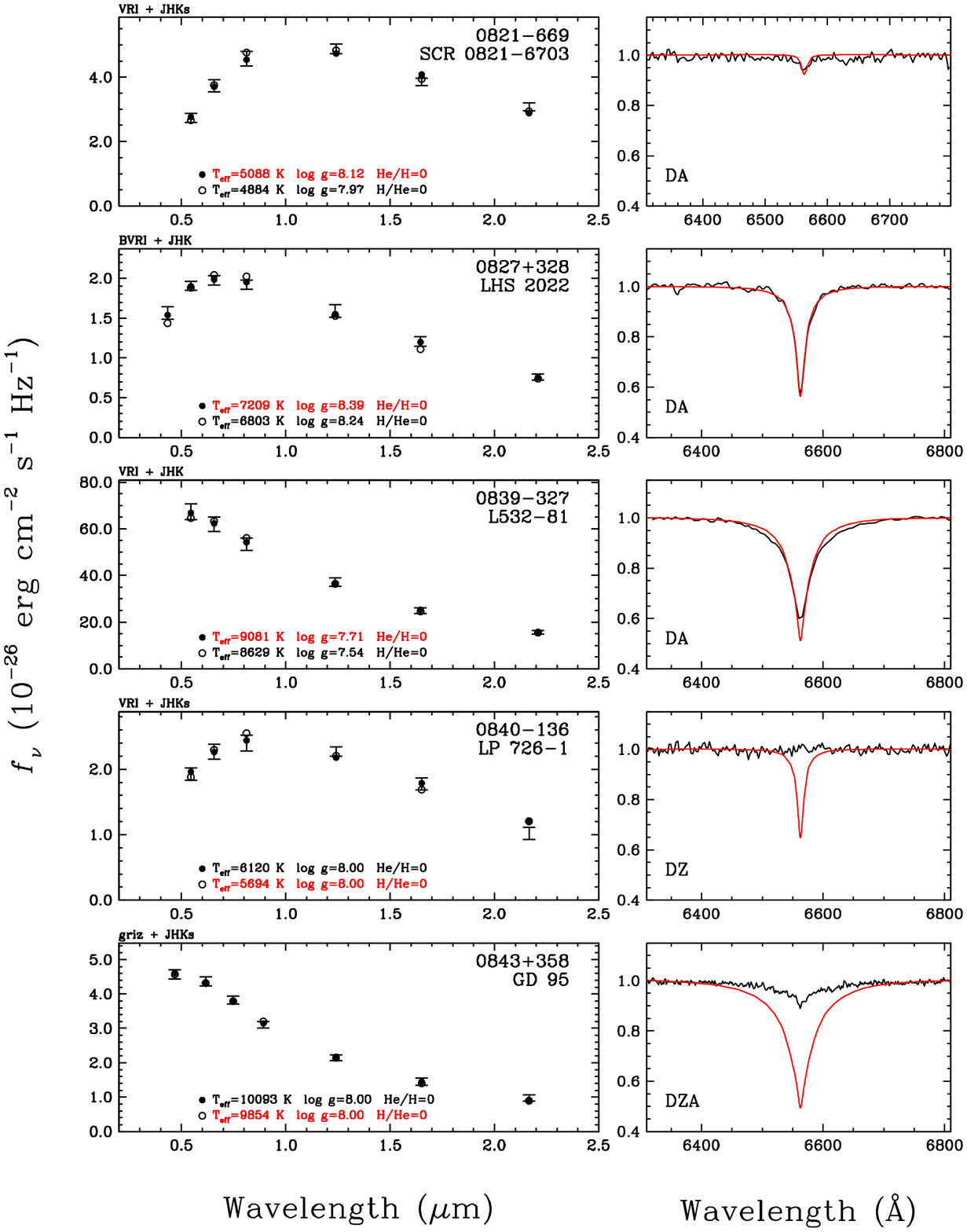}
\begin{flushright}
Figure \ref{fg:f11}k
\end{flushright}
\end{figure}

\clearpage
\begin{figure}[p]
\plotone{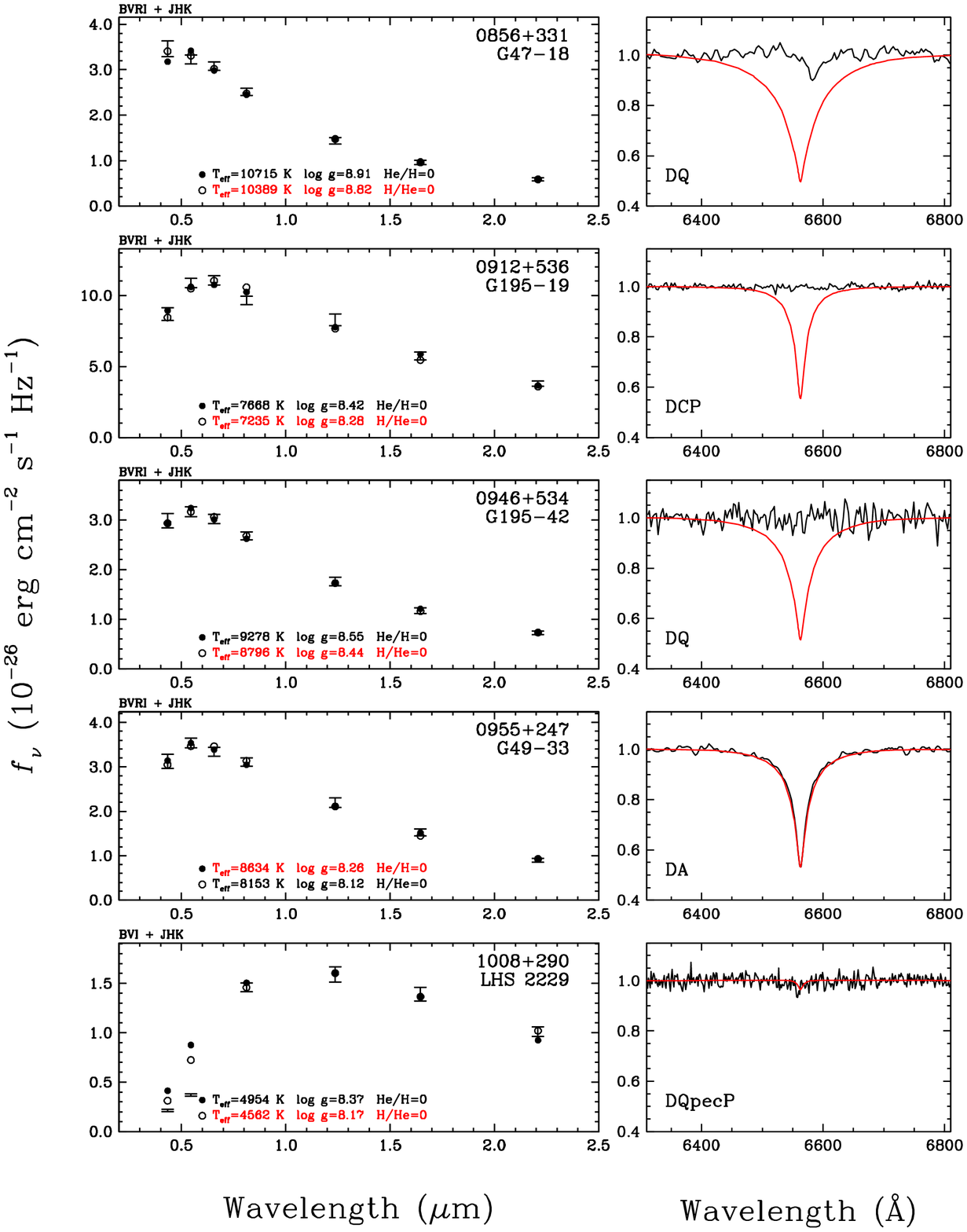}
\begin{flushright}
Figure \ref{fg:f11}l
\end{flushright}
\end{figure}

\clearpage
\begin{figure}[p]
\plotone{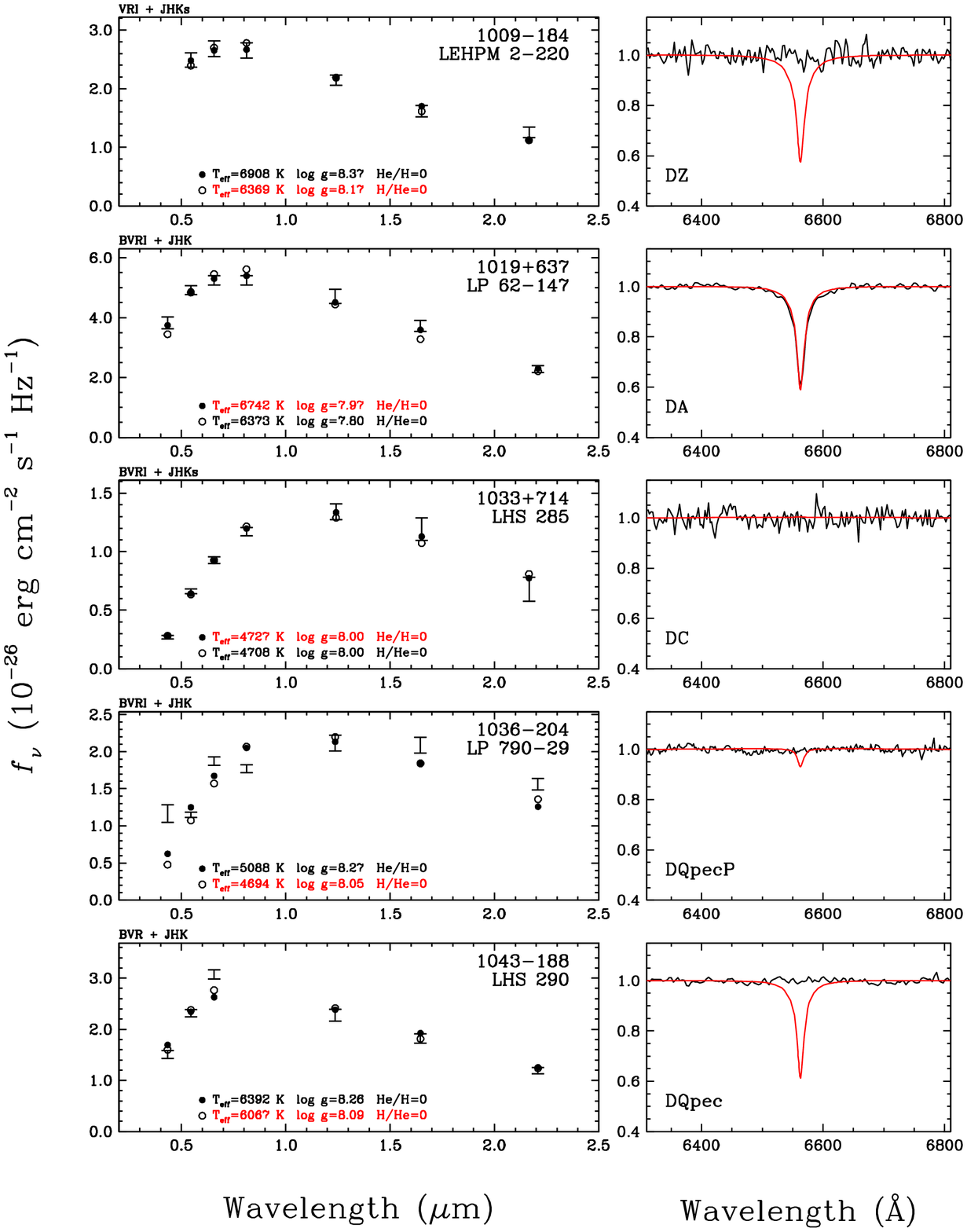}
\begin{flushright}
Figure \ref{fg:f11}m
\end{flushright}
\end{figure}

\clearpage
\begin{figure}[p]
\plotone{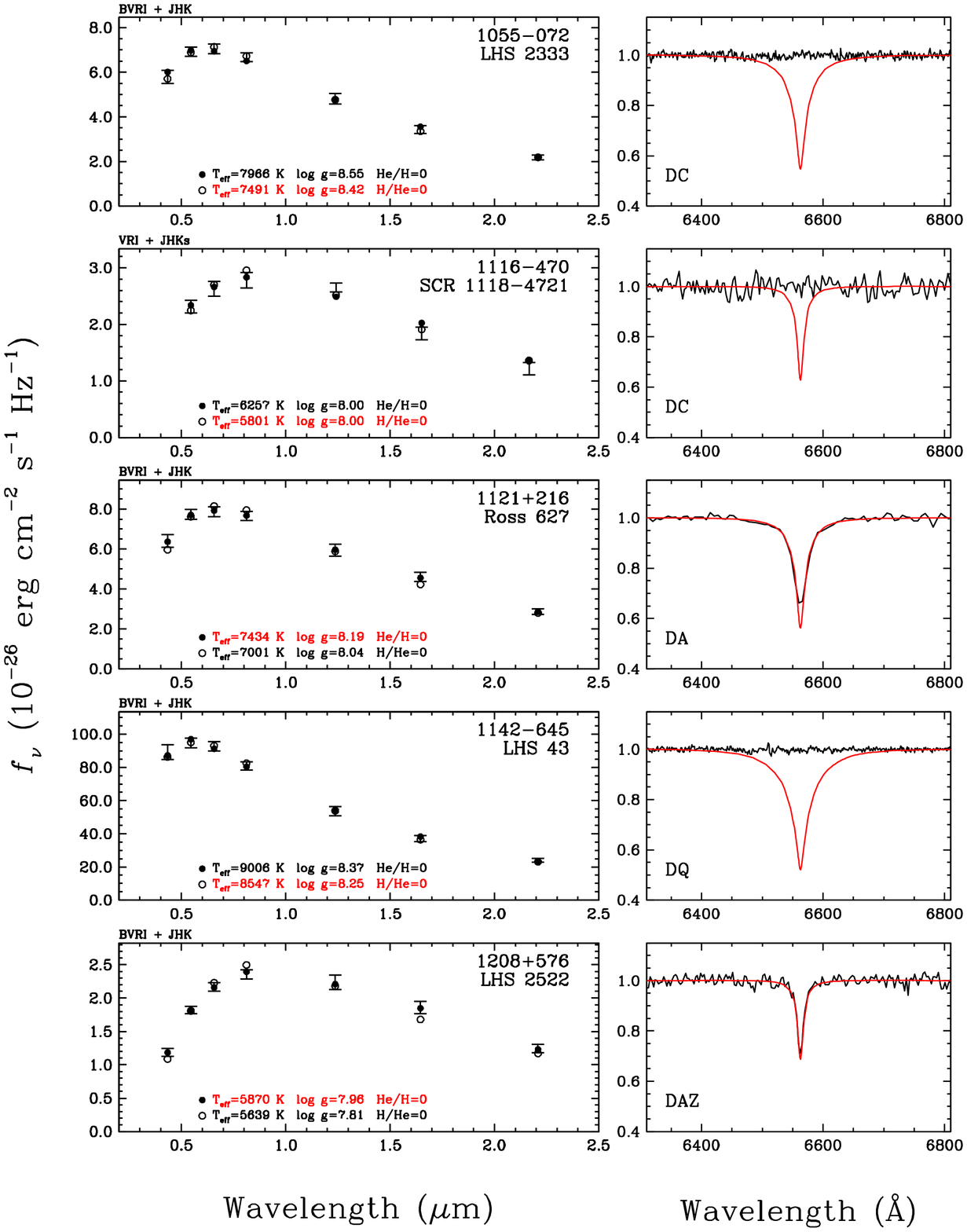}
\begin{flushright}
Figure \ref{fg:f11}n
\end{flushright}
\end{figure}

\clearpage
\begin{figure}[p]
\plotone{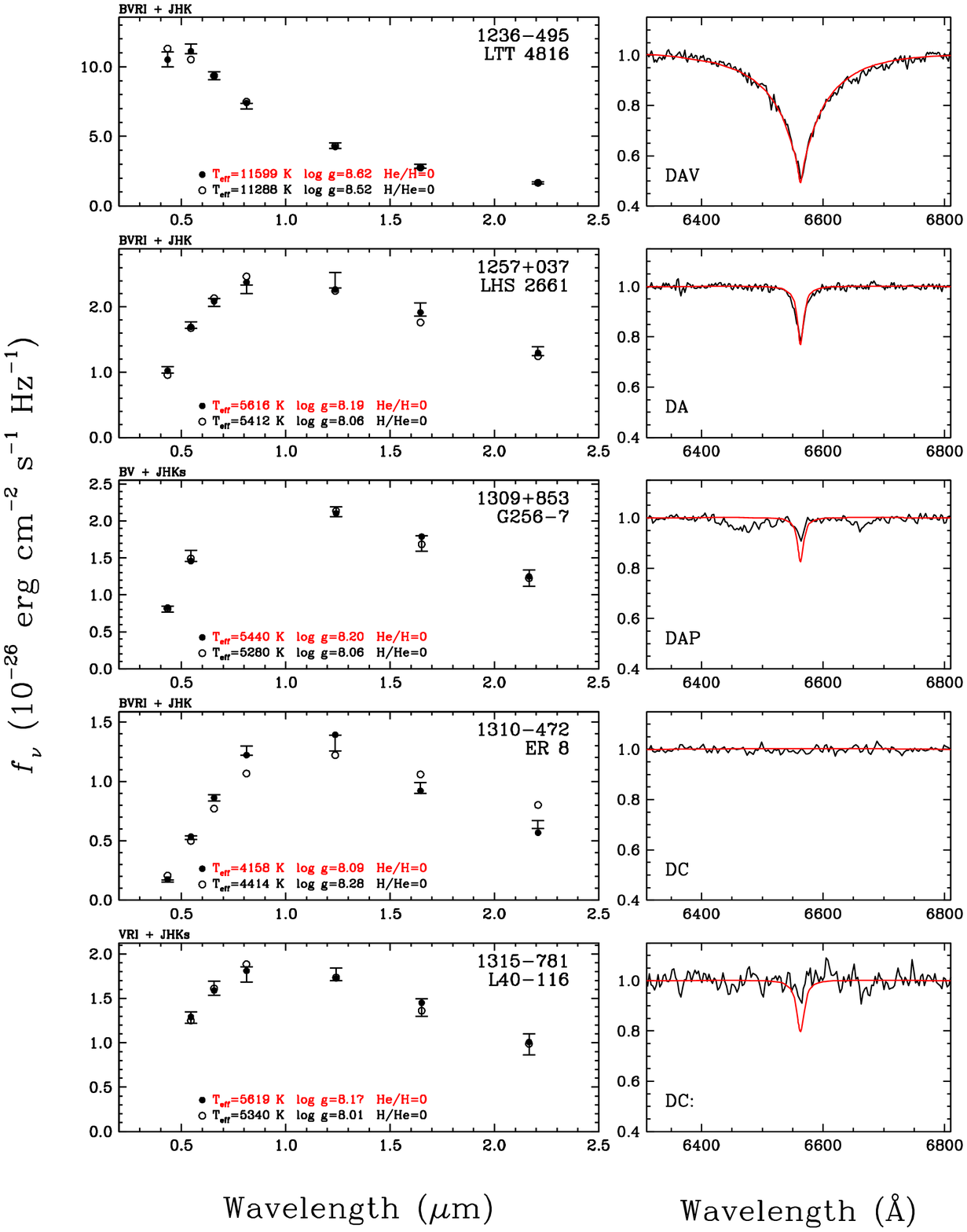}
\begin{flushright}
Figure \ref{fg:f11}o
\end{flushright}
\end{figure}

\clearpage
\begin{figure}[p]
\plotone{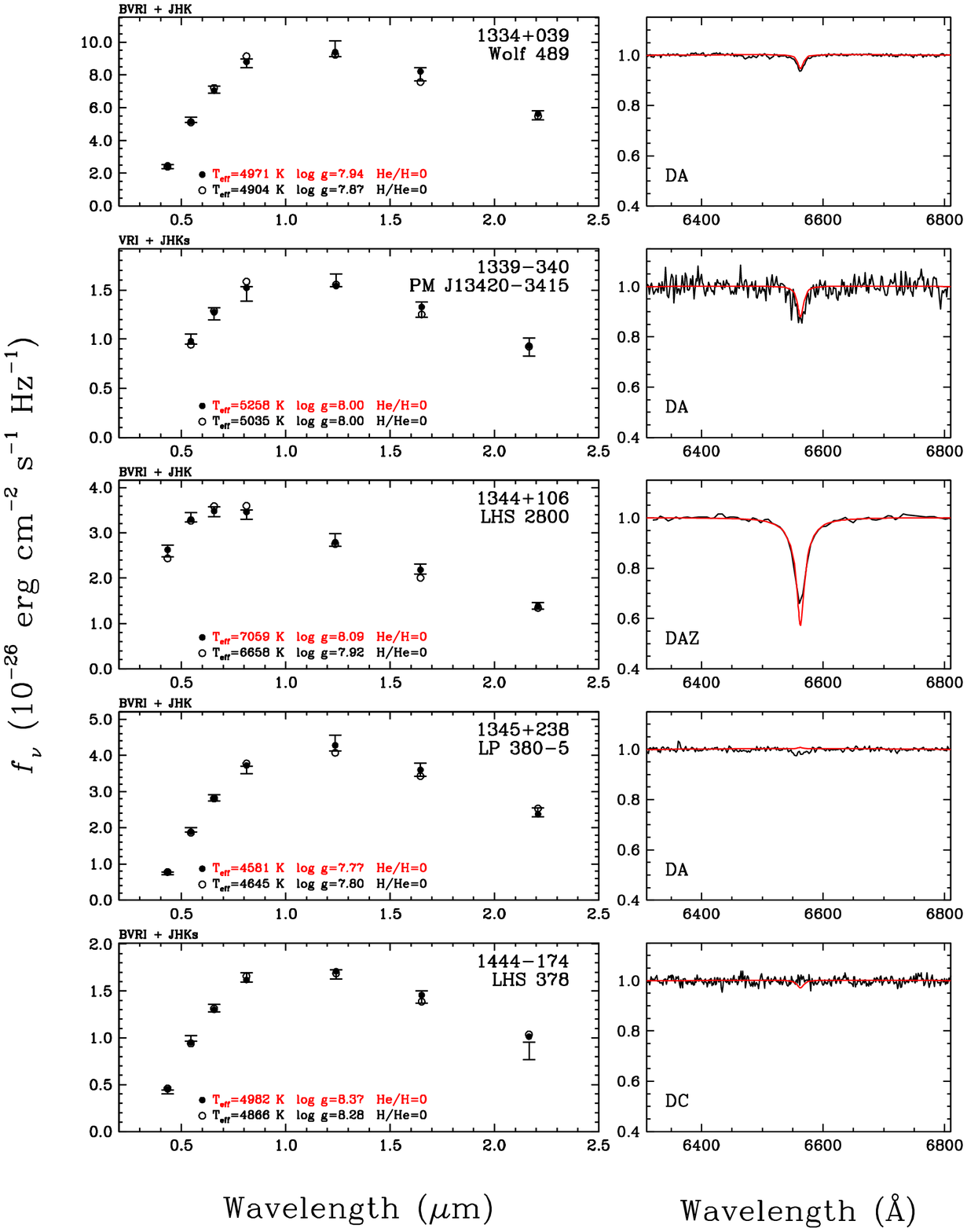}
\begin{flushright}
Figure \ref{fg:f11}p
\end{flushright}
\end{figure}

\clearpage
\begin{figure}[p]
\plotone{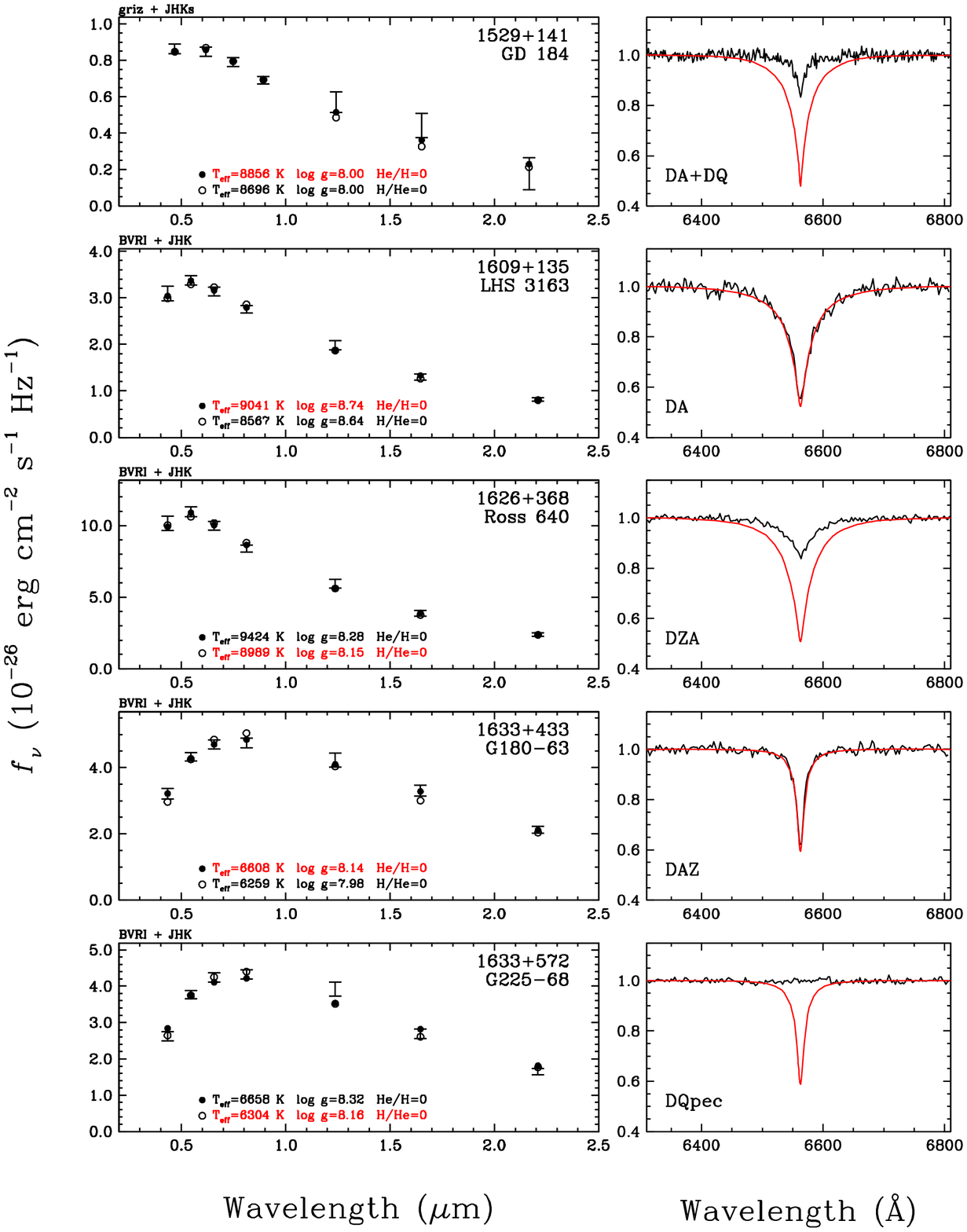}
\begin{flushright}
Figure \ref{fg:f11}q
\end{flushright}
\end{figure}

\clearpage
\begin{figure}[p]
\plotone{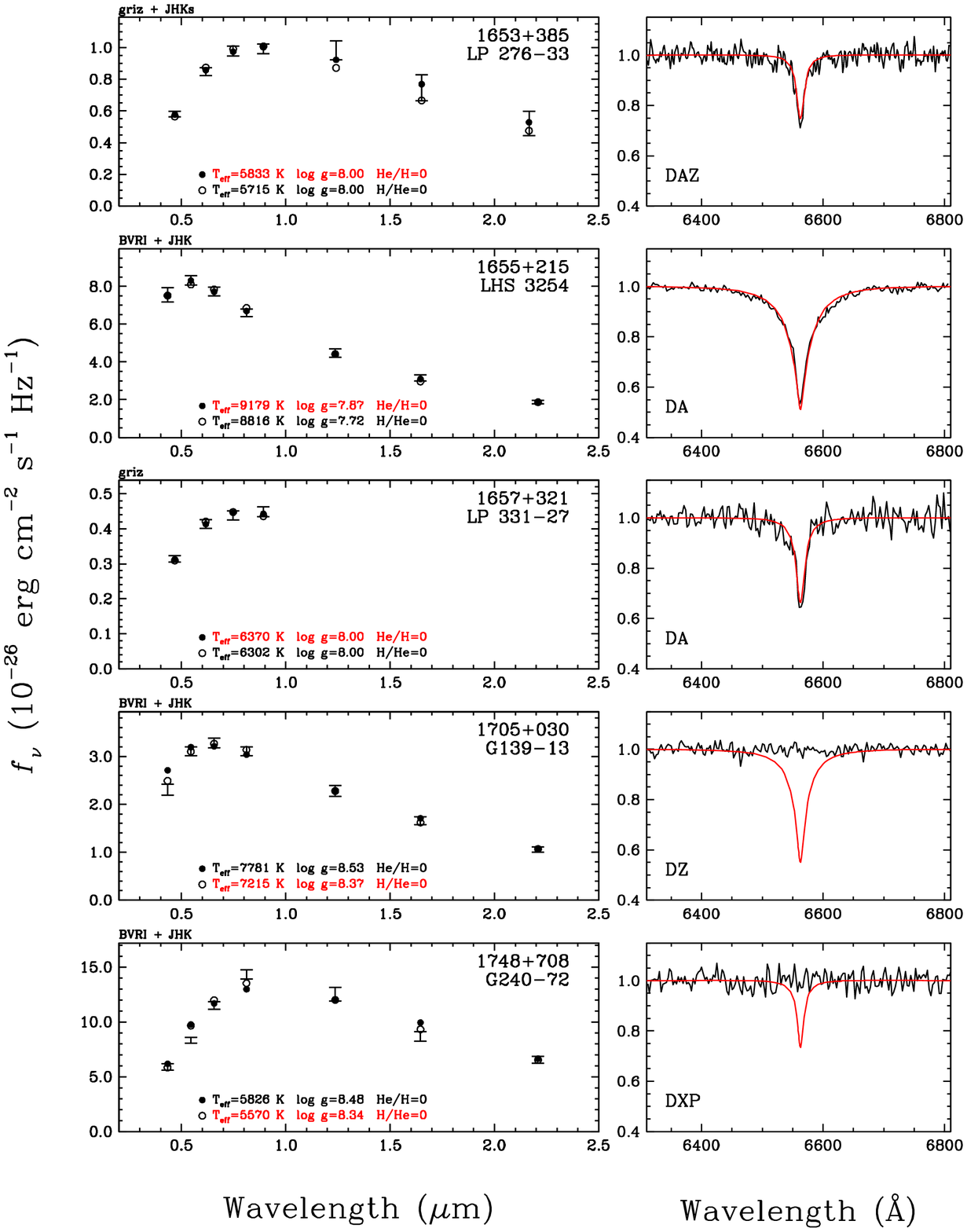}
\begin{flushright}
Figure \ref{fg:f11}r
\end{flushright}
\end{figure}

\clearpage
\begin{figure}[p]
\plotone{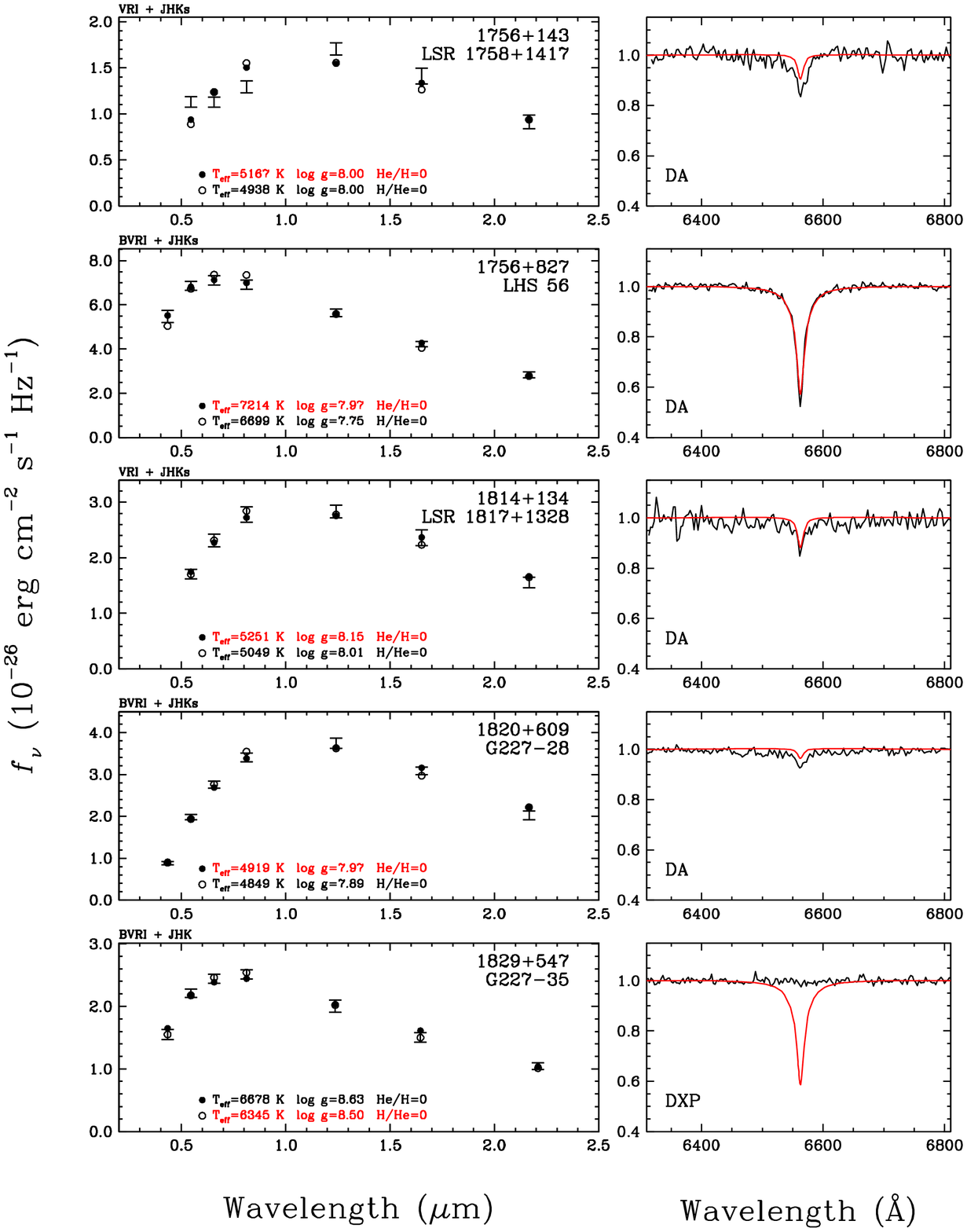}
\begin{flushright}
Figure \ref{fg:f11}s
\end{flushright}
\end{figure}

\clearpage
\begin{figure}[p]
\plotone{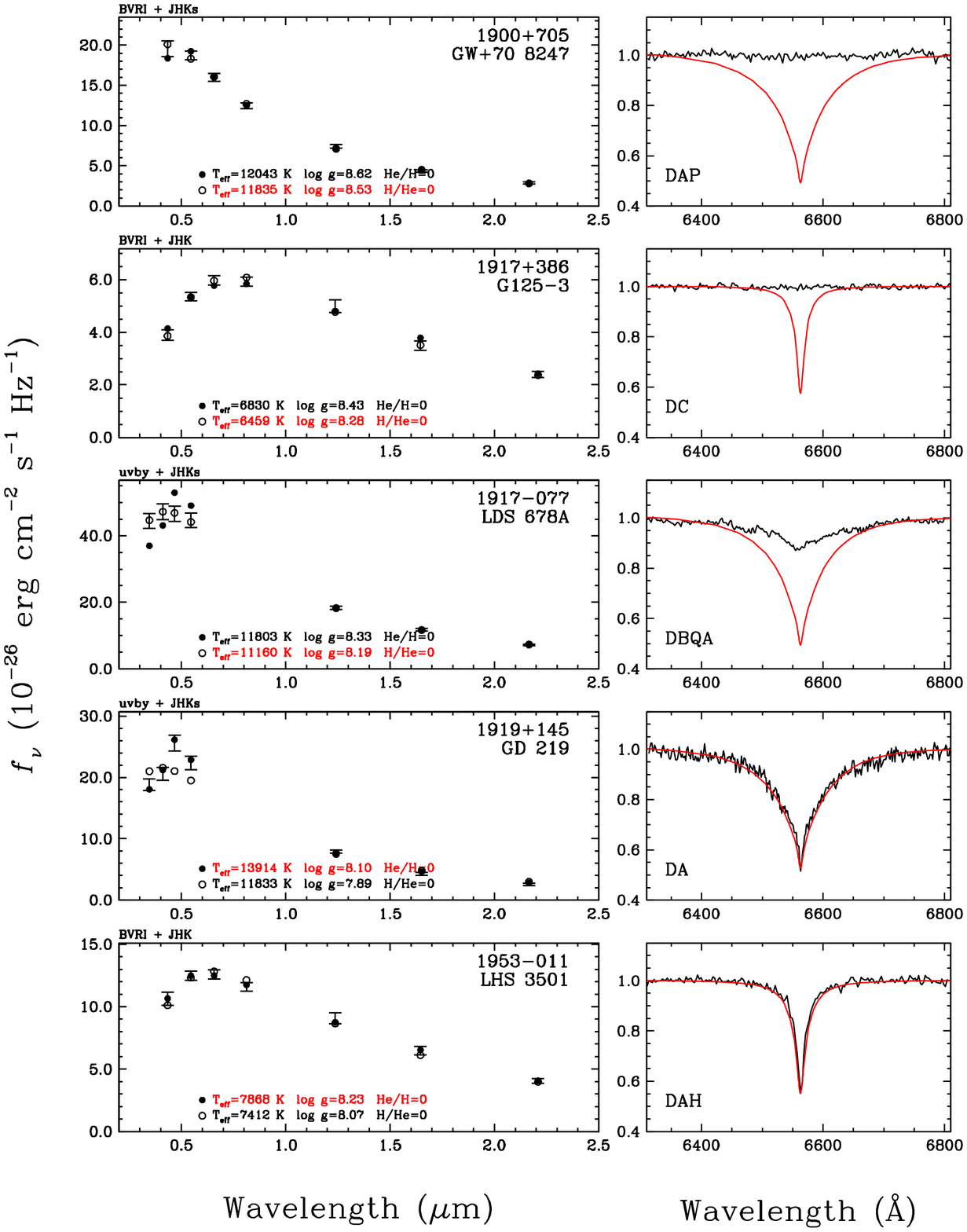}
\begin{flushright}
Figure \ref{fg:f11}t
\end{flushright}
\end{figure}

\clearpage
\begin{figure}[p]
\plotone{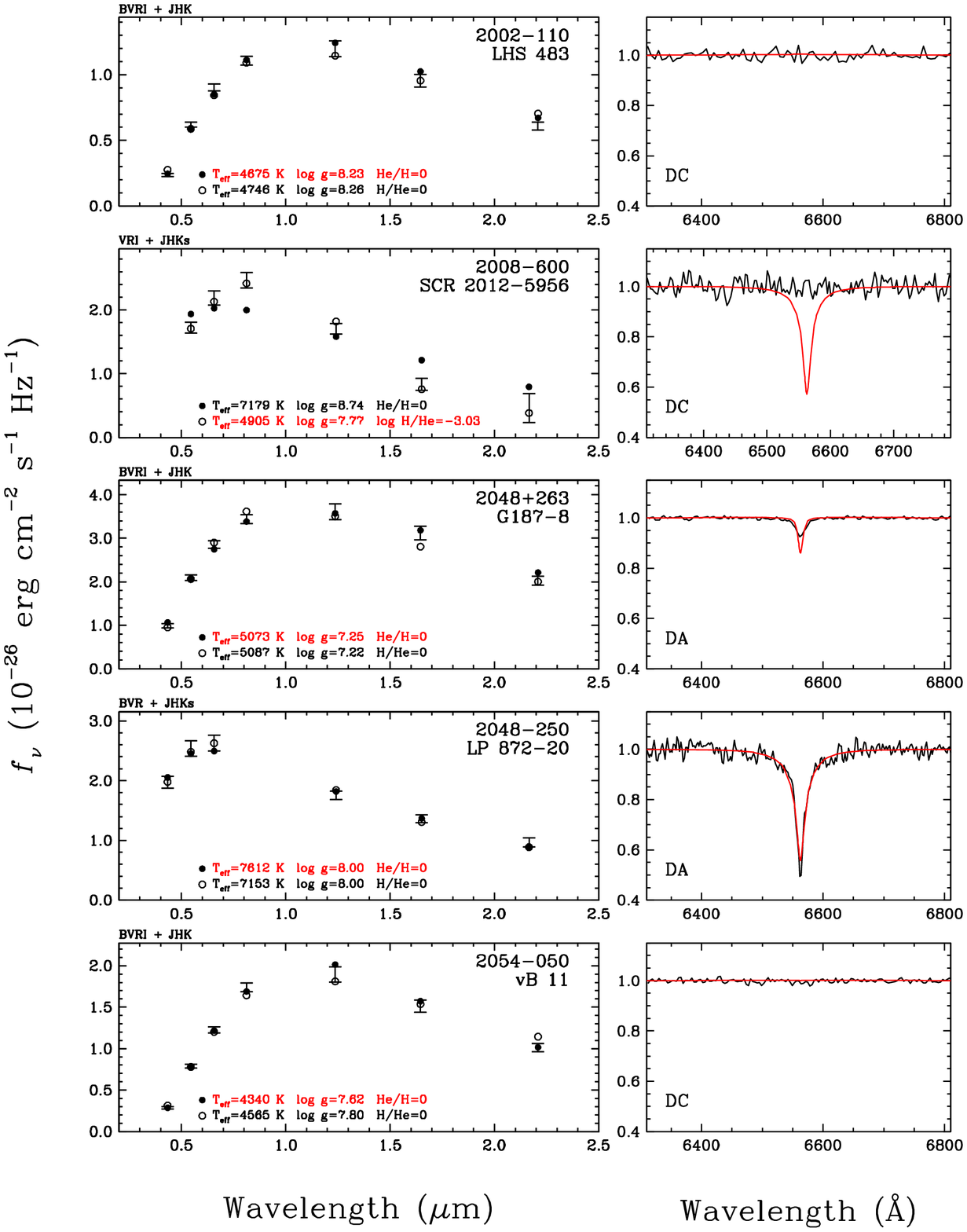}
\begin{flushright}
Figure \ref{fg:f11}u
\end{flushright}
\end{figure}

\clearpage
\begin{figure}[p]
\plotone{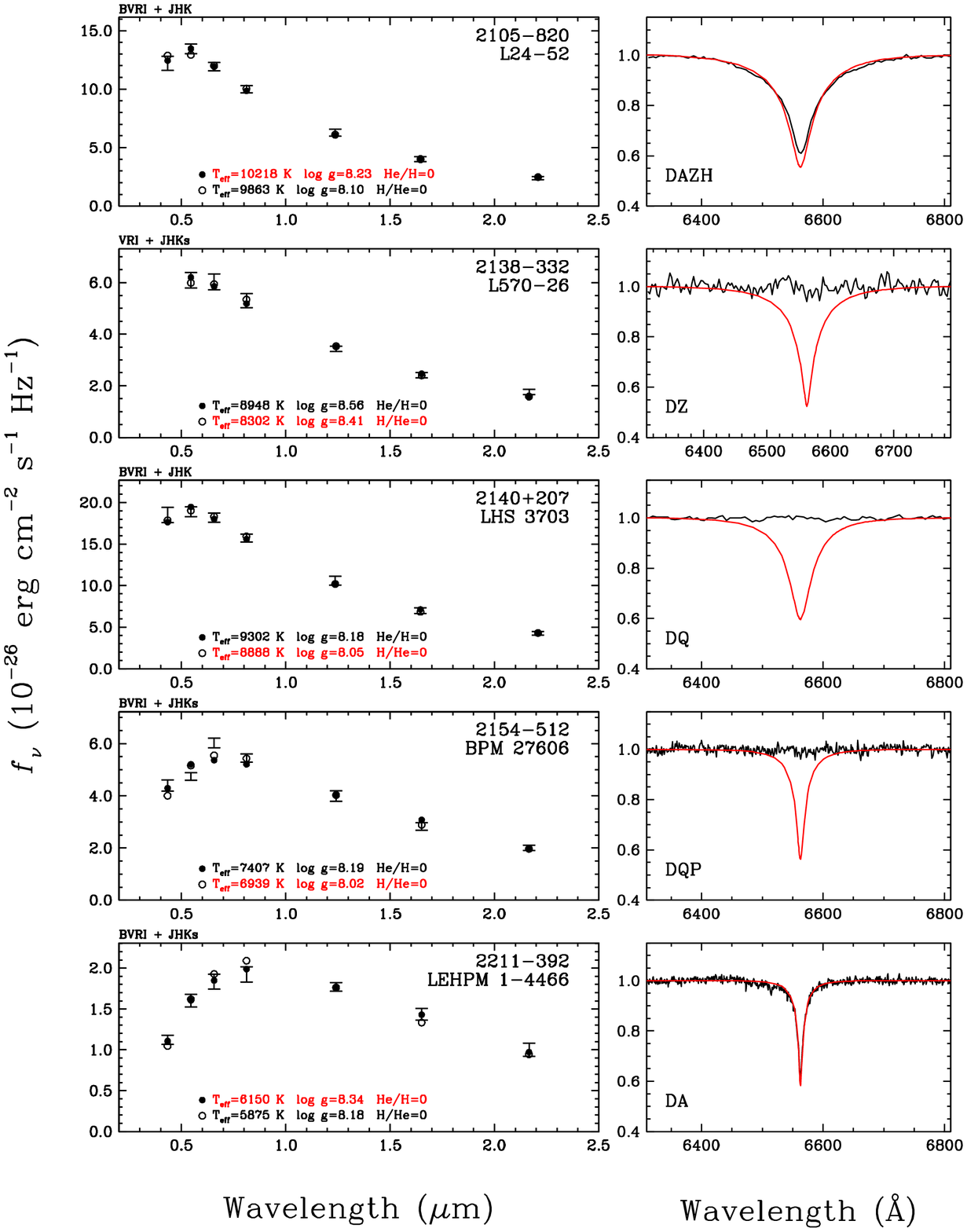}
\begin{flushright}
Figure \ref{fg:f11}v
\end{flushright}
\end{figure}

\clearpage
\begin{figure}[p]
\plotone{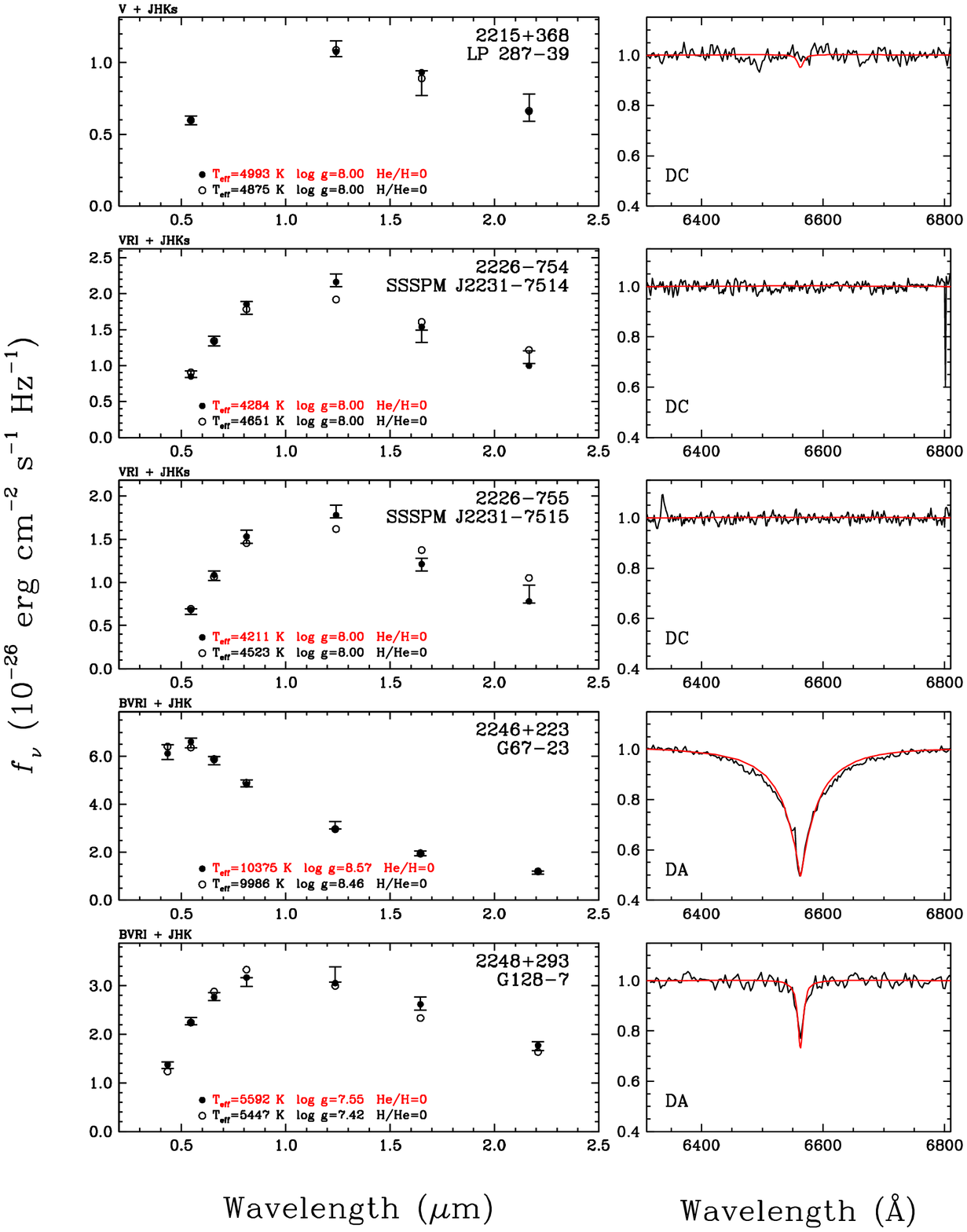}
\begin{flushright}
Figure \ref{fg:f11}w
\end{flushright}
\end{figure}

\clearpage
\begin{figure}[p]
\plotone{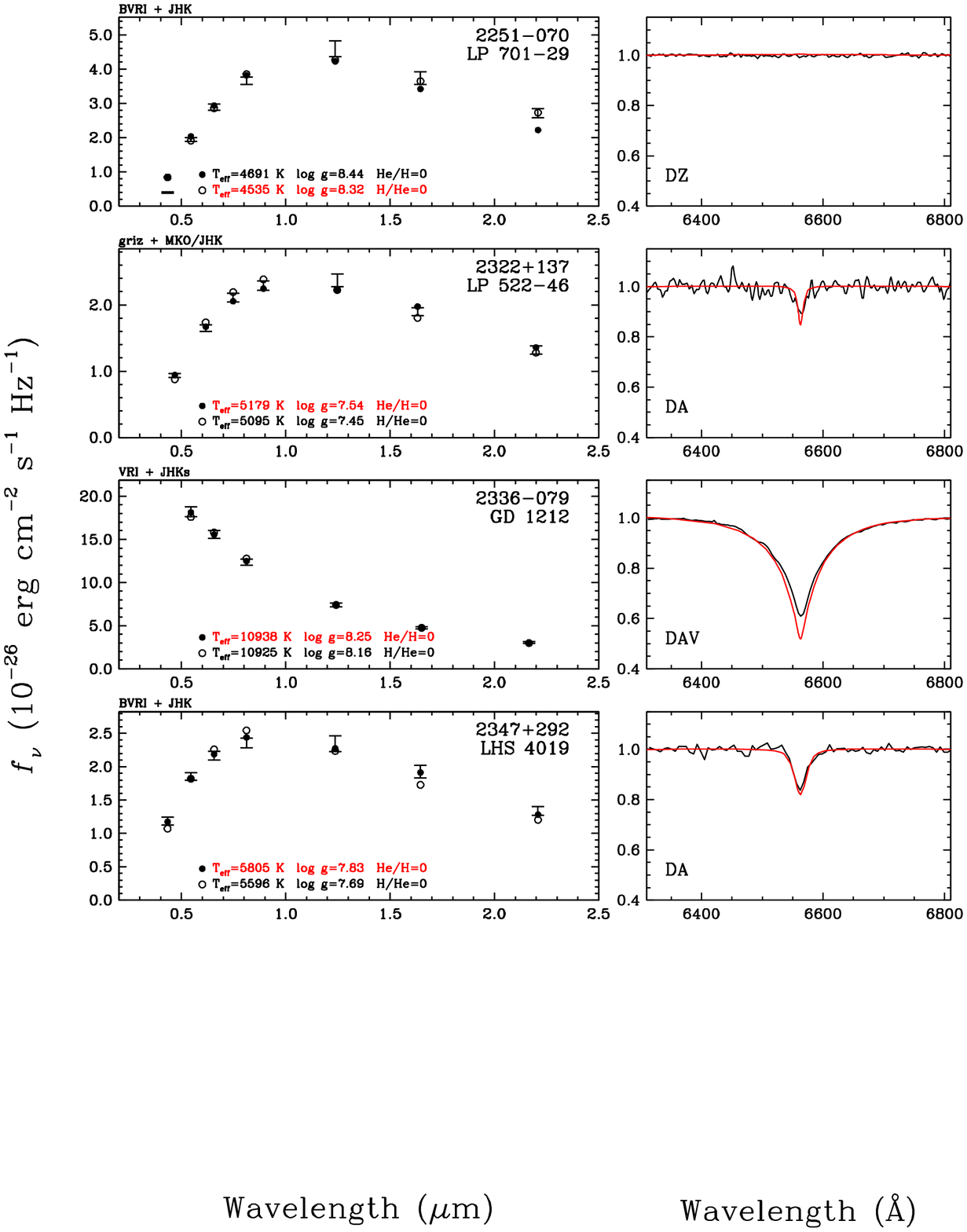}
\begin{flushright}
Figure \ref{fg:f11}x
\end{flushright}
\end{figure}

\clearpage
\begin{figure}[p]
\plotone{f12a}
\begin{flushright}
Figure \ref{fg:f12}a
\end{flushright}
\end{figure}

\clearpage
\begin{figure}[p]
\plotone{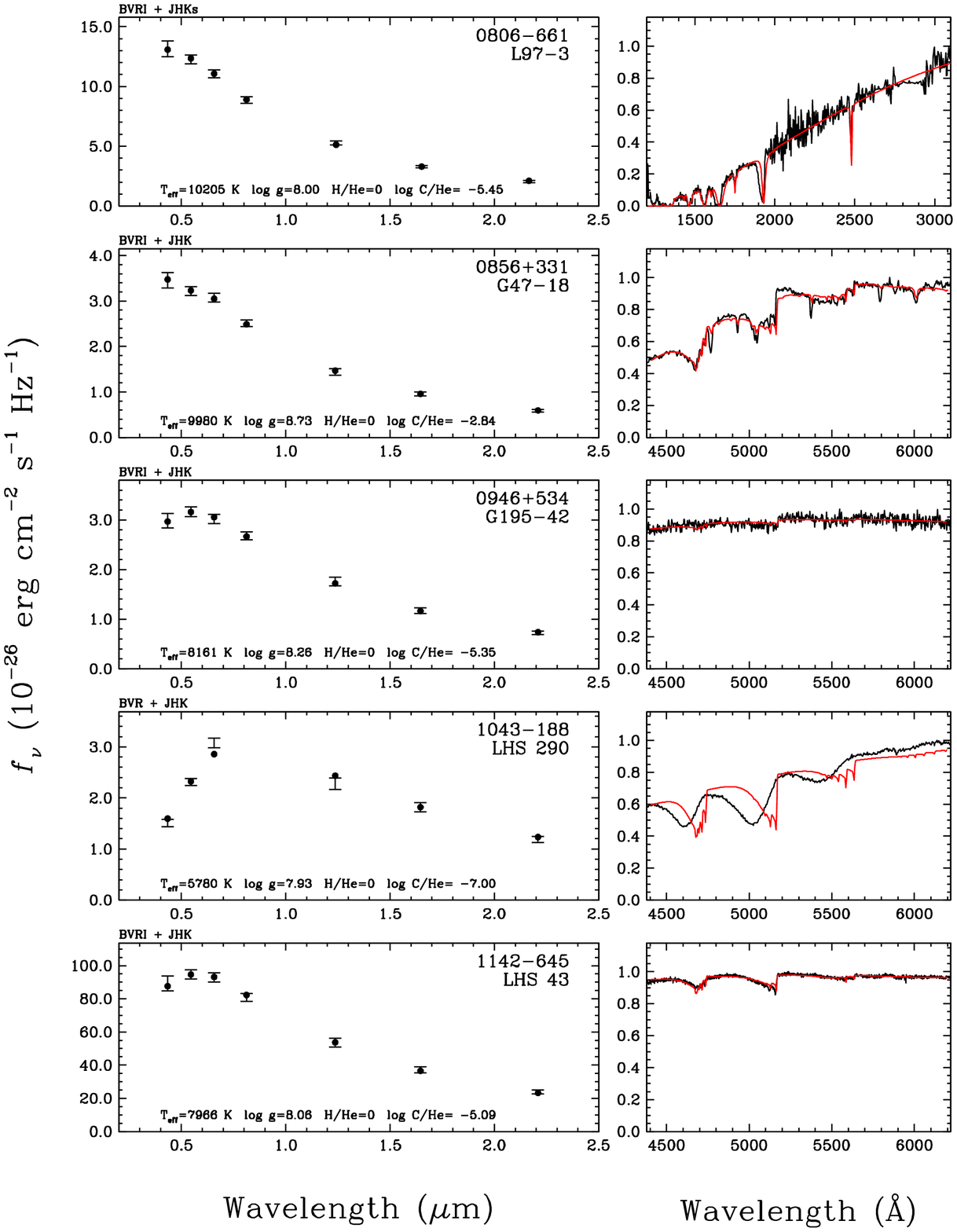}
\begin{flushright}
Figure \ref{fg:f12}b
\end{flushright}
\end{figure}

\clearpage
\begin{figure}[p]
\plotone{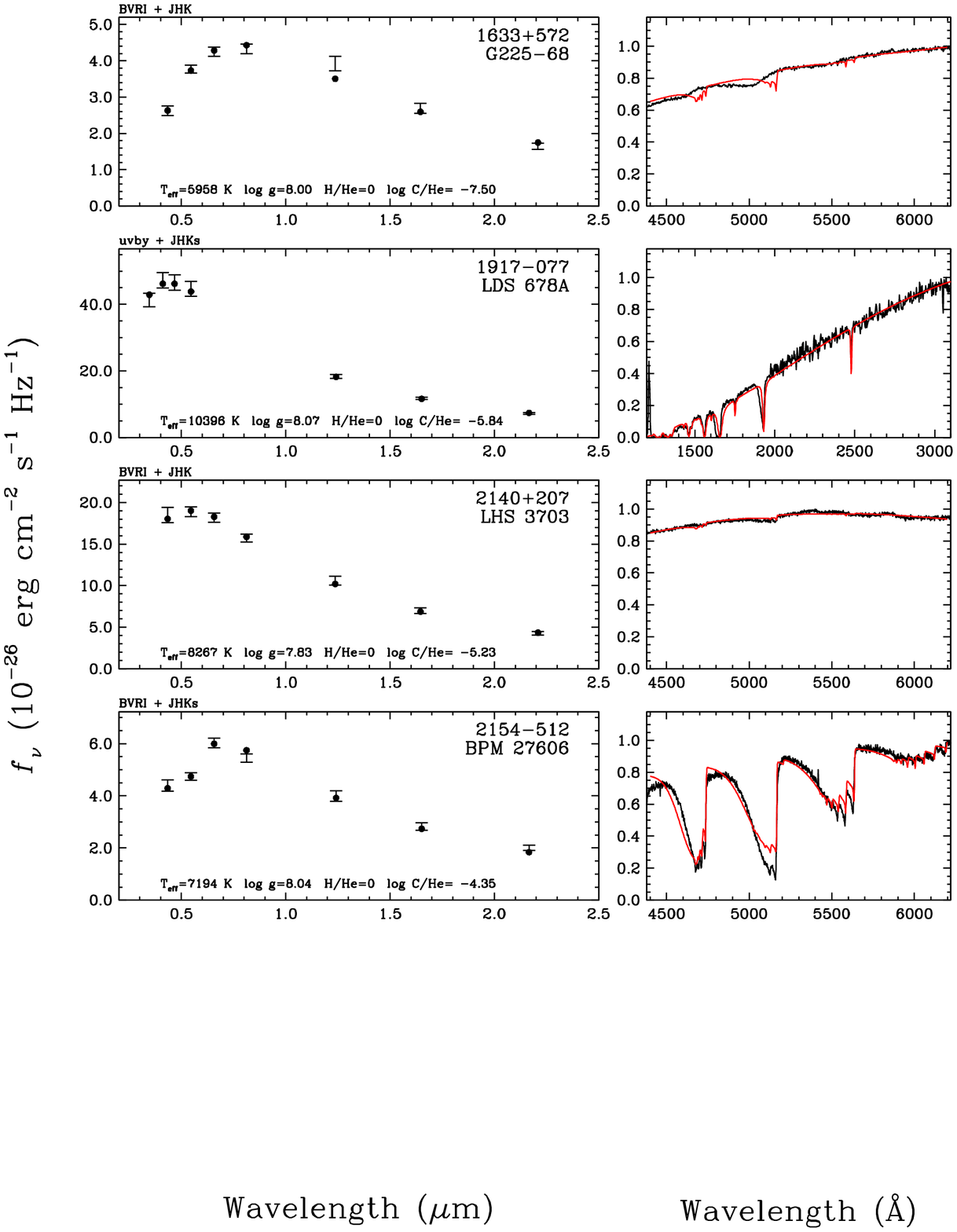}
\begin{flushright}
Figure \ref{fg:f12}c
\end{flushright}
\end{figure}

\clearpage
\begin{figure}[p]
\plotone{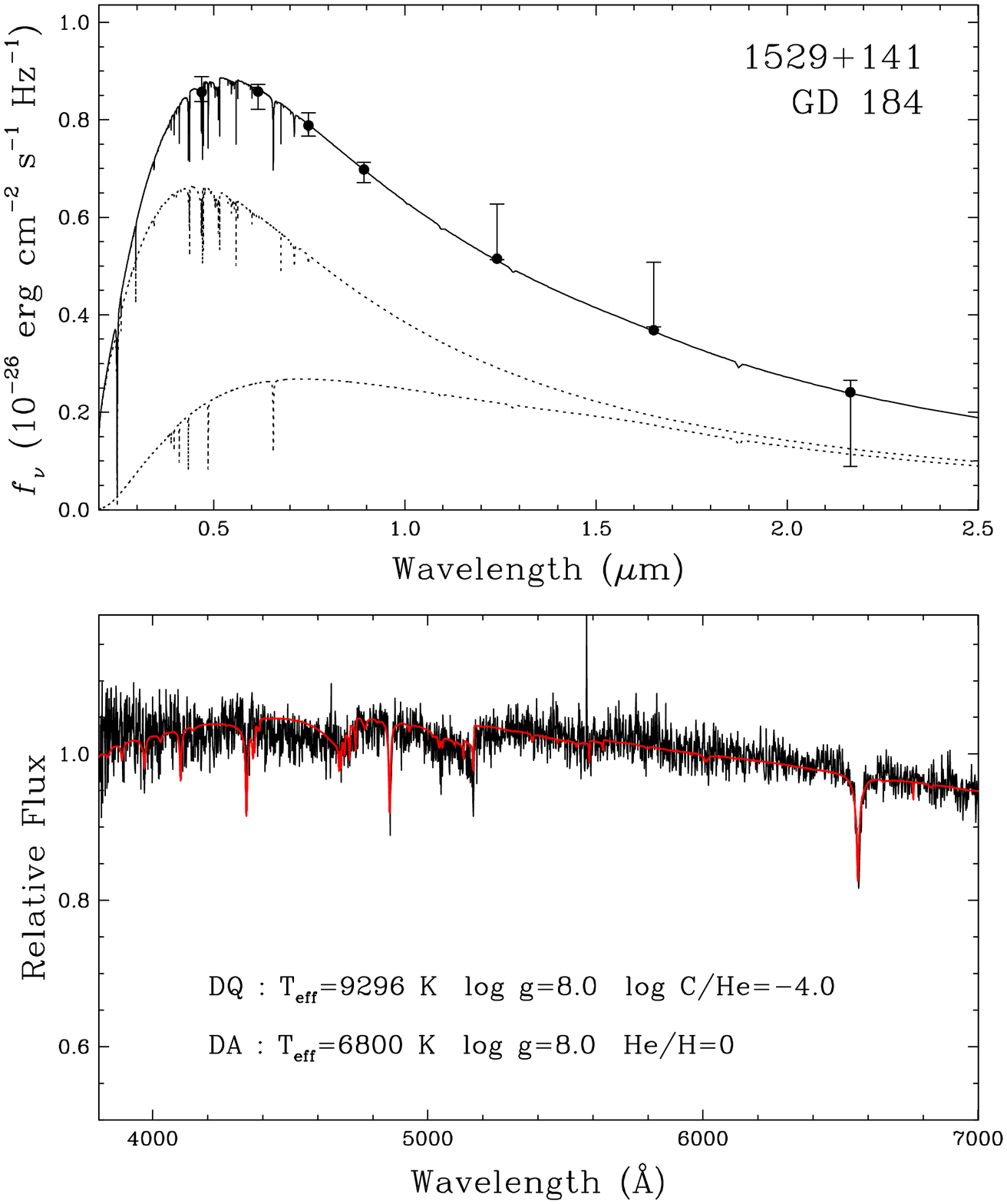}
\begin{flushright}
Figure \ref{fg:f13}
\end{flushright}
\end{figure}

\clearpage
\begin{figure}[p]
\plotone{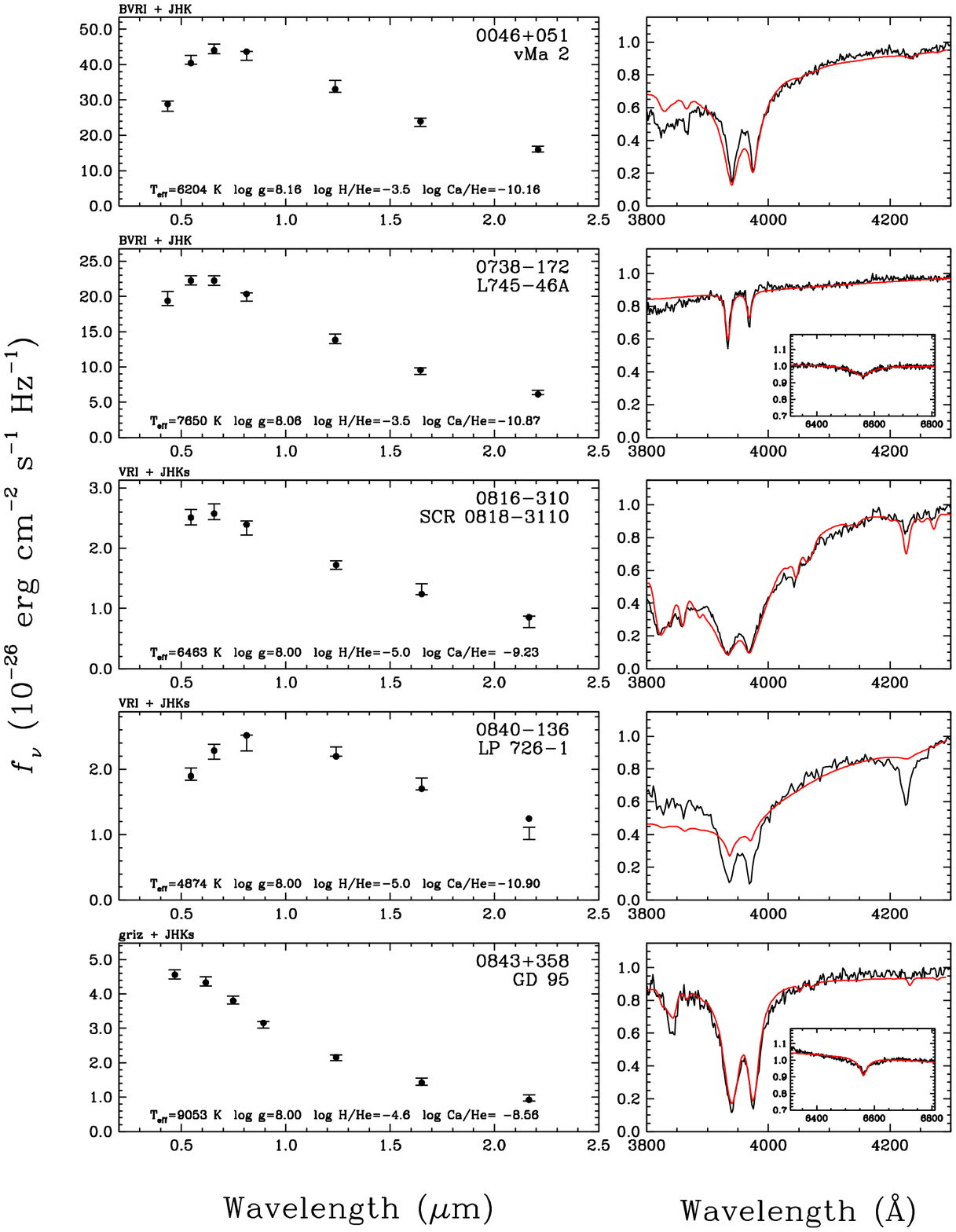}
\begin{flushright}
Figure \ref{fg:f14}a
\end{flushright}
\end{figure}

\clearpage
\begin{figure}[p]
\plotone{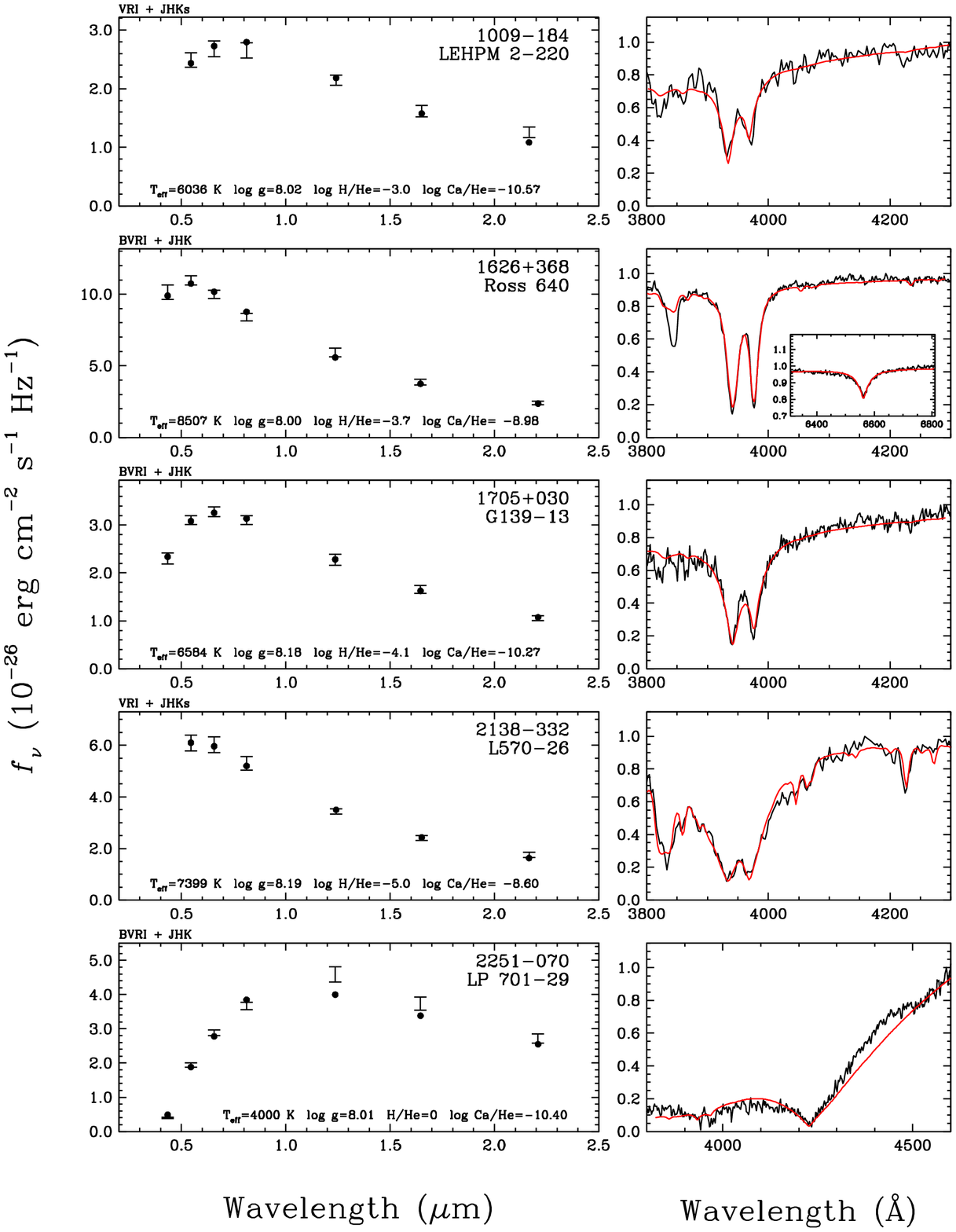}
\begin{flushright}
Figure \ref{fg:f14}b
\end{flushright}
\end{figure}

\clearpage
\begin{figure}[p]
\plotone{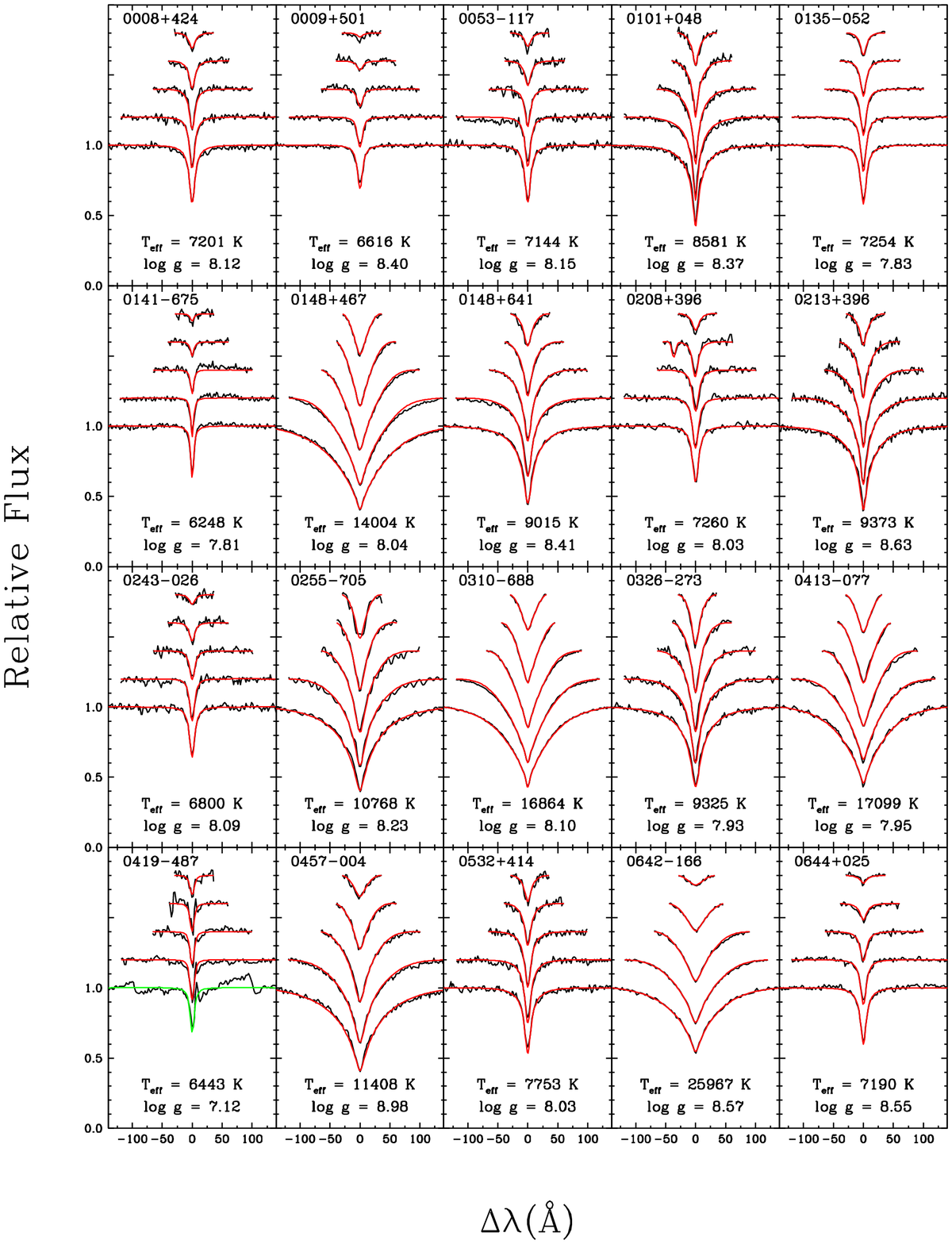}
\begin{flushright}
Figure \ref{fg:f15}a
\end{flushright}
\end{figure}

\clearpage
\begin{figure}[p]
\plotone{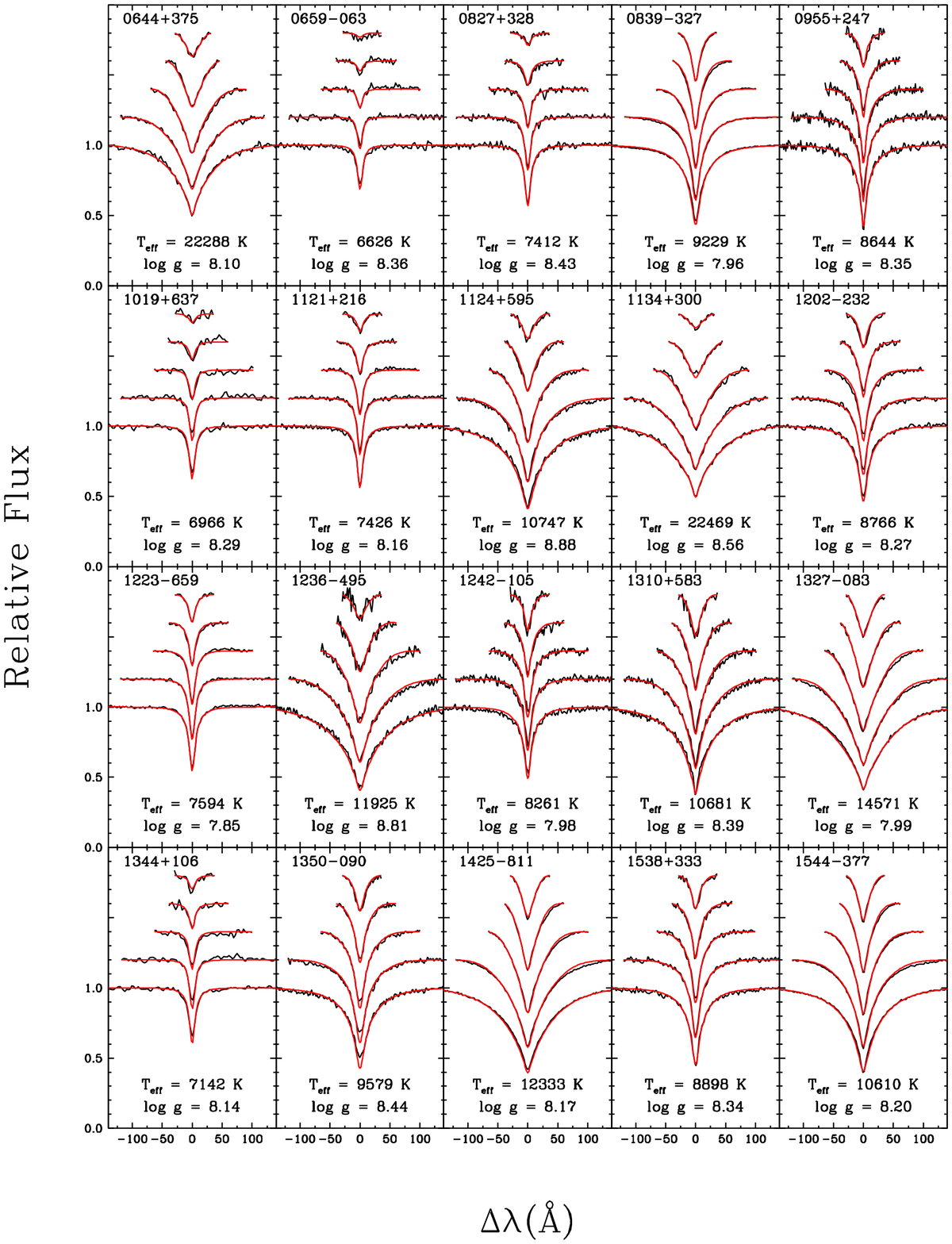}
\begin{flushright}
Figure \ref{fg:f15}b
\end{flushright}
\end{figure}

\clearpage
\begin{figure}[p]
\plotone{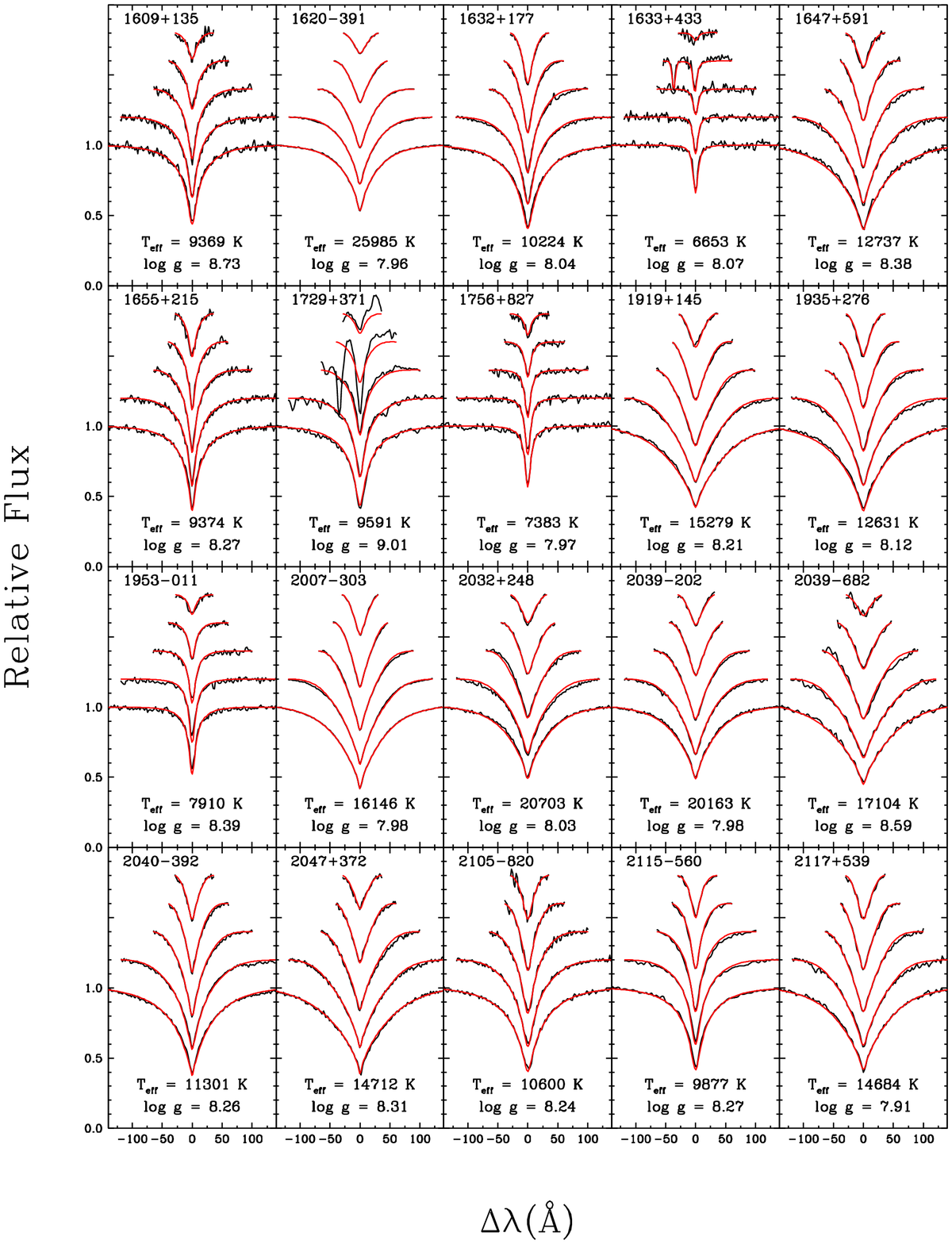}
\begin{flushright}
Figure \ref{fg:f15}c
\end{flushright}
\end{figure}

\clearpage
\begin{figure}[p]
\plotone{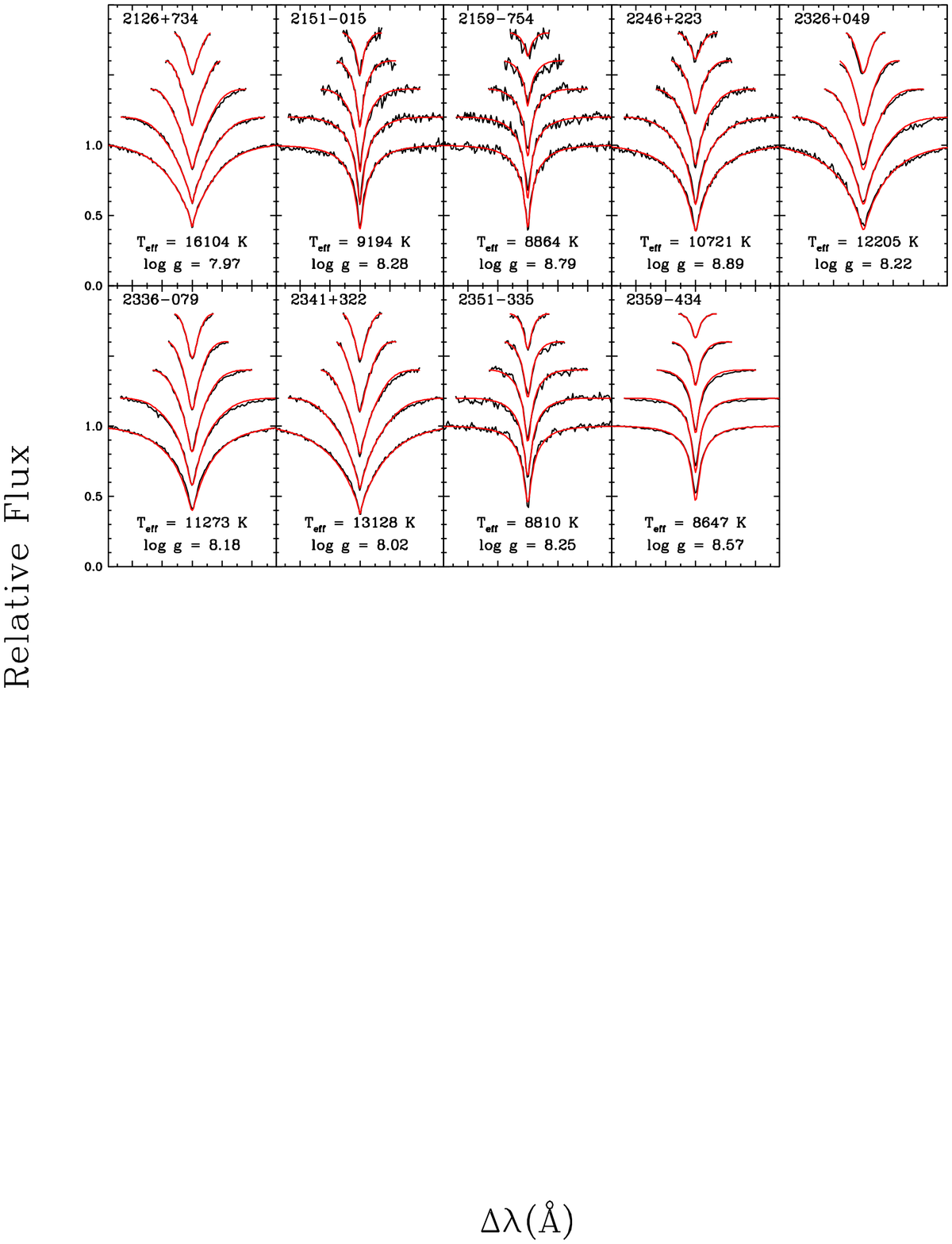}
\begin{flushright}
Figure \ref{fg:f15}d
\end{flushright}
\end{figure}

\clearpage
\begin{figure}[p]
\plotone{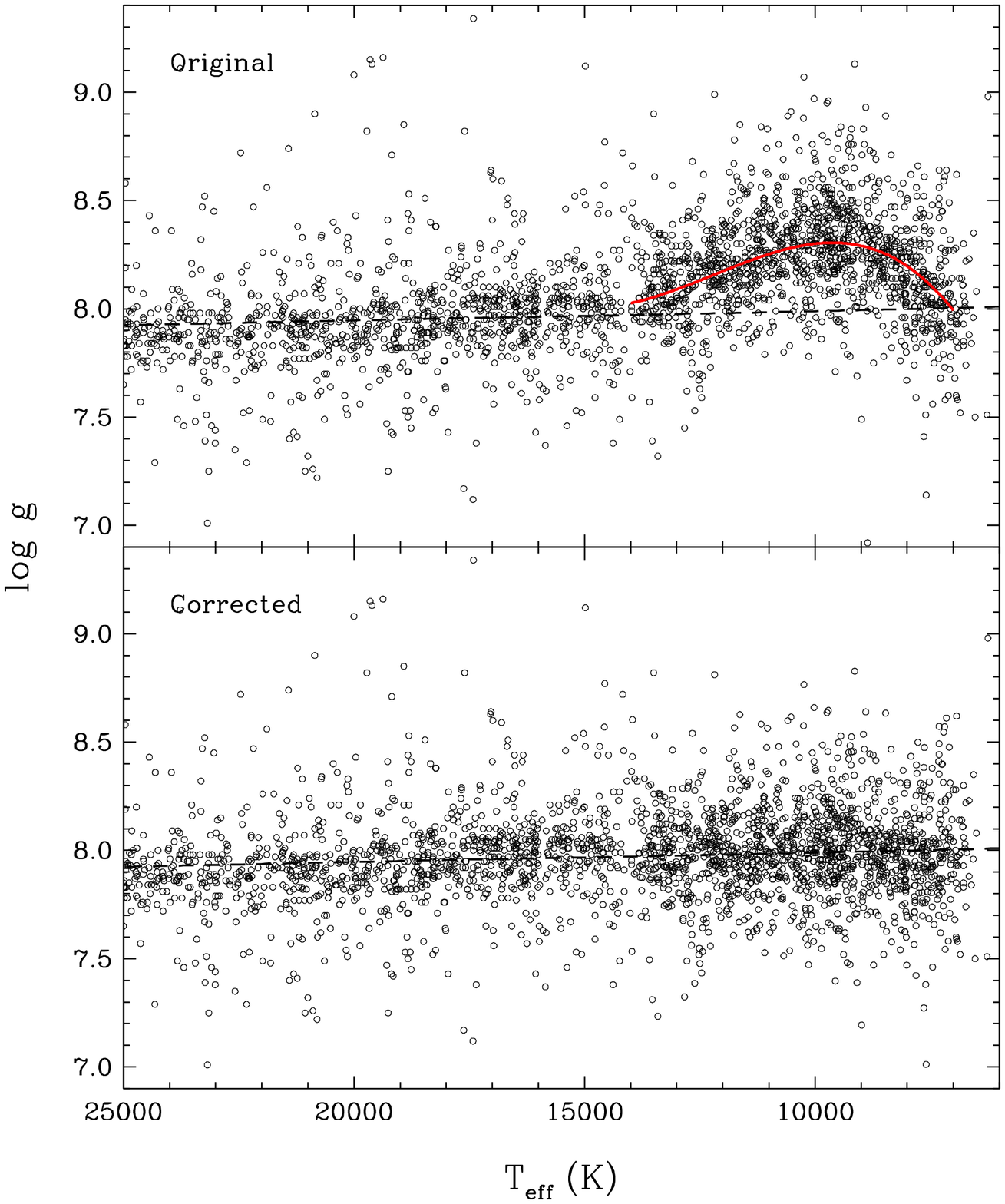}
\begin{flushright}
Figure \ref{fg:f16}
\end{flushright}
\end{figure}

\clearpage
\begin{figure}[p]
\plotone{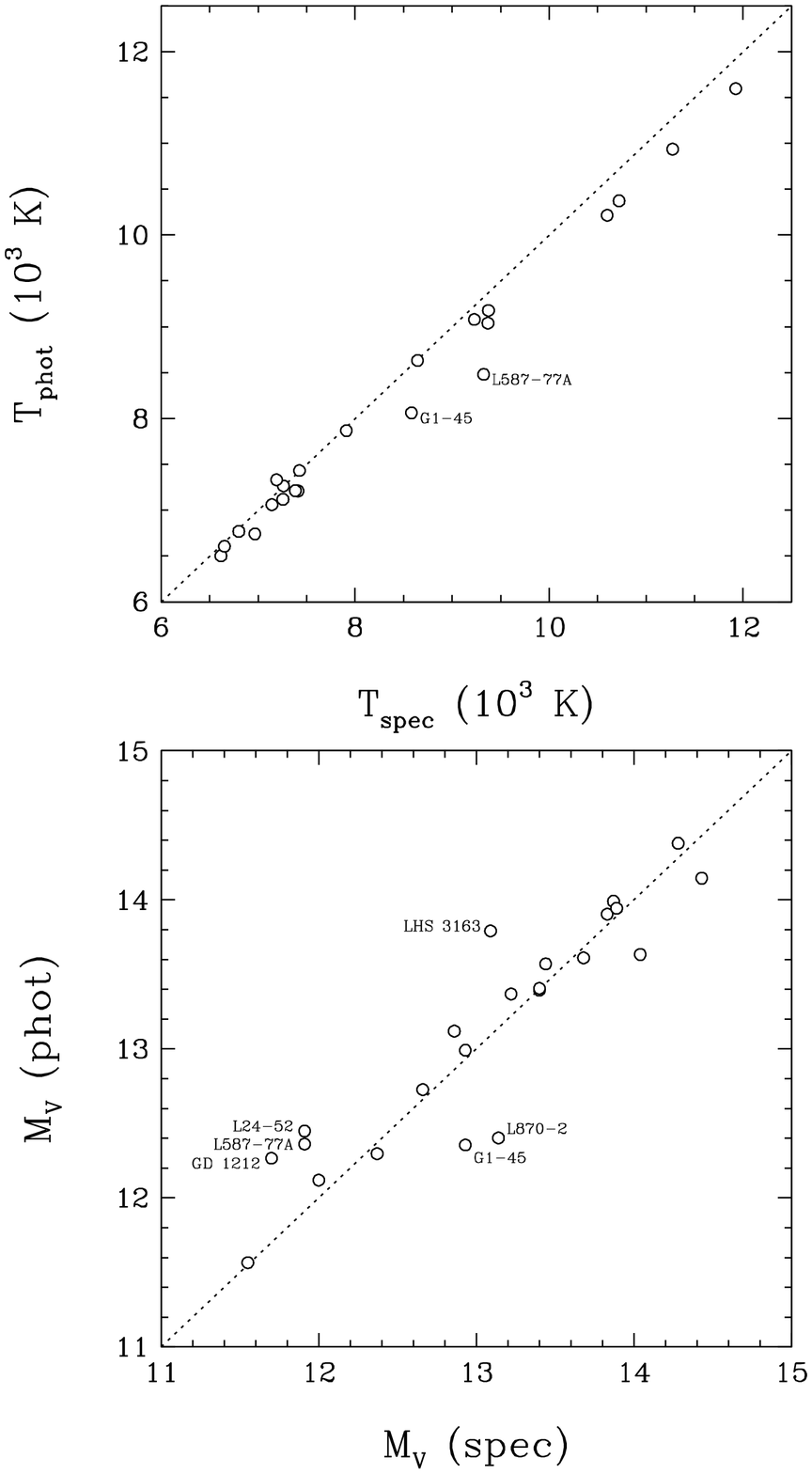}
\begin{flushright}
Figure \ref{fg:f17}
\end{flushright}
\end{figure}

\clearpage
\begin{figure}[p]
\plotone{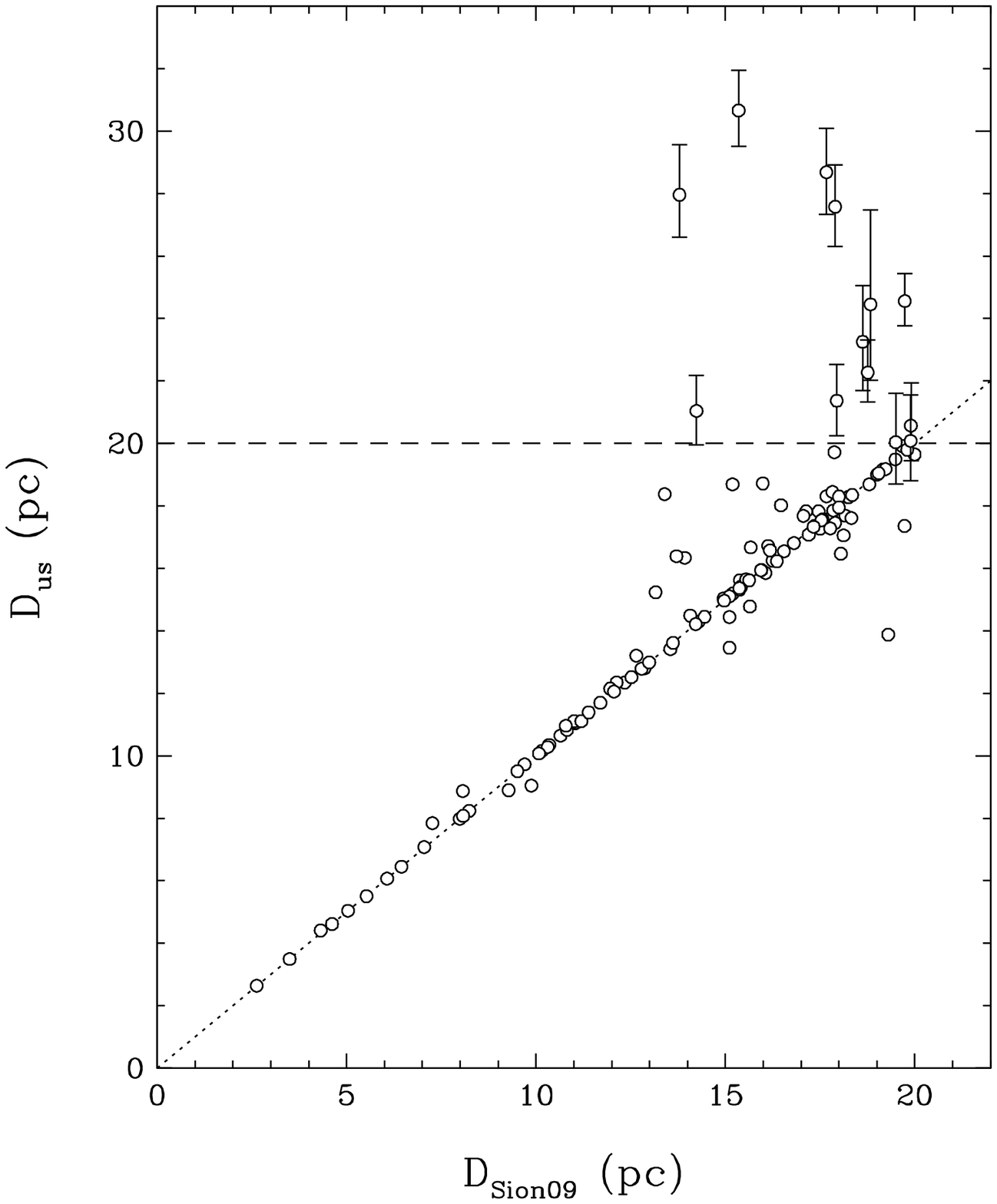}
\begin{flushright}
Figure \ref{fg:f18}
\end{flushright}
\end{figure}

\clearpage
\begin{figure}[p]
\plotone{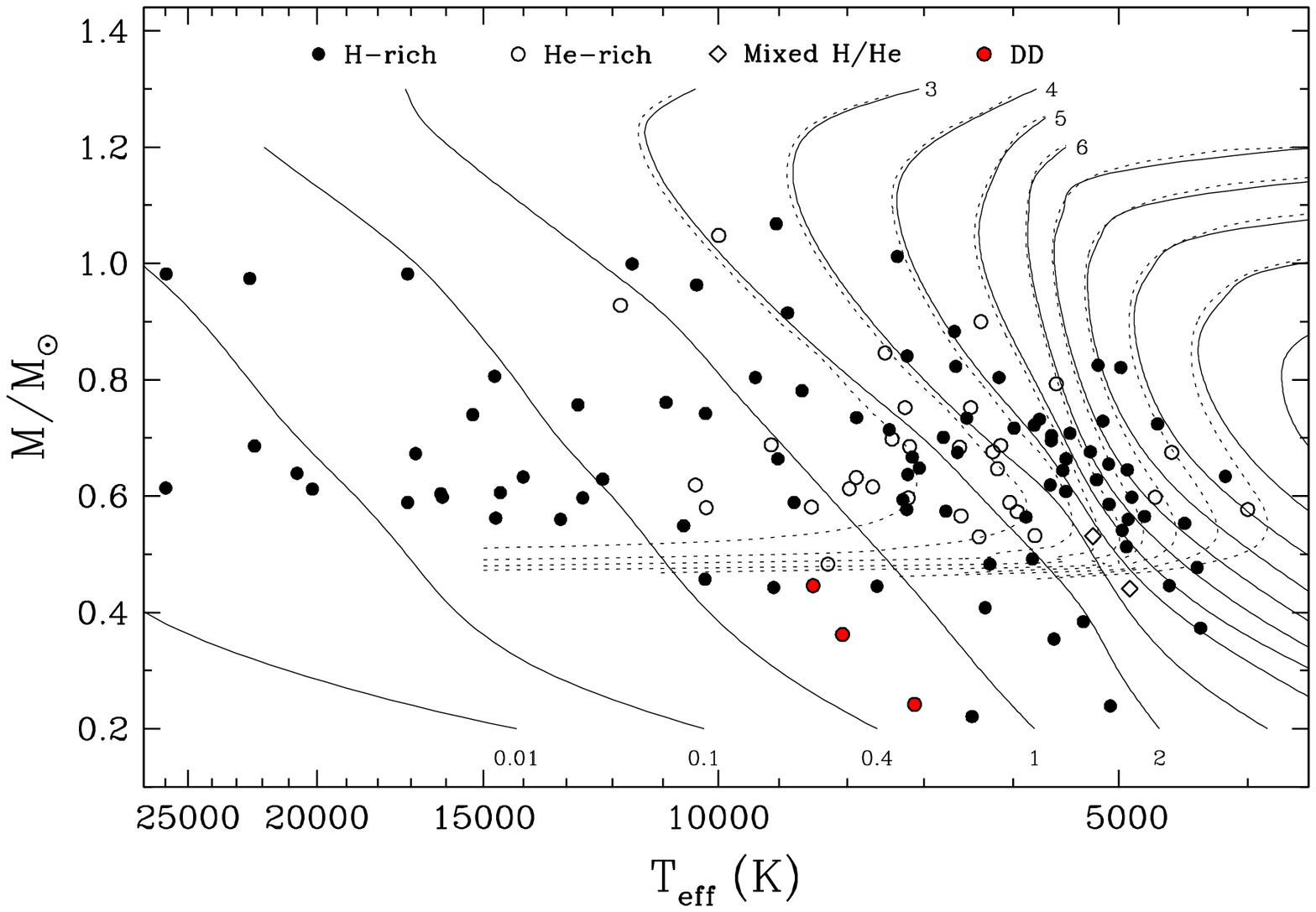}
\begin{flushright}
Figure \ref{fg:f19}
\end{flushright}
\end{figure}

\clearpage
\begin{figure}[p]
\plotone{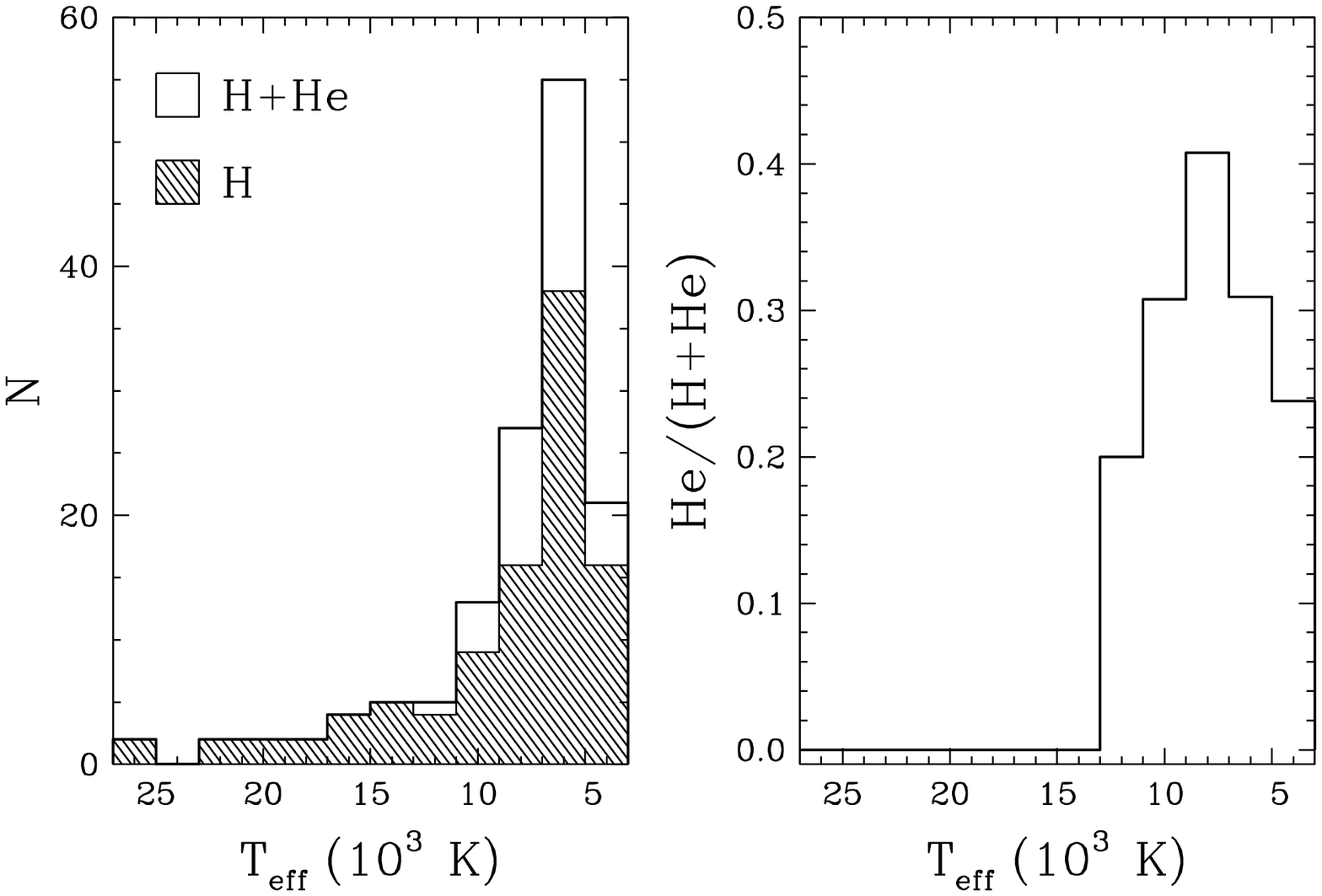}
\begin{flushright}
Figure \ref{fg:f20}
\end{flushright}
\end{figure}

\clearpage
\begin{figure}[p]
\plotone{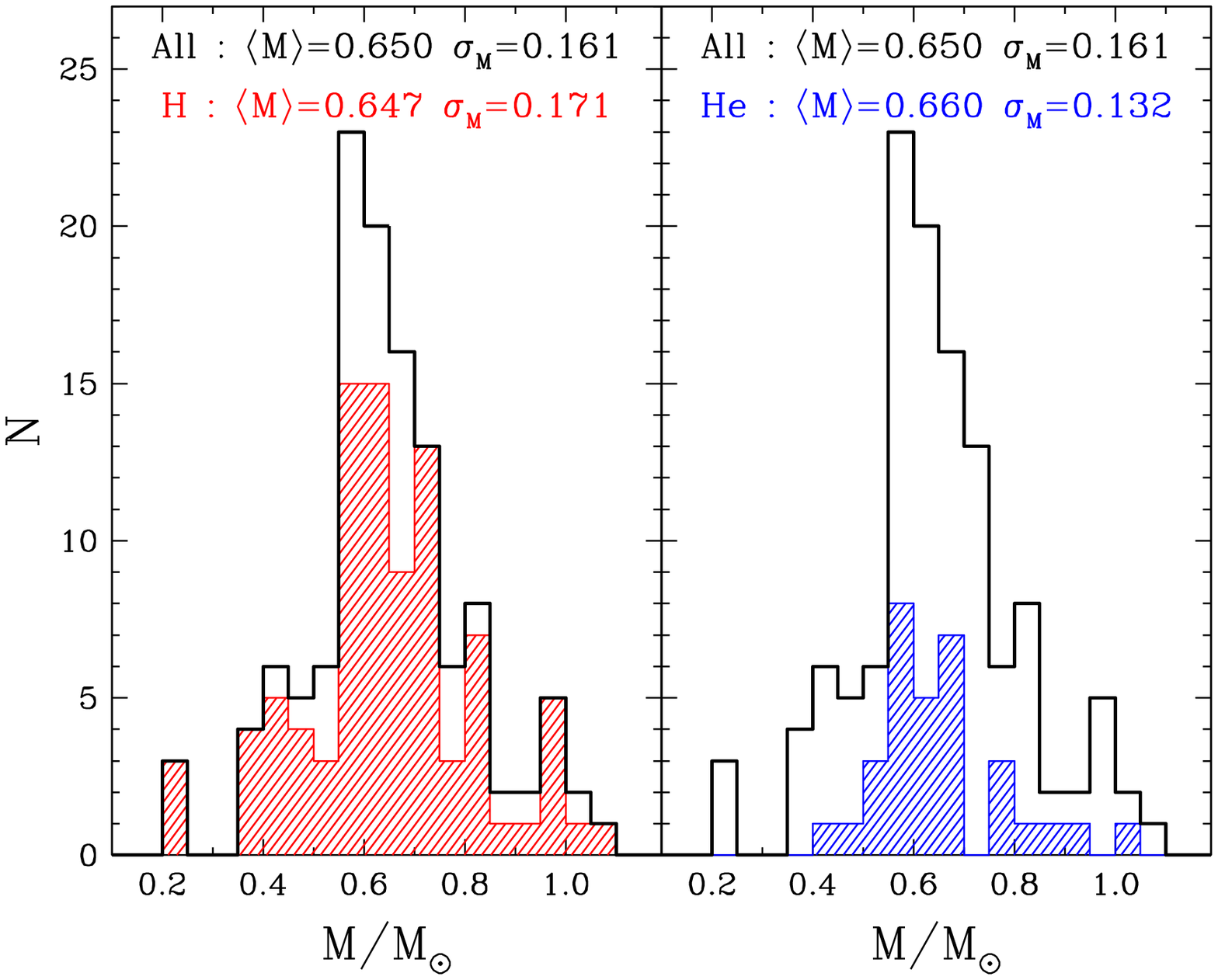}
\begin{flushright}
Figure \ref{fg:f21}
\end{flushright}
\end{figure}

\clearpage
\begin{figure}[p]
\plotone{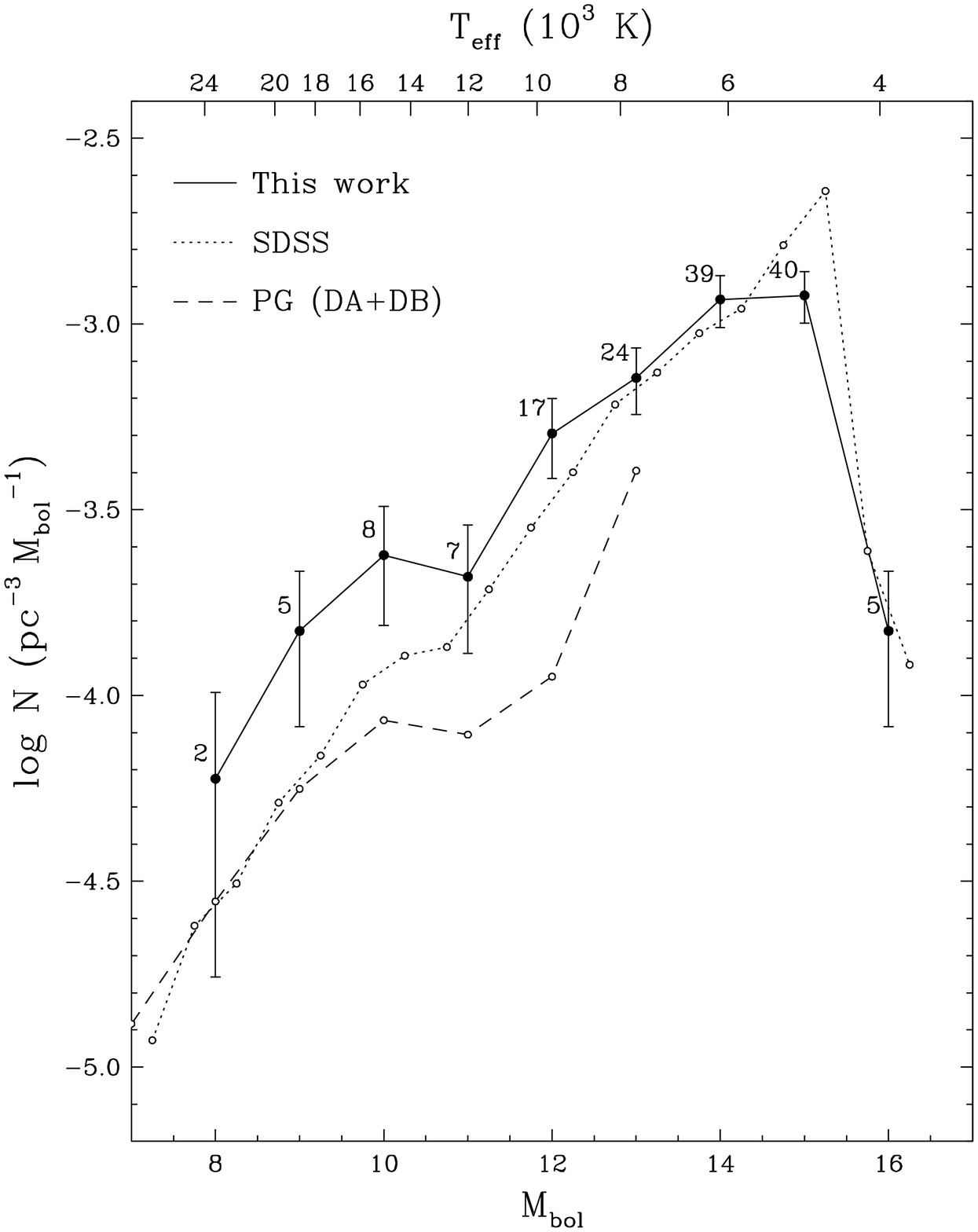}
\begin{flushright}
Figure \ref{fg:f22}
\end{flushright}
\end{figure}

\end{document}